\documentclass[useAMS,usenatbib,twocolumn]{mn2e}
\pdfoutput=1
\usepackage{amssymb}
\usepackage{amsmath}
\usepackage{natbib}
\usepackage[pdftex]{graphicx}
\usepackage{subfigure}
\usepackage{booktabs}

\begin{document}

\title[Waves and tidal dissipation in a solar-type star]
{On internal wave breaking and tidal dissipation near the centre of a solar-type star}
      \author[A.J. Barker \& G.I. Ogilvie]{Adrian J. Barker\thanks{E-mail:
	  ajb268@cam.ac.uk} and Gordon I. Ogilvie \\
	Department of Applied Mathematics and Theoretical
	Physics, University of Cambridge, Centre for Mathematical Sciences, \\ Wilberforce Road,
	Cambridge CB3 0WA, UK}

\date{Accepted 2010 January 21.  Received 2010 January 18; in original form 2009 December 14}

\pagerange{\pageref{firstpage}--\pageref{lastpage}} \pubyear{2009}

\maketitle

\label{firstpage}

\begin{abstract}
  We study the fate of internal gravity waves approaching the centre
  of an initially non-rotating solar-type star, primarily using
  two-dimensional numerical simulations based on a cylindrical model.
  A train of internal gravity waves is excited by tidal forcing at the
  interface between the convection and radiation zones of such a star.
  We derive a Boussinesq-type model of the central region of a star
  and find a nonlinear wave solution that is steady in the frame
  rotating with the angular pattern speed of the tidal forcing. We
  then use spectral methods to integrate the equations numerically,
  with the aim of studying at what amplitude the wave is subject to
  instabilities. These instabilities are found to lead to wave
  breaking whenever the amplitude exceeds a critical value. Below this
  critical value, the wave reflects perfectly from the centre of the
  star. Wave breaking leads to mean flow acceleration, which
  corresponds to a spin up of the central region of the star, and the
  formation of a critical layer, which acts as an absorbing barrier
  for subsequent ingoing waves. As these waves continue to be absorbed
  near the critical layer, the star is spun up from the inside out.

  Our results point to an important amplitude dependence of the
  (modified) tidal quality factor $Q^{\prime}$, since nonlinear effects
  are responsible for dissipation at the centre of the star. If the
  amplitude of the tidal forcing exceeds the critical amplitude for wave
  breaking to occur, then this mechanism produces efficient dissipation
  over a continuous range of tidal frequencies. This requires
  $\left(\frac{m_{p}}{M_{J}}\right)\left(\frac{P}{1
      \;\mathrm{day}}\right)^{\frac{1}{6}} \gtrsim 3.3$, for a planet of
  mass $m_{p}$ in an orbit of period $P$ around the current Sun,
  neglecting stellar rotation. However, this criterion depends
  strongly on the strength of the stable stratification at the centre
  of the star, and so it depends on stellar mass and main-sequence
  age.  If breaking occurs, we find $Q^{\prime} \approx 1.5\times
  10^{5} \left(\frac{P}{1 \; \mathrm{day}}\right)^{\frac{8}{3}}$, for
  the current Sun.  This varies by no more than a factor of 5
  throughout the range of solar-type stars with masses between
  $0.5-1.1 M_{\odot}$, for fixed orbital parameters.  This estimate of
  $Q^{\prime}$ is therefore quite robust, and can be reasonably
  considered to apply to all solar-type main-sequence stars, if this
  mechanism operates. The strong frequency dependence of the resulting
  dissipation means that this effect could be very important in
  determining the fate of close-in giant planets around G and K
  stars. We predict fewer giant planets with orbital periods of less
  than about 2 days around such stars if they cause breaking at the
  centre, due to the efficiency of this process. 
  
  Even if the waves are of too low amplitude to initiate breaking,
  radiative damping could, in principle, lead to a gradual spin-up of
  the stellar centre and to the formation of a critical layer. This
  process could provide efficient tidal dissipation in solar-type stars
  perturbed by less massive companions, but it may be prevented by
  effects that resist the development of differential rotation.
  
  These mechanisms would, however, be ineffective in stars with a
  convective core, such as WASP-18, WASP-12 and OGLE-TR-56, perhaps
  partly explaining the survival of their close planetary companions.  
\end{abstract}


\section{Introduction}

Tidal interactions are thought to be important in determining the
fate of short-period extrasolar planets and the spins of
their host stars, as well as in the sychronization and circularization
of close binary
stars. The extent of spin-orbit evolution that results from tides
depends on the dissipative properties of the bodies involved in the
interaction. It is standard to
parametrize our uncertainties in the mechanisms of tidal dissipation
in each body by defining a tidal quality factor $Q$, which is an
inverse measure of the dissipation. This is usually defined to be
proportional to the ratio
of the maximum energy stored in a tidal oscillation ($E_{0}$) to the energy dissipated
over one cycle, i.e.
\begin{eqnarray}
Q = 2\pi E_{0} \left(\oint -\dot E \,\mathrm{d}t\right)^{-1},
\end{eqnarray}
and we find it convenient to use the variant $Q^{\prime} =
\frac{3Q}{2k}$, where $k$ is the second--order potential Love number
of the body. This combination always appears together in the evolutionary
equations\footnote{$Q^{\prime}$ reduces to $Q$ for a homogeneous fluid body,
where $k=\frac{3}{2}$.}.

$Q^{\prime}$ is difficult to calculate from first principles in fluid
bodies, and uncertainties in the mechanisms of tidal dissipation
remain. The tidal disturbance can generally be decomposed into two
parts: an equilibrium tide and a dynamical tide. The equilibrium tide
is the quasi-hydrostatic ellipsoidal tidal bulge. In the frame
corotating with the fluid, the time-dependence of the equilibrium tide
is dissipated through its interaction with turbulent convection,
though the damping rate is uncertain, particularly when the convective
time exceeds the tidal period (\citealt{Zahn1966}; \citealt{Goldreich1977};
\citealt{GoodmanOh1997}; \citealt{Penev2007}). The dynamical tide
consists of internal waves that are excited by low-frequency tidal
forcing, and has received much recent interest with regard to its
possible contribution to $Q^{\prime}$ (\citealt{WS2002}; \citealt{Gio2004}, hereafter OL04;
\citealt{Wu2005b}; \citealt{PapIvanov2005}; \citealt{IvanovPap2007};
\citealt{Gio2007}, hereafter OL07; \citealt{GoodmanLackner2009}). This
is because if these waves have short wavelength, then they are more
easily damped than the large-scale equilibrium tide by radiative
diffusion (\citealt{Zahn1975}; \citealt{Zahn1977}), convective
viscosity \citep{Terquem1998} or nonlinear breaking
(\citealt{GoodmanDickson1998}; hereafter GD98).

The tidal frequency is typically much lower than the dynamical
frequency of the body, so the relevant internal waves must be
approximately incompressible, restored not by pressure but by buoyancy
or rotation. OL04 found that the
dissipation of tidally excited inertial waves, whose restoring force
is the Coriolis force, can
contribute significantly to the dissipation rate in a giant planet,
whose interior is mostly convective (see also \citealt{Wu2005b} and 
\citealt{IvanovPap2007}). These waves can also contribute to the
dissipation rate in convective regions of stars (OL07). They are
excited by tidal forcing of frequency $\hat{\omega}$, if
this is less than the Coriolis frequency ($2\Omega$), and this is true
for many astrophysically relevant circumstances. However,
these waves are not excited if the tidal frequency exceeds the
Coriolis frequency, so this process is then not effective at dissipating
the tide, and contributing to $Q^{\prime}$. In
this work we concentrate on waves that have buoyancy as the restoring
force, and which are commonly referred to as internal gravity waves
(hereafter IGWs)\footnote{Though note that they are often referred to
  as g-modes, or simply gravity waves, in the literature.}.

IGWs have been proposed to account for the efficient tidal dissipation
that has been inferred from the circularisation of early-type binary stars
(\citealt{Zahn1975}; \citealt{Zahn1977}; \citealt{Zahn2008};
\citealt{SavPap1983}; \citealt{PapSav1985}; \citealt{SavPap1995}; \citealt{SavPap1997};
\citealt{PapSav1997}), which
are massive enough to have a
convective core and an exterior radiative envelope. In these stars,
IGWs are excited near the boundary between these two regions, where
the buoyancy frequency (or Brunt-V\"ais\"al\"a frequency, see
\S~\ref{IGWtheory}) matches the tidal forcing frequency. These
propagate outwards into the stably stratified radiation zone, towards
the surface, where they are fully or partially damped by radiative
diffusion. In this
picture, these stars are tidally synchronized from the outside in, since
angular momentum is deposited in the regions of the star
where these waves damp (\citealt{Goldreich1989}a,b). 

The above model only works for stars with an exterior radiation zone, 
which is unlike that of the Sun and other stars of solar type, which have
radiative cores and convective envelopes. It is of particular
interest to study the efficiency of tidal dissipation in these stars,
since many have been found to harbour close-in planets, whose survival
is determined by the stellar $Q^{\prime}$. This is because a planet
with an orbital period shorter than the stellar spin period is
subject to tidally induced orbital decay, with an inspiral time
that depends on dissipation in the star (e.g. \citealt{Barker2009}; OL07)
In these stars, a train of IGWs are
again excited at the interface between the convective and radiative
zones, but here they propagate towards the stellar centre. If they can
coherently reflect from the centre, global standing modes can form in
the radiation zone. In this case, tidal dissipation is efficient only
when the tidal frequency matches that of a global standing mode
(which are commonly referred to as $g$-modes) (\citealt{Terquem1998};
\citealt{SW2002}). 
This would not contribute
appreciably to $Q^{\prime}$ because the system would evolve rapidly
through these resonances, unless resonance locking occurs
(\citealt{WS1999}; \citealt{WS2001}). On the other hand, if these
waves do not reflect coherently from the centre, and are either strongly
dissipated there, or are reflected with a perturbed phase, then
efficient dissipation is possible over a broad range of tidal
frequencies (GD98; OL07). The extent of
nonlinearity in the waves near the centre is likely to be the factor
that determines whether these waves reflect coherently, and this is
controlled by the amplitude and frequency of the tidal forcing, as well as the
properties of the stellar centre (OL07).

In this paper we study the problem of IGWs approaching the centre of
a solar-type star, primarily using two-dimensional numerical simulations. We first
derive a Boussinesq-type system of equations appropriate for the
stellar centre, which are ideal for integrating numerically using
spectral methods. An exact solution for tidally forced waves is
derived, and some of its properties are discussed. Our numerical
set-up is described and results are presented for both linear and
nonlinear forcing amplitudes, including an analysis of the
reflection coefficient and a study of the growth of different
azimuthal wavenumbers in the disturbance. This is followed by a
discussion of the results, especially their relevance to $Q^{\prime}$
for solar-type stars, and to the survival of close-in giant planets in
orbit around such stars.

\section{Internal gravity waves: elementary properties, wave breaking
  and critical layers}
\label{IGWtheory}

IGWs are a family of dispersive waves that are ubiquitous in
nature. They propagate in any
fluid with a stable density stratification, due to the restoring force
of buoyancy. Their influence can be observed in the oceans and atmosphere of the
Earth on a range of spatial and temporal scales, from the visual undulations
of striated cloud structures, to the complex interplay between these
waves and shearing flows, which produces the large-scale Quasi-Biennial
Oscillation in the equatorial stratosphere. It is widely recognised
that IGWs play a prominent role
in the transport of energy and angular momentum in geophysical and
astrophysical flows (\citealt{Buhler2009}; \citealt{Perspectives2000}; 
\citealt{Rogers2006}; \citealt{Kumar1999}). IGWs are also thought to
be important in stably stratified radiation
zones of stars. When excited by turbulent convection, they were at one
stage put forward as potential explanations for maintaining the solid body 
rotation of the radiative interior of the Sun
(\citealt{Schatzman1993}; \citealt{Zahn1997}). However, it was pointed
out that the ``antidiffusive'' nature of IGWs tends to enhance
local shear rather than reduce it \citep{Gough1998}. IGWs are
still thought to
produce angular velocity variations in the radiation zone \citep{Rogers2006}. They have also
been invoked to explain the Li depletion problem in F-stars \citep{Spruit1991}, affecting
solar neutrino production \citep{Press1981}, and possibly having an effect on the
solar cycle \citep{Kumar1999}.

Observations of oscillations on the solar surface are able to provide
information about the interior properties of the Sun \citep{Helio2002}.
IGWs in the radiative interior of the Sun can form global standing modes,
commonly referred to as $g$-modes, if their frequency matches that of a
free mode of oscillation. These are known to have their amplitude
largest close to the centre (see solution in \S~\ref{lineartheory}) 
and would therefore seem ideal probes of the deep
interior. Unfortunately for observers, the standing $g$-modes are effectively trapped 
in the radiative interior, where the stratification is stable, and are evanescent in the
convection zone, and so are unlikely to be visible at the solar
surface. Nevertheless, modes of sufficiently low degree, with high
enough amplitude, may have already been observed at the surface by
\cite{Garcia2007}, though it must be noted that thus far there is no
undisputed evidence for observations of $g$-modes
\citep{QuestSolargmodes2009}. 

The frequencies of the largely incompressible internal waves lie in
ranges controlled by the buoyancy frequency (or Brunt-V\"ais\"al\"a
frequency) $N$ and the Coriolis frequency $2\Omega$. The square of the
buoyancy frequency in a spherically symmetric star is defined by
\begin{eqnarray}
N^{2}(r) = -\frac{1}{\rho}\frac{dp}{dr}\left(\frac{1}{\Gamma_{1}}\frac{d \ln p}{d r} -\frac{d \ln\rho}{dr}\right),
\end{eqnarray}
where $\rho, p$ are the density and pressure, and $\Gamma_{1} =
\left(\frac{\partial \ln p}{\partial \ln \rho}\right)_{s}$ (at constant
specific entropy $s$). The local dispersion relation for linear noncompressive internal waves
in a fluid body rotating with angular velocity $\mathbf{\Omega}$ is
\begin{eqnarray}
\label{dispersionrelation}
\omega^{2} = N^{2}\sin^{2}\alpha + 4\Omega^{2}\cos^{2}\beta,
\end{eqnarray}
where $\alpha$ is the angle between the wavevector $\mathbf{k}$ and the
gravitational acceleration $\mathbf{g}$, and $\beta$ is the angle
between $\mathbf{k}$ and $\mathbf{\Omega}$. The
frequency of these waves is independent of wavelength (in the absence
of viscosity or thermal conduction), and only depends on the direction
of the wavevector. This is
different from waves whose restoring force is due to
compressibility, whose frequency is inversely proportional to
wavelength. When $N=0$, Eq.\ref{dispersionrelation} describes inertial
waves, which have frequencies in the range $(0,2\Omega)$. If the body
is non-rotating ($\mathbf{\Omega} = 0$), then these waves are IGWs,
and possess frequencies in the range $(0,N)$. In the presence of nonzero
$\mathbf{\Omega}$ and $N$, these waves are intermediate
between inertial waves and IGWs and are referred to as inertia-gravity
waves. For waves in a spherical star, at a given latitude there is a
minimum frequency for inertia-gravity wave propagation. 
Near the equator, waves can propagate with arbitrarily low
frequency. From here on we neglect the
bulk rotation, and assume that $\mathbf{\Omega} = 0$, i.e., we consider only IGWs. The local dispersion
relation for IGWs can be rewritten
\begin{eqnarray}
\label{dispersionrelation}
\omega^{2} = N^{2}\frac{k^{2}_{h}}{k^{2}_{r} + k^{2}_{h}},
\end{eqnarray}
where $k_{h}$ and $k_{r}$ are the horizontal and radial
wavenumbers.

The phase and group velocities of these waves can be calculated from
Eq.~\ref{dispersionrelation} to give
\begin{eqnarray}
\label{phasegroupvelocities}
\mathbf{c}_{p} &=& \frac{\omega}{|\mathbf{k}|}\frac{\mathbf{k}}{|\mathbf{k}|} = \frac{Nk_{h}}{(k^{2}_{r}
    + k^{2}_{h})^{\frac{3}{2}}}\left(k_{r}\mathbf{e}_{r}
    + k_{h}\mathbf{e}_{h}\right) \\ 
&\approx&
    \frac{Nk_{h}}{k^{3}_{r}}\left(k_{r}\mathbf{e}_{r} +
    k_{h}\mathbf{e}_{h}\right) \\
\mathbf{c}_{g} &=& \nabla_{\mathbf{k}}\omega = -\frac{N k_{r}}{(k^{2}_{r}
    + k^{2}_{h})^{\frac{3}{2}}}\left(k_{h}\mathbf{e}_{r} -
    k_{r}\mathbf{e}_{h}\right) \\ 
&\approx& -\frac{N}{k_{r}^{2}}\left(k_{h}\mathbf{e}_{r} -
    k_{r}\mathbf{e}_{h}\right),
\end{eqnarray}
in the tidally relevant limit that the radial wavelength of the waves is much shorter
than the horizontal wavelength, i.e., $k_{h}\ll k_{r}$ (which is true
except near turning points or within the last few wavelengths from the centre of a star). In this limit,
$\mathbf{c}_{g}\cdot \mathbf{e}_{r} = -N\frac{k_{h}}{k_{r}^{2}} =
-\mathbf{c}_{p}\cdot \mathbf{e}_{r}$, i.e., the radial wave pattern moves in
the opposite direction to the radial energy flux. Since $\omega$ is
independent of $|\mathbf{k}|$, $\mathbf{c}_{g}\cdot \mathbf{k} = 0$, meaning
that the energy in IGWs propagates along surfaces of constant phase.

For waves on a non-zero horizontal background shear flow $\mathbf{U}$,
Eq.~\ref{dispersionrelation} still applies if the Richardson number $Ri =
\frac{N^{2}}{|\partial \mathbf{U} / \partial r|^{2}} \gg 1$, if we
replace $\omega$ by the Doppler-shifted frequency $\hat{\omega}$, and
similarly for the phase and group velocity of the waves
\begin{eqnarray}
\hat{\omega} = \omega - \mathbf{k}\cdot \mathbf{U}, \;\;\;\; 
\hat{\mathbf{c}}_{p} = \mathbf{c}_{p} - \mathbf{U}, \;\;\;\;
\hat{\mathbf{c}}_{g} = \mathbf{c}_{g} - \mathbf{U}.
\end{eqnarray}
We now have the possibility of the wave frequency being
Doppler-shifted upwards to $N$, in which case $\hat{\mathbf{c}}_{g}\cdot
\mathbf{e}_{r}$ reverses, giving total internal reflection
\citep{Perspectives2000}. 

The other extreme, that $\hat{\omega} \rightarrow 0$, occurs when the horizontal
velocity in the shear matches that of the horizontal phase
velocity. This occurs at a so-called critical layer, which is defined
as the layer at which the wavelength of the waves would be Doppler-shifted to zero,
if they were ever to reach it. Note though, that
$\hat{\mathbf{c}}_{g}\cdot\mathbf{e}_{r} \rightarrow 0$ as
$\hat{\omega}\rightarrow 0$, so in linear theory the waves never reach
the critical layer in a finite time. Early work on IGWs in a
background shear, including a study of critical layers, can be found in
\cite{BookerBretherton1967} and \cite{Hazel1967}. They find that an IGW
propagating through a critical layer is attenuated by a factor $\sim \exp
\left(-2\pi(Ri-1/4)^{1/2}\right)$. If $Ri \gg 1/4$, the wave is
fully absorbed, and irreversibly transfers its energy to the mean
flow. However, it must be noted that at the critical layer, linear theory predicts that the wave steepness
\begin{eqnarray}
s = \frac{\mathrm{max}(\mathbf{u}\cdot\mathbf{e}_{h})}{\hat{\mathbf{c}}_{p}\cdot\mathbf{e}_{h}},
\end{eqnarray}
where $\mathbf{u}$ is the velocity perturbation, of these waves goes to
infinity, as the Doppler-shifted horizontal
phase speed $\hat{\mathbf{c}}_{p}\cdot\mathbf{e}_{h}$ goes to
zero, i.e., the waves become strongly nonlinear at the critical layer
(which \citealt{Perspectives2000} refers to as linear theory predicting its own
breakdown). Since wave breaking is expected to occur whenever $s>1$,
the waves are likely to break before they reach the critical
layer. Nonlinear effects have been studied in early simulations by
\cite{Winters1994}, who find that the of initial wave energy on encountering 
a critical layer, roughly one third
reflects, one third results in mean flow acceleration and
the remainder cascades to small scales where it is dissipated. This
implies that wave absorption by the mean flow need not be complete, in
contrast with the prediction from linear theory. This is discussed
further in \S~\ref{refcoeffresults} in relation to the results of our simulations.

Wave breaking is defined as a wave-induced process that leads to the
rapid and irreversible deformation of otherwise wavy material contours
\citep{Perspectives2000}, and it leads to the production of turbulence and
irreversible energy dissipation. The breaking process results from the growth of an 
instability upon a basic state composed of a wave with $s > 1$.
The susceptibility of a wave to breaking can be enhanced
by resonant triad interactions, in which a primary wave resonantly 
interacts with a pair of low-amplitude secondary waves. This process 
transfers energy to the secondary waves, whose steepness can then grow
beyond the critical value required for breaking to occur, even though
the primary steepness may not be sufficient for breaking on its own 
\citep{Staquet2002}. Previous work has shown that the
process leading to the wave steepening is two-dimensional, but that
breaking is a three-dimensional process (\citealt{Klostermeyer1991};\citealt{Winters1994}). 
Nevertheless, the mechanisms responsible for breaking and the final
outcome of the breaking process are likely to be similar in 2D. 

Here we are interested in studying what
happens when IGWs excited by tidal forcing approach the centre of a star 
with an inner radiation zone, i.e. G-type stars such as the Sun, which
do not possess a convective core. We are primarily concerned with the tidal
dissipation efficiency for solar-type stars that may result from
breaking of these waves near the centre. Throughout the rest of this paper we restrict our problem to 2D,
and postpone study of any 3D effects. We now describe our basic
problem in more detail.

\section{Basic description of the problem}

\subsection{Tidal potential}

Consider a star and a planet in a mutual Keplerian orbit (though we make no
assumptions about the relative masses at this stage). The tidal 
potential experienced by the star can be written as a sum of
rigidly rotating spherical harmonics (e.g. OL04). For the simplest case of a planet on a circular orbit, that is coplanar with the
equatorial plane of the star, we can consider a two-dimensional
restriction of the problem to this plane. This allows us to write the time-dependent part of
the quadrupolar ($l=2$) tidal potential in the equatorial plane of the
star as
\begin{eqnarray}
\Psi(r,\phi,t) = -\frac{3}{4}\frac{Gm_{p}}{a^{3}}r^{2}\cos(2\phi - \hat{\omega}t),
\end{eqnarray}
in spherical polar coordinates $(r,\theta,\phi)$ with origin at the
centre of the star, in the plane $\theta = \pi/2$. Here $m_{p}$ is the
planet mass, $a$ is the orbital semi-major axis, and $n>0$ is the mean
motion. The relevant tidal
frequency is $\hat{\omega} = 2n-2\Omega$ for a star rotating with
angular velocity $\Omega$. From here on we assume that the star is
non-rotating, i.e. we assume that $\Omega = 0$, which is a reasonable
assumption if the star is spinning much slower than the orbit. This is
justified for the problem of a close-in gas giant planet with a
several-day orbital period around a solar-type star that has been spun down by magnetic braking for the
duration that it has spent on the main sequence, to rotate with a spin
period of several tens of days.

We consider a restriction of the full three-dimensional problem, in
which the tidal potential is composed of many different spherical harmonic
components for an orbit of arbitrary eccentricity and inclination, to
instead consider a simplified two dimensional model of a star, forced by this
single component of the tide. The main motivation for this decision is
simplicity, since this is a first attack on the problem, though it is
likely that this is a reasonable approximation. Extensions
can be studied in the future.

\subsection{Central regions of a star}

\begin{figure}
    \subfigure{
      \includegraphics[width=0.42\textwidth]{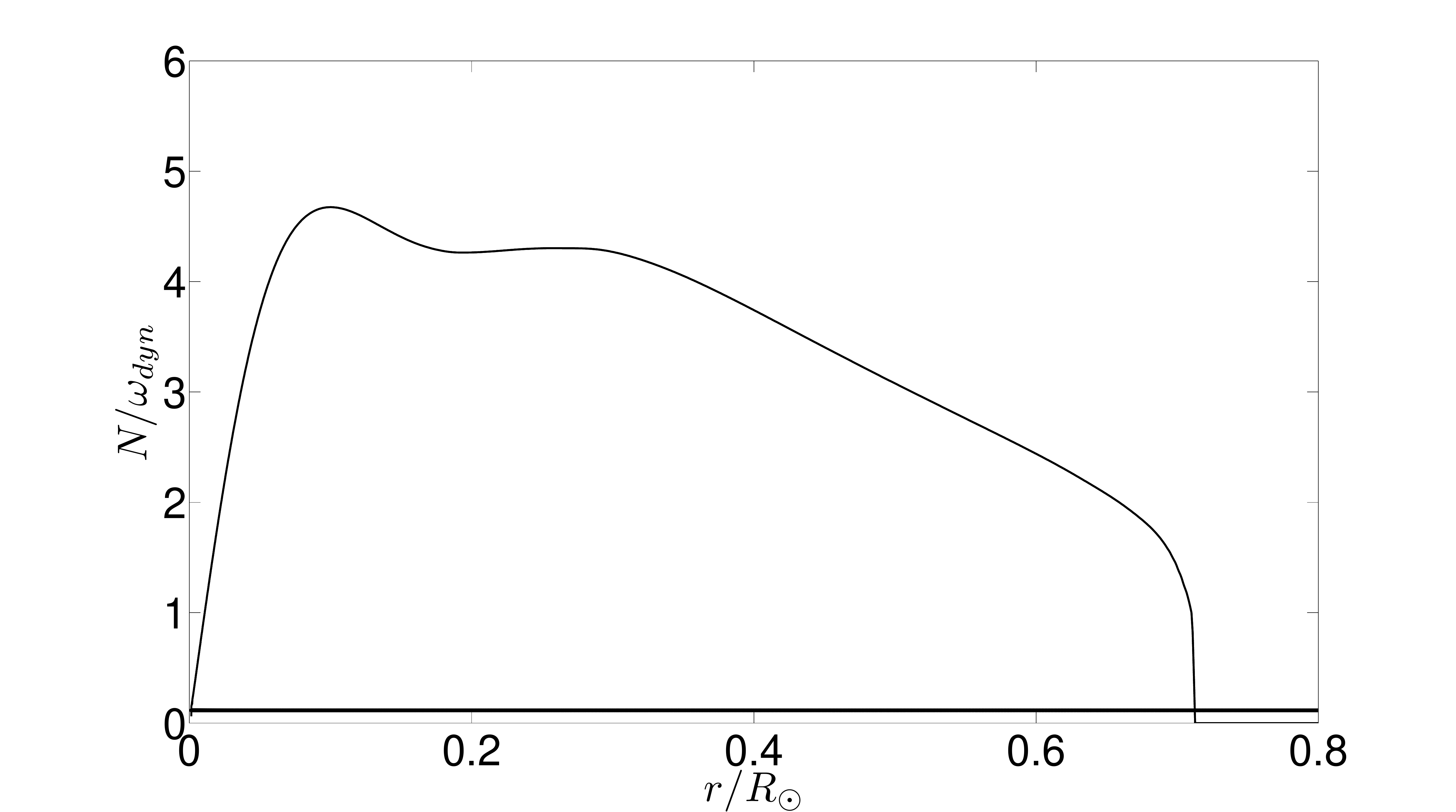}
    }
    \subfigure{
      \includegraphics[width=0.42\textwidth]{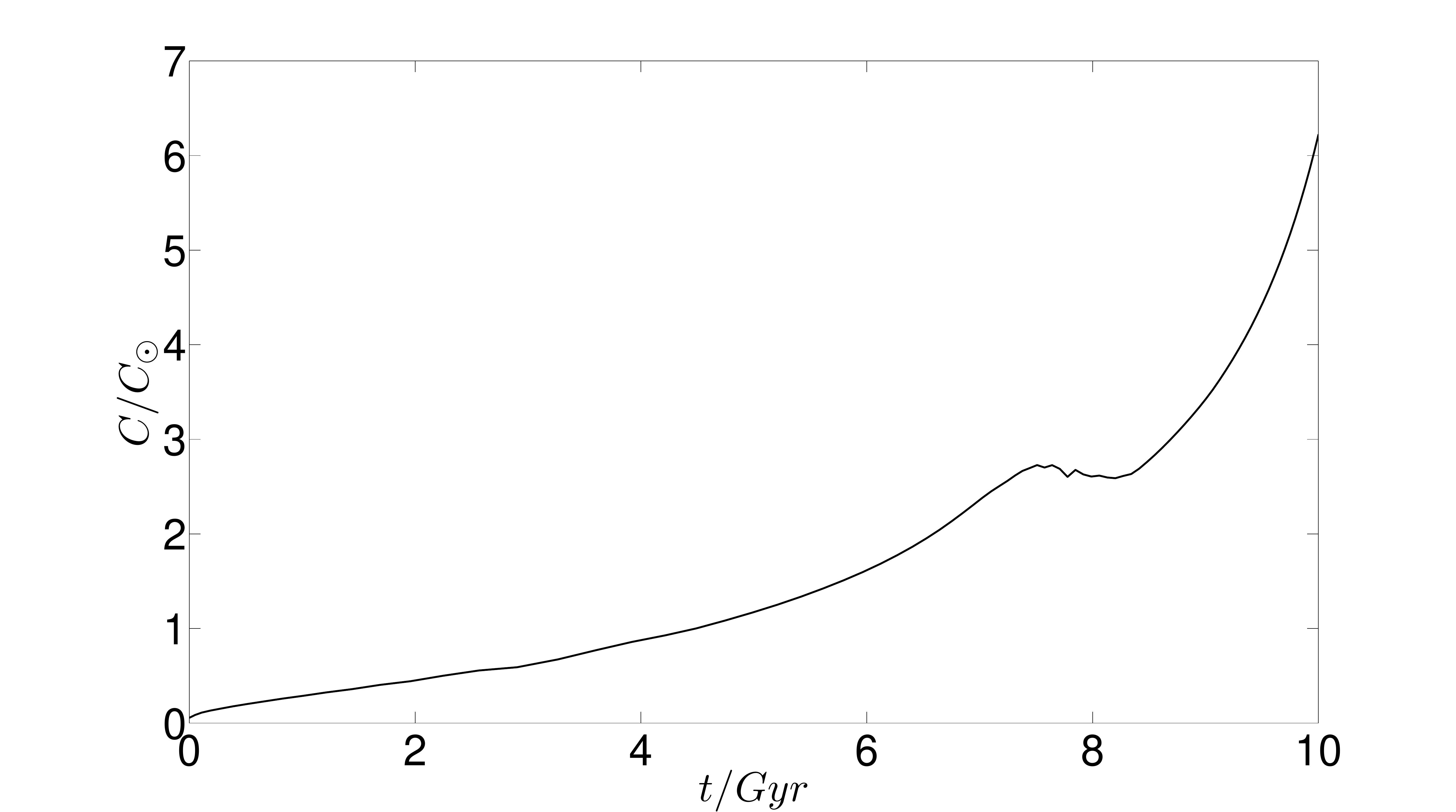}
    }
  \caption{Top: Buoyancy frequency $N$ normalised to dynamical frequency
    $\omega_{dyn}$ versus radius, based on Model S of the current Sun. 
    Also plotted is a frequency $2\pi / 1 \mathrm{day} \sim 0.1 \omega_{dyn}$, corresponding
    with the orbital frequency for a one day
    orbit -- note that this is half the tidal frequency, so
    $\hat \omega \ll N$ throughout the bulk of the radiation zone. Only
    near the centre and at the interface between the convection and
    radiation zones, does $\hat \omega  \sim N$. Bottom: Coefficient
    $C$ in the expansion $N \approx Cr$ near the centre of the Sun normalised
    to its current value versus main-sequence age for a sequence of solar models that pass through
    Model S of the current Sun. The stratification steepens
    as the star evolves.}
  \label{Fig:NvsrModelS}
\end{figure}

The buoyancy frequency is real and 
comparable with the dynamical frequency of the star
\begin{eqnarray}
\omega_{dyn} = \left(\frac{Gm_{\star}}{R_{\star}^{3}}\right)^{\frac{1}{2}},
\end{eqnarray}
throughout the bulk of the radiation zone, where $m_{\star}$ and
$R_{\star}$ are the stellar mass and radius, respectively. In
Fig.~\ref{Fig:NvsrModelS} we plot $N$ normalised to $\omega_{dyn}$ in
the radiation zone for Model S of the current Sun \citep{CD1996}.
For our problem, the tidal frequencies of interest $\hat{\omega} \ll
N$, which implies $k_{r} \gg 2\pi/r$, except near the
centre. IGWs are 
excited at the top of the radiation zone by a combination of
tidal forcing in that region, together with the pressure of inertial
waves acting at the interface if $|\hat{\omega}| < 2\Omega$ (OL07). It is within this
transition region that $N$ increases linearly with distance into the
radiation zone, so there is a point at which $N\sim \hat{\omega}$, and
IGWs are efficiently excited. These propagate towards the centre
with radial wavelengths $\lambda_{r} \lesssim 10^{-3}-10^{-2}
R_{\odot}$, for typical tidal frequencies.

Expanding the standard equations of
stellar structure about $r=0$, we obtain the density stratification 
\begin{eqnarray}
\rho(r) = \rho_{0} + \rho_{2}r^{2} + O(r^{4}).
\end{eqnarray}
For sufficiently small $r$, $g$ is linear
in $r$, and $N = C r$ is also linear in $r$, where $C$ is a constant
that varies with stellar model and main-sequence
age. For the current Sun, $C \equiv C_{\odot} \approx 8.0
\times 10^{-11} \mathrm{m}^{-1}\mathrm{s}^{-1}$. This is valid throughout only the 
inner $\lesssim 3\%$ of the Sun. However, even with the largest
expected radial wavelength, this region contains multiple wavelengths. In Fig.~\ref{Fig:NvsrModelS} we plot the variation in
$C$ versus main-sequence age, normalised to its value in the current
Sun, from a sequence of solar models that pass through Model S. The
increase in $C$ with age is due both to the increasing central
condensation, and the build-up of a gradient in the hydrogen
abundance (there is a small drop around $8$ Gyr when hydrogen is
nearly used up and the contribution from the
composition gradient decreases, after which the central
density increases rapidly).

\section{Derivation of a Boussinesq-type system of equations}

The ideal compressible fluid equations in 2D plane polar ($r,\phi$) coordinates are
\begin{eqnarray}
&& Du_{r} - \frac{u_{\phi}^{2}}{r} = -\frac{1}{\rho}\partial_{r}p -
\partial_{r} \Phi, \\
%%%%%%%%%%%%%%%%%%%%%%%%%%%%%%%%%%%%%%%
&& Du_{\phi} + \frac{u_{r}u_{\phi}}{r} = -\frac{1}{\rho r}\partial_{\phi}p -
\frac{1}{r}\partial_{\phi} \Phi, \\
%%%%%%%%%%%%%%%%%%%%%%%%%%%%%%%%%%%%%%%
&& D\rho + \rho\left[\frac{1}{r}\partial_{r}(ru_{r}) +
  \frac{1}{r}\partial_{\phi}u_{\phi}\right] = 0, \\
%%%%%%%%%%%%%%%%%%%%%%%%%%%%%%%%%%%%%%%
&& Dp - \frac{\gamma p}{\rho}D\rho = 0, \\
%%%%%%%%%%%%%%%%%%%%%%%%%%%%%%%%%%%%%%%
&& D = \partial_{t} + u_{r}\partial_{r} +
 \frac{u_{\phi}}{r}\partial_{\phi}, \\
%%%%%%%%%%%%%%%%%%%%%%%%%%%%%%%%%%%%%%%
&& \nabla^{2}\Phi = 4\pi G \rho.
\end{eqnarray}
The basic state is static and circularly symmetric. Near the centre,
we pose the expansion
\begin{eqnarray}
\rho = \rho_{0} + \rho_{2}r^{2} + \rho_{4}r^{4} + O(r^{6}),
\end{eqnarray}
and similarly for $p$ and $\Phi$. The coefficients are related by the
condition for the background to be in hydrostatic equilibrium,
together with Poisson's equation, giving
\begin{eqnarray} && p_{2} = -\rho_{0} \Phi_{2}, \;\; p_{4} = -\rho_{0}\Phi_{4} +
\frac{p_{2}\rho_{2}}{2\rho_{0}}, \\
&& \Phi_{2} = \pi G \rho_{0},
\;\; \Phi_{4} = \frac{\pi G \rho_{2}}{4},
\end{eqnarray}
and so on for the coefficients of higher-order terms.

We are interested in a region where $r/R_{\star} = O(\epsilon)$, where
$\epsilon \ll 1$, so let $r = \epsilon x$. Now
introduce a slow time $\tau = \epsilon t$, then the
solution including the basic state and a slow nonlinear density perturbation
has the form
\begin{eqnarray}
&& \rho = \underbrace{\rho_{0} + \epsilon^{2}\rho_{2}x^{2} +
\epsilon^{4}\rho_{4}x^{4} + ...}_{\mathrm{basic\;state}} +
\underbrace{\epsilon^{2}\rho^{\prime}_{2}(x,\phi,\tau) +
  ...}_{\mathrm{nonlinear\;perturbation}}, \\
%%%%%%%%%%%%%%%%%%%%%%%%%%%%%%%%%%%%%%%
&& p = \underbrace{p_{0} + \epsilon^{2}p_{2}x^{2} +
\epsilon^{4}p_{4}x^{4} + ...}_{\mathrm{basic\;state}} +
\underbrace{\epsilon^{4}p^{\prime}_{4}(x,\phi,\tau) +
  ...}_{\mathrm{nonlinear\;perturbation}}, \\
%%%%%%%%%%%%%%%%%%%%%%%%%%%%%%%%%%%%%%%
&&\Phi = \underbrace{\Phi_{0} + \epsilon^{2}\Phi_{2}x^{2} +
\epsilon^{4}\Phi_{4}x^{4} + ...}_{\mathrm{basic\;state}} +
\underbrace{\epsilon^{4}\Phi^{\prime}_{4}(x,\phi,\tau) +
  ...}_{\mathrm{nonlinear\;perturbation}}, \\
%%%%%%%%%%%%%%%%%%%%%%%%%%%%%%%%%%%%%%%
&&u_{r} = \underbrace{\epsilon^{2}u_{r2}(x,\phi,\tau) + ...}_{\mathrm{nonlinear\;perturbation}}, \\
&&u_{\phi} = \underbrace{\epsilon^{2}u_{\phi2}(x,\phi,\tau)+ ...}_{\mathrm{nonlinear\;perturbation}},
\end{eqnarray}
In the Boussinesq approximation, the fractional pressure perturbation
is small compared with the fractional density perturbation because
this is a low-frequency perturbation, with $\omega \ll \omega_{dyn}$, for
which acoustic effects are negligible. For such short-wavelength
perturbations, the gravitational perturbation is also small \citep{Helio2002}. These two points explain the absence of
terms in the above solutions proportional to $\epsilon^{2}$ for $p$
and $\Phi$ in the nonlinear perturbation. This corresponds to
looking for small density constrasts and weak material accelerations
compared with gravity.

Substituting these expansions into the basic equations, and
subtracting terms that arise only in the basic state, we obtain at
leading order
\begin{eqnarray}
&&D_{1}u_{r2} - \frac{u_{\phi2}^{2}}{x} =
-\frac{1}{\rho_{0}}\partial_{x}p^{\prime}_{4} +
\frac{2xp_{2}\rho_{2}^{\prime}}{\rho_{0}^{2}} - \partial_{x}
\Phi^{\prime}_{4}, \\
%%%%%%%%%%%%%%%%%%%%%%%%%%%%%%%%%%%%%%%
&&D_{1}u_{\phi2} + \frac{u_{r2}u_{\phi2}}{x} = -\frac{1}{\rho_{0} x}\partial_{\phi}p^{\prime}_{4} -
\frac{1}{x}\partial_{\phi} \Phi^{\prime}_{4}, \\
%%%%%%%%%%%%%%%%%%%%%%%%%%%%%%%%%%%%%%%
&&\rho_{0}\left[\frac{1}{x}\partial_{x}(xu_{r2}) +
  \frac{1}{x}\partial_{\phi}u_{\phi2}\right] = 0, \\
%%%%%%%%%%%%%%%%%%%%%%%%%%%%%%%%%%%%%%%
&&2xp_{2}u_{r2} - \frac{\gamma
  p_{0}}{\rho_{0}}\left(D_{1}\rho^{\prime}_{2} +
  2x\rho_{2}u_{r2}\right) = 0, \\
%%%%%%%%%%%%%%%%%%%%%%%%%%%%%%%%%%%%%%%
&&D_{1} = \partial_{\tau} + u_{r2}\partial_{x} +
\frac{u_{\phi2}}{x}\partial_{\phi}, \\
%%%%%%%%%%%%%%%%%%%%%%%%%%%%%%%%%%%%%%%
&&\nabla^{2}\Phi^{\prime}_{4} = 4\pi G \rho_{2}^{\prime}.
\end{eqnarray}
Note that Poisson's equation is no longer required, since we have
separated out the gravitational potential perturbation in
this approximation, which is equivalent to Cowling's approximation \citep{Cowling1941}.

We can rewrite these equations in a more natural notation, removing the
asymptotic scalings to find
\begin{eqnarray}
&&Du_{r} - \frac{u_{\phi}^{2}}{r} =
-\partial_{r}q + rb, \\
%%%%%%%%%%%%%%%%%%%%%%%%%%
&&Du_{\phi} + \frac{u_{r}u_{\phi}}{r} = -\frac{1}{\rho r}\partial_{\phi}q, \\
%%%%%%%%%%%%%%%%%%%%%%%%%%%%
&&\frac{1}{r}\partial_{r}(ru_{r}) +
  \frac{1}{r}\partial_{\phi}u_{\phi} = 0, \\
%%%%%%%%%%%%%%%%%%%%%%%%%%%%
&&Db + C^{2}r u_{r} = 0, \\
%%%%%%%%%%%%%%%%%%%%%%%%%%
&&D = \partial_{t} + u_{r}\partial_{r} +
\frac{u_{\phi}}{r}\partial_{\phi},
\end{eqnarray}
where
\begin{eqnarray}
&&b = \frac{2p_{2}}{\rho_{0}^{2}}\rho^{\prime}_{2}, \\
%%%%%%%%%%%%%%%%%%%%%%%%%%%%%%%%%%%%%%%%%%%%%%%%
&&q = \frac{1}{\rho_{0}}p_{4}^{\prime} + \Phi_{4}^{\prime},
\end{eqnarray}
are a buoyancy variable and a modified pressure variable, and 
\begin{eqnarray}
C^{2} = 4\Phi_{2}\left(\frac{p_{2}}{\gamma p_{0}} -
\frac{\rho_{2}}{\rho_{0}}\right)
\end{eqnarray}
is related to the buoyancy frequency $N$ by $N=Cr$, with $C^{2} > 0$
for a stably stratified centre of a star. 

We now write them in the vector-invariant form
\begin{eqnarray}
\label{MainEqs1}
&&D \mathbf{u} = -\nabla q + \mathbf{r}b, \\
%%%%%%%%%%%%%%%%%%%%%%%%%%
&&D b + C^{2} \mathbf{r}\cdot\mathbf{u}  = 0, \\
\label{MainEqs2}
%%%%%%%%%%%%%%%%%%%%%%%%%%
&&\nabla \cdot \mathbf{u} = 0, \\
\label{MainEqs3}
%%%%%%%%%%%%%%%%%%%%%%%%%%
&&D = \partial_{t} + \mathbf{u}\cdot \nabla.
\end{eqnarray}

These equations are similar to the standard Boussinesq system for a
slab of fluid in Cartesian geometry (see e.g. \citealt{Buhler2009}, Ch. 6) with a
uniform stratification, with the exception that our problem is in
cylindrical geometry, with $g$ and $N$ proportional to $r$. Note also that the buoyancy
variable defined here is related, but not identical, to that of the
standard Boussinesq approximation used in atmospheric sciences and
oceanography (see e.g. \citealt{Buhler2009}). The buoyancy variable is proportional to the
density and entropy perturbation.

An energy equation for our system can be derived by contracting Eq.~\ref{MainEqs1} with $\mathbf{u}$:
\begin{eqnarray}
\label{BoussEnergy}
\partial_{t}
\left(\frac{1}{2}|\mathbf{u}|^{2}+\frac{b^{2}}{2C^{2}}\right)+
\nabla \cdot
\left[\left(\frac{1}{2}|\mathbf{u}|^{2} + \frac{b^{2}}{2C^{2}} +
  q\right)\mathbf{u}\right] = 0.
\end{eqnarray}
Thus $E = \frac{1}{2}\rho_{0}|\mathbf{u}|^{2}+\rho_{0}\frac{b^{2}}{2C^{2}}$ is the energy
density per unit volume, and $\mathbf{F}_{E} = \rho_{0}\left(\frac{1}{2}|\mathbf{u}|^{2} + \frac{b^{2}}{2C^{2}} +
q\right)\mathbf{u}$ is the energy flux density.

If the fluid is at rest, with $b=0$, then the stratification surfaces
are circles (spheres in 3D), and $C^{2}$ measures the strength of the
stable stratification. If we disturb the fluid from rest, then a
positive (negative) buoyancy is associated with an inward (outward) radial
displacement of particles, resulting in an outward (inward)
acceleration of the fluid due to buoyancy to restore the system to
equilibrium. From the energy equation Eq.~\ref{BoussEnergy}, we can
see that the state $b=0$ is the state of minimum gravitational
potential energy, since the available potential energy density
$\rho_{0}\frac{b^{2}}{2C^{2}}$ is minimised for this state. This makes sense,
since this corresponds to having a background state with no wave-like disturbance.

\section{Linear theory of IGWs approaching the stellar centre}
\label{lineartheory}

\subsection{Linear solution steady in a frame rotating with the pattern speed
 of forcing}

If the radiation zone is forced from above then this will excite
waves, which will propagate to the centre of
the star. If the incoming wave has frequency $\omega$ and azimuthal wavenumber
$m$, then it seems reasonable to assume that the response is steady in
a frame rotating with the angular pattern speed of the forcing
$\Omega_{p} = \omega/m$, in the absence of instabilities. The
dependence on $\phi$ and $t$ is then only through the combination 
\begin{eqnarray}
\xi = m\phi - \omega t = m(\phi - \Omega_{p}t).
\end{eqnarray}
We can choose $\Omega_{p}/C$ as a unit of length, and
$\Omega_{p}^{-1}$ as a unit of time (note that these units are only used in this
section and in Appendix \ref{refcoeff}), to allow us to write the equations
in the dimensionless form
\begin{eqnarray}
&& Du_{r} - \frac{u_{\phi}^{2}}{r} = -\partial_{r}q + rb, \\
%%%%%%%%%%%%%%%%%%%%%%%%%%%%%%%%%%%%%%%%%%%%%%%%%%%%%%%
&& Du_{\phi} + \frac{u_{r}u_{\phi}}{r} = -\frac{m}{\rho r}\partial_{\xi}q, \\
%%%%%%%%%%%%%%%%%%%%%%%%%%%%%%%%%%%%%%%%%%%%%%%%%%%%%%
&& \frac{1}{r}\partial_{r}(ru_{r}) +
  \frac{m}{r}\partial_{\xi}u_{\phi}, \\
%%%%%%%%%%%%%%%%%%%%%%%%%%%%%%%%%%%%%%%%%%%%%%%%%%%%%%%%%
&& Db + ru_{r} = 0, \\
%%%%%%%%%%%%%%%%%%%%%%%%%%%%%%%%%%%%%%%%%%%%%%%%%%%%%%%%%
&& D = u_{r}\partial_{r} + m\left(\frac{u_{\phi}}{r}-1\right)\partial_{\xi}.
\end{eqnarray}
The energy equation (Eq.~\ref{BoussEnergy}) allows us to infer
that the radial energy flux
\begin{eqnarray}
F^{E}_{r} = \rho_{0}\int_{0}^{2\pi} \left[\frac{1}{2}\left(|\mathbf{u}|^{2} +
  b^{2}\right) + q\right]r u_{r} d\xi
\end{eqnarray}
is independent of $r$ for disturbances steady in this frame of
reference, since the solutions are periodic with period $2\pi$
starting at $\xi = 0$.

The radial angular momentum flux is
\begin{eqnarray}
F^{L}_{r} = \frac{m}{\omega}F^{E}_{r}.
\end{eqnarray}

We can obtain a solution to these equations by linearization as follows,
assuming that the solution is proportional to $e^{i \xi}$. Then we
obtain (where real parts are assumed to
be taken)
\begin{eqnarray}
&&-imu_{r} = -\partial_{r}q + rb, \\
%%%%%%%%%%%%%%%%%%%%%%%%%%%%%%%%%
&&-imu_{\phi} = -\frac{imq}{r}, \\
&&\frac{1}{r}\partial_{r}(ru_{r}) +\frac{im}{r}u_{\phi} = 0, \\
%%%%%%%%%%%%%%%%%%%%%%%%%%%%%%%%
&&-imb + ru_{r} = 0.
\end{eqnarray}

The incompressibility constraint allows us to express the velocity in
terms of the streamfunction $\psi(r,\phi)$, which is defined by
\begin{eqnarray}
\mathbf{u} = \nabla \times \left(\psi \mathbf{e}_{z}\right) = \left(\frac{1}{r}\partial_{\phi}\psi\right)\mathbf{e}_{r}
+ \left(-\partial_{r}\psi\right)\mathbf{e}_{\phi},
\end{eqnarray}
so we can write
\begin{eqnarray}
\label{streamfn}
u_{r} = \mathrm{Re}\left[\frac{im}{r}\psi\right], \\
u_{\phi} = \mathrm{Re}\left[-\partial_{r}\psi\right].
\end{eqnarray}
This enables us to reduce the system to Bessel's equation of order $m$,
\begin{eqnarray}
  L_{m} \psi = \partial_{r}(r\partial_{r}\psi) +
  r\left(1-\frac{m^{2}}{r^{2}}\right)\psi = 0
\end{eqnarray}
with solution regular at the origin $\psi \propto J_{m}(r)$. This
represents a wave that approaches from infinity, reflects perfectly
from the centre and goes out to infinity. Pure ingoing and outgoing
wave solutions are described by $J_{m}(r) \pm iY_{m}(r)$ respectively. 

\subsection{Properties of the (non-)linear solution}

The general solution can be written in the form of a
sum of ingoing and outgoing waves, with complex amplitudes $A_{in}$ and
$A_{out}$, as follows:
\begin{eqnarray}
\psi_{in}(r,\xi) = [J_{m}(r) + iY_{m}(r)]e^{i\xi}, \\
%%%%%%%%%%%%%%%%%%%%%%%%%%%%%%%%%
\psi_{out}(r,\xi) = [J_{m}(r) - iY_{m}(r)]e^{i\xi}, \\
%%%%%%%%%%%%%%%%%%%%%%%%%%%%%%%%%
\psi(r,\xi) = A_{in}\psi_{in}(r,\xi) + A_{out}\psi_{out}(r,\xi).
\end{eqnarray}

We can check that $\psi_{in}$ corresponds to an ingoing wave by
calculating its phase and group velocities. A simple calculation shows
that the radial phase velocity is directed outward if we adopt the
convention that $\omega = m\Omega_{p} > 0$, and the group velocity is
directed inward. This highlights one of the peculiarities of IGWs --
that the phase and group velocities are oppositely directed, as
discussed in \S~\ref{IGWtheory}.  For these
linear waves, we have
\begin{eqnarray}
F^{E}_{r} = \rho_{0}\pi m r \mathrm{Im}[\psi\partial_{r}\psi^{*}] =
2m\rho_{0}\left( |A_{out}|^{2} - |A_{in}|^{2}\right).
\end{eqnarray}
The solution for a wave that perfectly reflects from
the centre is 
\begin{eqnarray} 
\label{analytic}
\psi(r,\xi) = 2A_{in} J_{m}(r)e^{i\xi},
\end{eqnarray}
and has $F^{E}_{r}=0$, since $A_{in} = A_{out}$.

If we take the curl of Eq.~\ref{MainEqs1}, we
obtain
\begin{eqnarray}
\partial_{t} (\nabla \times \mathbf{u}) = \nabla \times (\mathbf{r}b) -
\nabla \times (\mathbf{u}\cdot \nabla \mathbf{u}),
\end{eqnarray}
which has eliminated the modified pressure perturbation $q$. The
$z$-component of this equation expressed in terms of the
streamfunction is
\begin{eqnarray}
\partial_{t} (-\nabla^{2} \psi) = - \partial_{\phi}
b + J(\psi,-\nabla^{2}\psi),
\end{eqnarray}
and the buoyancy equation is
\begin{eqnarray}
\partial_{t} b = - C^{2}\partial_{\phi}\psi + J(\psi,b).
\end{eqnarray}
The nonlinear terms take the form of Jacobians,
\begin{eqnarray}
  J(A,B) &=& \frac{\partial (A,B)}{\partial (x,y)} = \frac{1}{r}\frac{\partial (A,B)}{\partial
  (r,\phi)} \\
  &=& (\partial_{r}A)(\frac{1}{r}\partial_{\phi}B) -
  (\frac{1}{r}\partial_{\phi}A)(\partial_{r}B).
\end{eqnarray} 
The Jacobians of the solution derived above are
\begin{eqnarray}
J(\psi,-\nabla^{2}\psi)= J(\psi,b) = 0,
\end{eqnarray}
which expresses the surprising result that the solutions derived are exact
nonlinear solutions of the system. This
follows from the fact that $-\nabla^{2}\psi=b=\psi$ for these
waves. This arises because although the nonlinear terms $\mathbf{u}\cdot \nabla \mathbf{u}
\ne 0$, they are balanced by the modified pressure term in the equations
of motion. We also have $\mathbf{u}\cdot \nabla b = 0$.

This is distinct from, but analogous to, the result that a single propagating plane IGW
in a uniform stratification is a nonlinear solution of the standard Boussinesq
system (\citealt{Drazin1977}; \citealt{Klostermeyer1982}). This is a consequence of the fact
that $\mathbf{k} \cdot \mathbf{u} = 0$ for these waves, which implies that the
advective operator $\mathbf{u} \cdot \nabla$ annihilates any 
disturbance belonging to the same plane wave. A useful consequence of
this is that a stability analysis can be performed on finite-amplitude
propagating IGWs, allowing detailed understanding of the initial stages of the
breaking process for these waves (\citealt{Drazin1977};
\citealt{Klostermeyer1982}; \citealt{Klostermeyer1991}). Such
studies have shown that a single propagating IGW solution is \textit{always} unstable whatever
its amplitude, since it undergoes resonant triad interactions
\citep{Drazin1977}. One important difference in our problem is that the
nonlinearity is spatially localized to the innermost wavelengths, whereas the nonlinearity is present everywhere in the
plane IGW problem. We postpone a study of the stability of our nonlinear standing wave 
until a subsequent paper, though we expect waves of sufficiently large
amplitude to be unstable if they overturn the stratification.

The amplitude required to overturn the stratification can be derived
from our solution Eq.~\ref{analytic}. The entropy (or more precisely,
a quantity proportional to the entropy) is $s = b + (1/2)r^{2}$, in
these units. Overturning the stratification
means that the entropy profile, perturbed by a wave with buoyancy
perturbation $b$, must satisfy $\partial_{r} s < 0$, which implies
$(1/r)\partial_{r} b < -1$ is a condition for overturning. Since
$b=\psi$ for these nonlinear waves, this can be expressed in terms of
the streamfunction as $(1/r)\partial_{r} \psi < -1$. Reintroducing
dimensional variables, and substituting for $u_{\phi}$, modifies this
criterion to $(u_{\phi}/r) > \Omega_{p}$, i.e., overturning occurs if
the angular velocity of the wave exceeds the angular pattern
speed. Equivalently, wave breaking occurs if
\begin{eqnarray}
\label{convectiveinstability}
\mathrm{max}(u_{\phi}) \gtrsim \frac{\omega^{2}}{4C},
\end{eqnarray} 
whose largest value occurs where the amplitude of $\partial_{r}J_{2}(r)$ is largest, which
is one wavelength from the centre.

In 3D the equivalent linear solution (written down in the Appendix of OL07),
is not an exact nonlinear solution. However, the results of a weakly
nonlinear analysis find that the reflection is close to perfect, with
a reflection coefficient that is extremely close to unity for moderate
amplitudes.

\section{Numerical methods}
\label{numericalmethods}

\subsection{Snoopy spectral code}
\label{Snoopy}
We solve the system of equations (\ref{MainEqs1})-(\ref{MainEqs3})
using a Cartesian spectral code, Snoopy
(\citealt{Lesur2005}; \citealt{Lesur2007}). 
It is advantageous to use a Cartesian code over one in the
more natural (for the problem) cylindrical geometry, because of the
absence of a coordinate singularity at the origin, near to which is
the region of the flow that we are most interested in. We also avoid the
timestep issues close to the centre that would be present in a
time-explicit cylindrical code. These arise from the CFL condition,
because the grid spacing becomes very small near the origin.

Since this is a Fourier spectral code, the problem must be periodic
in space. We solve our non-periodic problem using this code by
setting up a region near the outer boundary, in which
the fluid variables are smoothed to zero as we approach the boundary, 
using a parabolic smoothing function. We find it is quite
acceptable to do this over a region about $1/10$ of the total box
size. For this value there is negligible interaction between
neighbouring boxes. This approach is one that may be useful in many applications
which would benefit from the use of spectral methods, but have
non-cartesian geometry and/or non-periodic boundary conditions. The
obvious drawback of such an approach is the slight increase in computational
cost, since the smoothing region is additional to the flow in the
region of interest. Interior to this we have a thin
ring in which we implement a forcing term in the radial momentum
equation of the form $f_{r}\cos(2\phi - \omega t)$. This is designed
to excite IGWs with $m=2$, but is not designed to accurately describe
the excitation of IGWs at the top of the radiation zone, since we are
only interested in the dynamics of the central region. Our forcing is
non-potential, which reflects the fact that the tidal forcing of waves
is indirect (OL04; \citealt{Ogilvie2005}). A potential force would be
absorbed in this model by a hydrostatic adjustment of $q$.

We solve the equations
\begin{eqnarray}
\label{SnoopyEqs1}
&&  D \mathbf{u} = -\nabla q + \mathbf{r}b  + \nu \nabla^{2}\mathbf{u} + \Big\{
  \begin{array}{ll} 0, & 0 \leq r < r_{force}, \\
    \mathbf{f}, & r_{force} \leq r < r_{smooth}, \\
  \end{array} \\
%%%%%%%%%%%%%%%%%%%%%%%%%%
&& D b + C^{2} \mathbf{r}\cdot\mathbf{u} = 0, \\ %\kappa \nabla^{2} b \\
\label{SnoopyEqs2}
%%%%%%%%%%%%%%%%%%%%%%%%%%
&&\nabla \cdot \mathbf{u} = 0, \\
\label{SnoopyEqs3}
%%%%%%%%%%%%%%%%%%%%%%%%%%
&& D = \partial_{t} + \mathbf{u}\cdot \nabla,
\end{eqnarray}
where $\mathbf{f} =  - f_{r}\cos(2\phi-\omega t) \, \mathbf{e}_{r}$.
We use a parabolic smoothing function 
$d(r) = \left(\frac{r-r_{smooth}}{r_{box}-r_{smooth}}\right)^{2}$, to 
instantaneously smooth $u_{r}$,$u_{\phi}$ and $b$ to zero as we
approach the outer boundary. We do this by multiplying the variables in
the region $r_{smooth} \leq r < r_{box}$ by $d(r)$ during every
timestep.

Our choice of units for length and time are arbitrary, but we choose
the following. We study a region 
$-1.5 < x < 1.5$, $-1.5 < y < 1.5$ and set $r_{force} = 0.85 \; r_{box}$, 
$r_{smooth} = 0.9 \; r_{box}$, with $r_{box} = 1.5$. We
choose a typical IGW radial wavelength of $\lambda_{r} = 0.1$, so that
we are resolving $\sim 12$ wavelengths within the box. The radial
wavelength of these waves is not strictly constant, but its variation 
for large $r$ is small, and can be reasonably approximated by a
constant value in that region. We then choose a forcing frequency
$\omega = 1$. These choices are arbitrary, and are made to ensure that 
we are resolving a sufficient number of wavelengths within the box.  

Explicit viscosity is added to
Eq.~\ref{MainEqs1}, since the code has
no intrinsic dissipation. This is necessary for stability -- to ensure that we
have no unphysical growth of energy at small scales. The value of the 
viscosity is chosen such that it dissipates disturbances on the grid
scale, and a value of $\nu = 2\times 10^{-6}$ is chosen for all
simulations. Viscous terms are implemented in a
time-implicit manner. We do not include thermal diffusion in the
buoyancy equation since this was found to be unnecessary for
stability. We solve the relevant Poisson equation for the
modified pressure during each timestep. 

We normalise the velocity components with respect to a typical radial phase
velocity of the wave $\omega\lambda_{r}/2\pi$ and thus set
\begin{eqnarray}
 u_{r,\phi}= \tilde{u}_{r,\phi} \left(\frac{\omega \lambda_{r}}{2\pi}\right),
\end{eqnarray} 
in which $\tilde{u}_{\phi}$ is equivalent to the wave steepness $s$,
and is a measure of the nonlinearity in
the wave. This allows us to write the condition for overturning the stratification in
Eq.~\ref{convectiveinstability} as 
\begin{eqnarray}
\label{convinstcrit}
\tilde{u}_{\phi} > \frac{1}{2}.
\end{eqnarray}

Since we have chosen to specify
the radial wavelength $\lambda_{r}$ and frequency $\omega$ of the waves that we wish to
study, we have already constrained the stratification
\begin{eqnarray}
C = \frac{\pi \omega}{\lambda_{r}}.
\end{eqnarray}

There is now only one further parameter, $f_{r}$ (except for viscous damping and
smoothing terms), to fully specify the problem. 
We set
\begin{eqnarray}
f_{r} = \tilde{f}_{r} \left(\frac{\omega \lambda_{r}}{2\pi}\right)\omega 
\end{eqnarray}
and we vary the normalised amplitude $\tilde{f}_{r}$ to model the
effects of different tidal
forcing amplitudes. From preliminary investigation, we find that it is
appropriate to choose values between $10^{-2}$ and $10^{3}$, since these
result in central amplitudes that range from $\tilde{u}_{r} \ll 1$ to $\tilde{u}_{r}
= O(1)$ (higher central amplitudes are not observed, as is described
in the results, owing to wave breaking above a critical amplitude). This
represents a vast range of amplitudes of tidal forcing, from cases in which
the secondary body is a low-mass planet to a solar-mass binary
companion in a close orbit.

Our background is a hydrostatic equilibrium with no wave, with $b=0$ in the initial state. 
We use a resolution of $512\times 512$ for most simulations, though
several higher resolution runs have been performed using $1024\times
1024$ and $1536\times 1536$. We confirm that the results are not dependent on the numerical method
(and that our system Eqs.~\ref{MainEqs1}-\ref{MainEqs3} correctly
describes the relevant physics) by reproducing the basic results using
ZEUS-2D \citep{Stone1992}. We describe our implementation of the problem in
this code in Appendix~\ref{ZEUS}. ZEUS reproduces the same
basic results as the Snoopy code, which indicates that the
effects of nonzero compressibility are unimportant. In
light of this, we only discuss the Snoopy results below.

\section{Numerical results}
\label{Results}

We use the set-up described in \S~\ref{Snoopy} for a set of
simulations with a variety of forcing amplitudes
$\tilde{f}_{r}$. The typical radial group velocity and wave crossing
time are, respectively,
\begin{eqnarray}
c_{g,r} = \frac{C\lambda_{r}^{2}}{2\pi^{2}}, \\
t_{c} = \frac{r_{box}}{c_{g,r}}.
\end{eqnarray}
For the initial conditions described in the previous section, $t_{c} \sim 90$.
We define for the purposes of the following, a ``wave'' to be a
non-axisymmetric oscillatory flow represented by a single azimuthal
wavenumber $m \ne 0$, whereas a ``mean flow'' is an axisymmetric
azimuthal flow with $m=0$. 

We perform several different quantitative analyses of
the results. We separate the amplitudes of the waves into an ingoing
wave (IW) and an outgoing wave (OW), and calculate a reflection coefficient, using
the method described in Appendix~\ref{refcoeff}. The reflection
coefficient $\mathcal{R}$ is defined as the ratio of the absolute amplitudes of the
outgoing ($A_{out}$) and ingoing waves ($A_{in}$) for a given radial ring,
\begin{eqnarray}
\label{refcoeffdef}
\mathcal{R} = \left| \frac{A_{out}}{A_{in}}\right|,
\end{eqnarray}
and measures the amplitude decay for a wave travelling from $r$ to the
centre, and back to $r$. We can relate it to the phase change on reflection ($\Delta \phi$) by
\begin{eqnarray}
\Delta \phi = i\left[ \ln \mathcal{R} - \ln \left(\frac{A_{out}}{A_{in}}\right) \right]
\end{eqnarray}
For perfect standing waves, $A_{in} = A_{out}$, and
$\mathcal{R}=1$. If the ingoing wave is entirely absorbed at the
centre, then $\mathcal{R}=0$. Thus, $\mathcal{R}$ is a measure of how
much the wave has been attenuated on reflection from the centre. 

We also Fourier analyse the solution, to study the temporal evolution
of different azimuthal wavenumbers in the flow. This is done by selecting a ring of cells
in the grid at a particular radius, which is chosen to be at $r=0.1$,
since this is probably close enough to the centre to detect the effects
of nonlinear wave couplings, if they occur. Since this is a cartesian
grid, we do this by selecting all cells within a particular radial
ring to within a tolerance width comparable with the size of a grid cell. We then
compute the Discrete Fourier
Transform of the velocity components, and from this
calculate the power spectral density,
\begin{eqnarray}
P_{m} = \frac{1}{N}\left|\sum_{k=0}^{N-1}u_{r,k} \exp{\left(-im
      k\left(\frac{2\pi}{N}\right)\right)}\right|^{2}
\end{eqnarray}
and similarly for $u_{\phi}$, where $N$ is the number of grid points
in the ring -- which depends on $r$ and the resolution, though $10^{2}
< N < 10^{3}$ for all resolutions at $r=0.1$. From this we can determine which components of the
solution grow or decay as a result of viscous damping, instabilities or 
nonlinear wave-wave interactions. Note that this is only a rough
approximation to the azimuthal power spectral density because the
points are irregularly spaced around the ring, yet we have assigned an
even weighting to each point. Nevertheless, this is
justified in practice because this method works well when tested on 
low-amplitude solutions that are well described by
the standing wave in \S~\ref{lineartheory}, for which we know that the
solution is an $m=2$ wave for both $u_{r}$ and $u_{\phi}$. Note also
that $P_{m} = P_{-m}$ since $\mathbf{u}$ is real, so we cannot
distinguish between waves with wavenumbers $m$ and $-m$ without also
observing the time-dependence of the flow.  

The main result that will be discussed in more detail
below is that we find that there exists a critical wave amplitude beyond
which wave breaking occurs near the centre. Below this amplitude, the waves reflect
coherently from the centre, and a steady state is reached 
in the reference frame rotating with
$\Omega_{p}$, consisting of an $m=2$ standing wave solution. This is the outcome
inferred from linear theory (see \S \ref{lineartheory}), and we will
discuss these cases first, followed by those in which nonlinear
effects start to become important. We refer to the former as
``low-amplitude'' cases, and the latter as ``high-amplitude'' cases.

\section{Low-amplitude forcing: coherent reflection}
\label{linearsims}

\begin{figure}
  \begin{center}
    \subfigure{
      \includegraphics[width=0.45\textwidth]{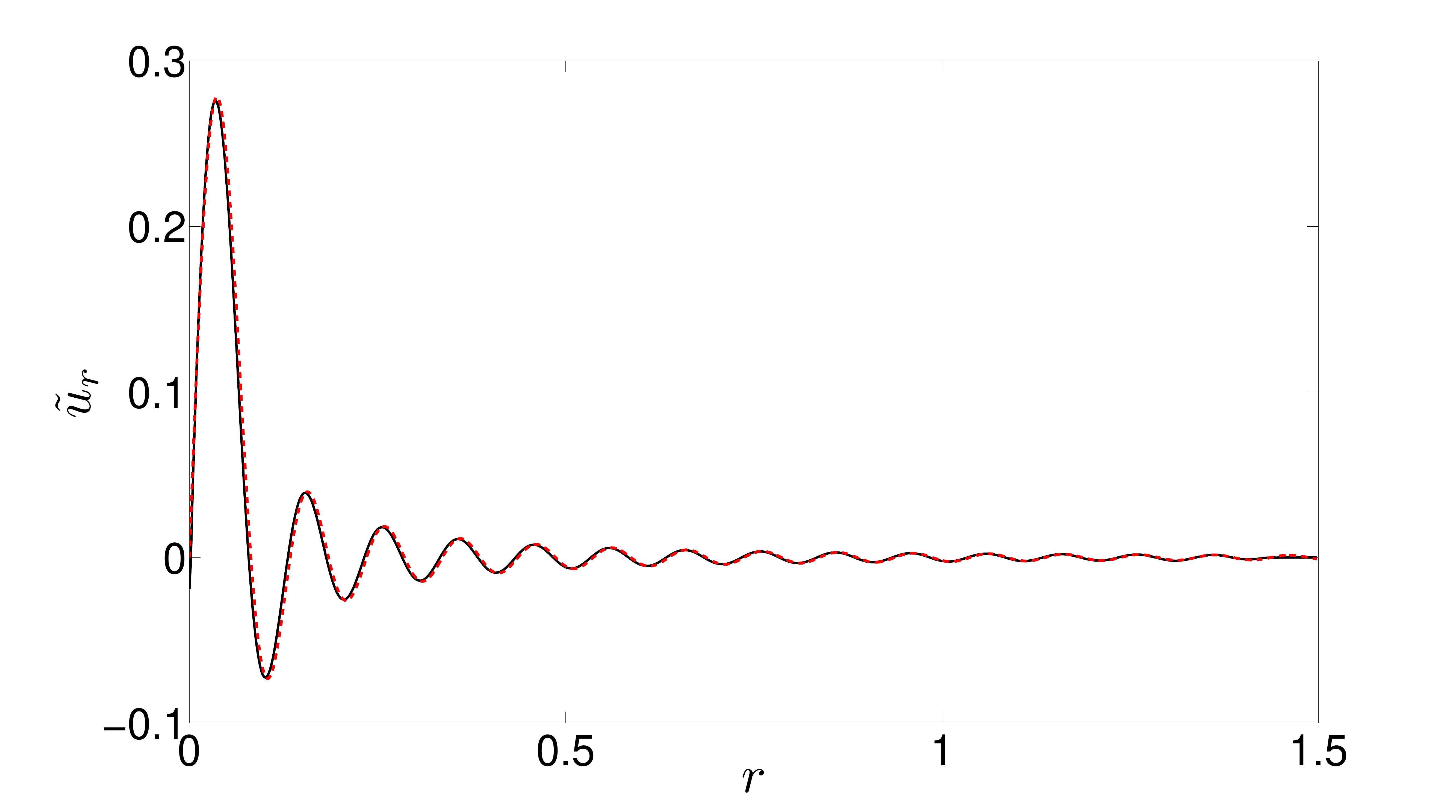}
      }
  \end{center}
  \caption{Radial velocity along the $x$-axis for a simulation with forcing amplitude
  insufficient to cause breaking. The amplitude of
  this wave is largest in the centre. Also plotted is the
  corresponding analytic
  standing wave Eq.\ref{analytic}, converted into a radial velocity
  using Eq.\ref{streamfn}, 
  showing that our simulations accurately describe
  the waves for the case in which the waves reflect coherently from
  the centre.}
  \label{Fig:Comparisonwithanalytic}
\end{figure}

\begin{figure}
  \begin{center}
    \subfigure{
      \includegraphics[width=0.4\textwidth]{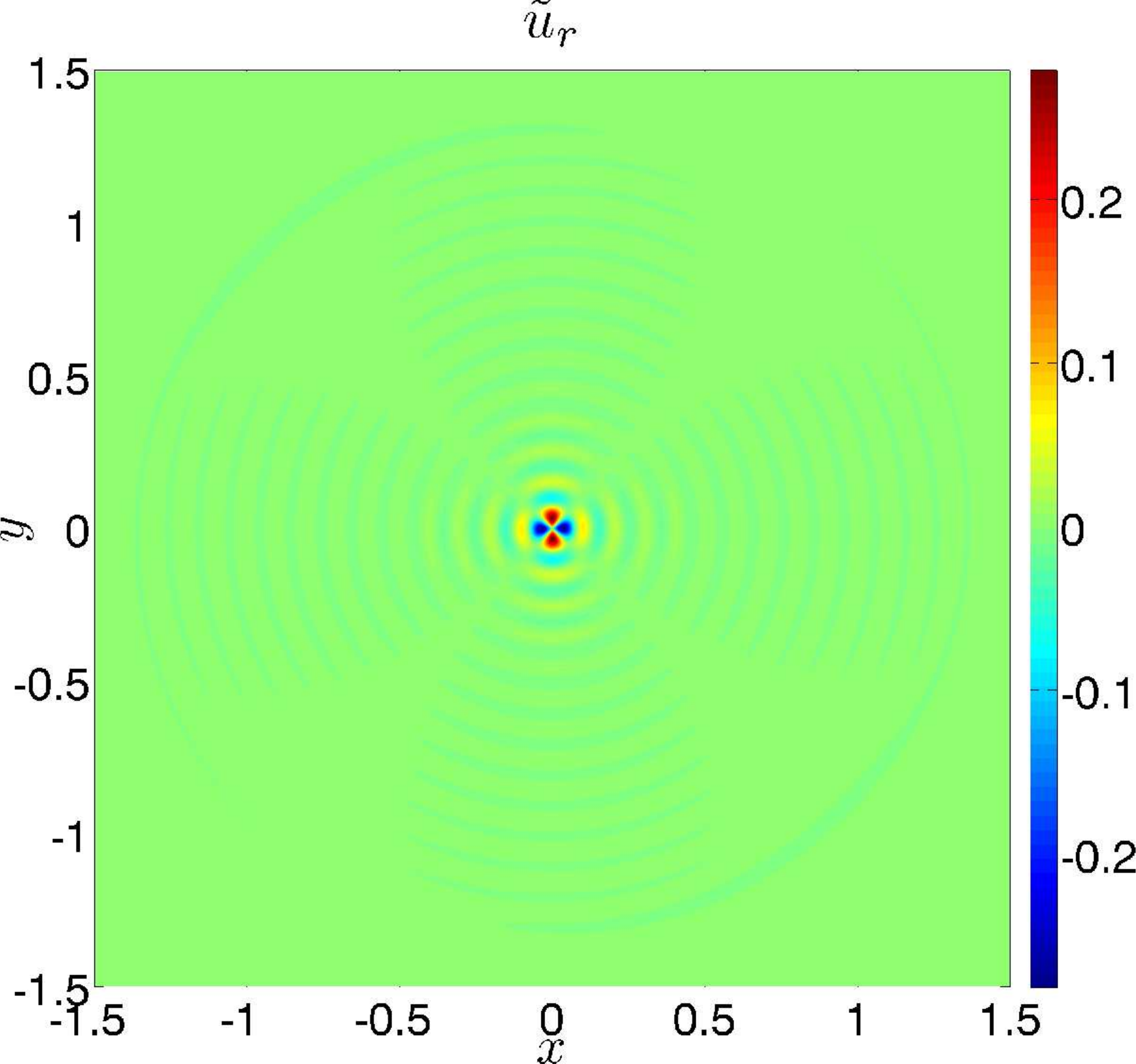}
    }
    \subfigure{
      \includegraphics[width=0.4\textwidth]{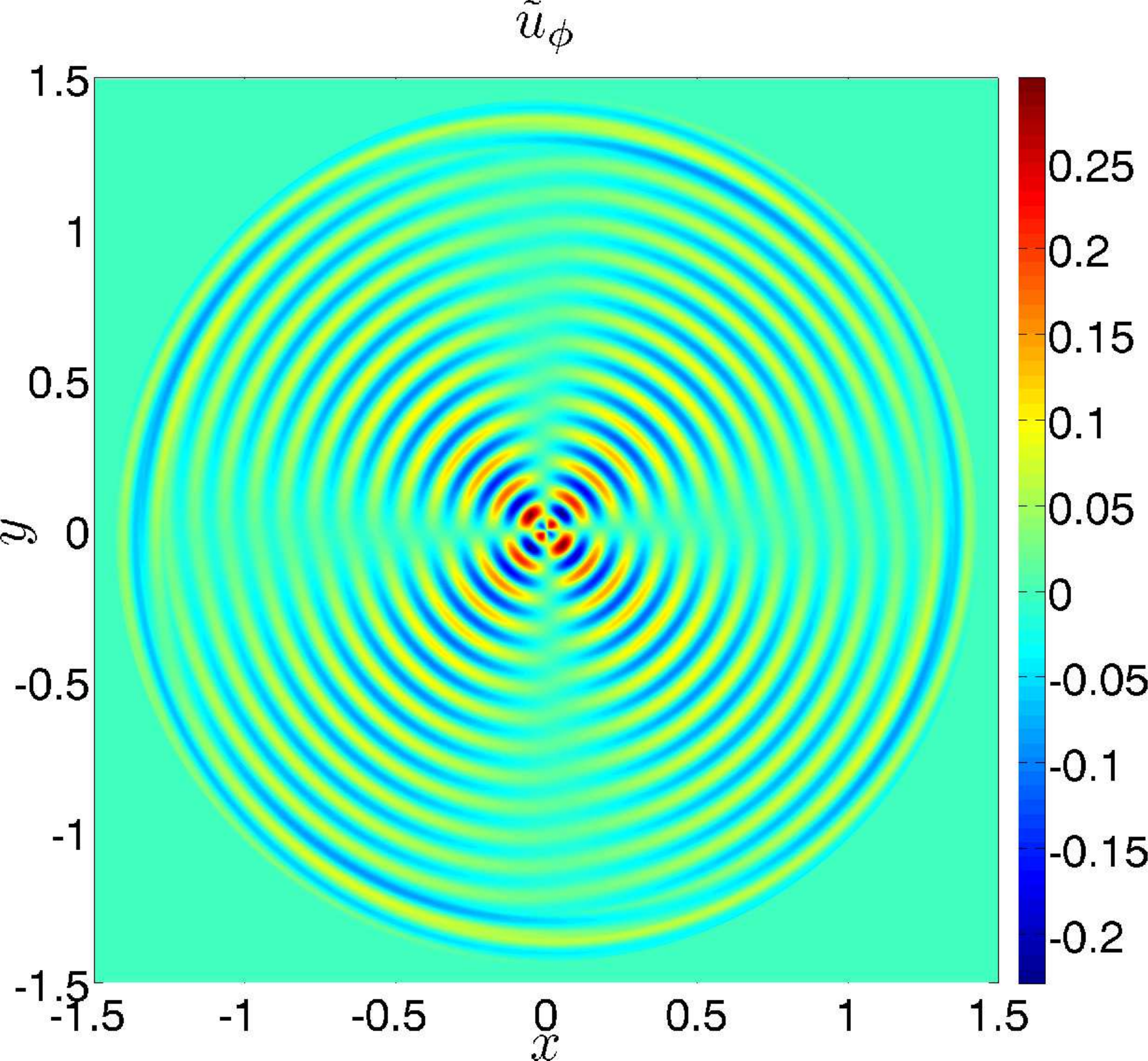}
    }
  \end{center}
  \caption{2D plot of the radial (top) and
  azimuthal (bottom) velocity in the equatorial plane of the star for
  small-amplitude waves. This is at a time $t=36t_{c}$, once
  standing waves have formed, in a simulation
  with $\mathrm{max}(\tilde{u}_{\phi}) < 0.3$, in which we have coherent reflection from the
  centre. In the outer part
  of the grid, the solution is smoothed to zero to satisfy periodic boundary conditions.}
  \label{Fig:2Dlinearsim}
\end{figure}

When the simulations are started, transients are excited by the forcing at
many different frequencies (and radial wavelengths), centred around
$\omega=1$ in frequency space. As more inward propagating waves
are excited by the forcing, an ingoing wave train
propagates toward the centre. At this stage in the time evolution, the
solution is composed of many different frequencies, so our
decomposition of the solution into a single IW and OW does not work
well. As more transients escape the region
and are damped, the primary response of the fluid is in the form
of waves with frequency $\omega=1$ and azimuthal wavenumber $m=2$.

\begin{figure}
  \begin{center}
    \subfigure{\label{linref1}
      \includegraphics[width=0.41\textwidth]{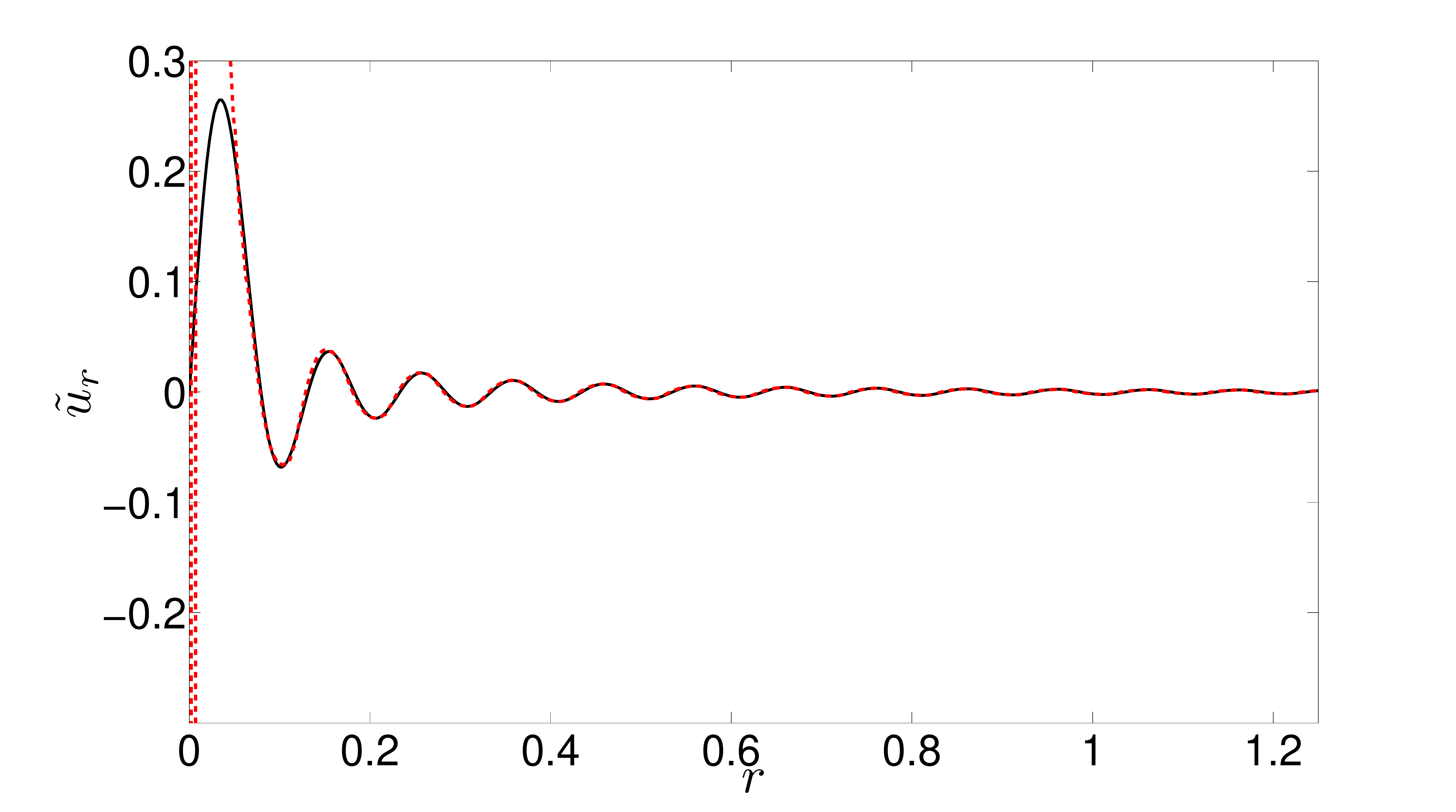}
    }
    \subfigure{
      \includegraphics[width=0.41\textwidth]{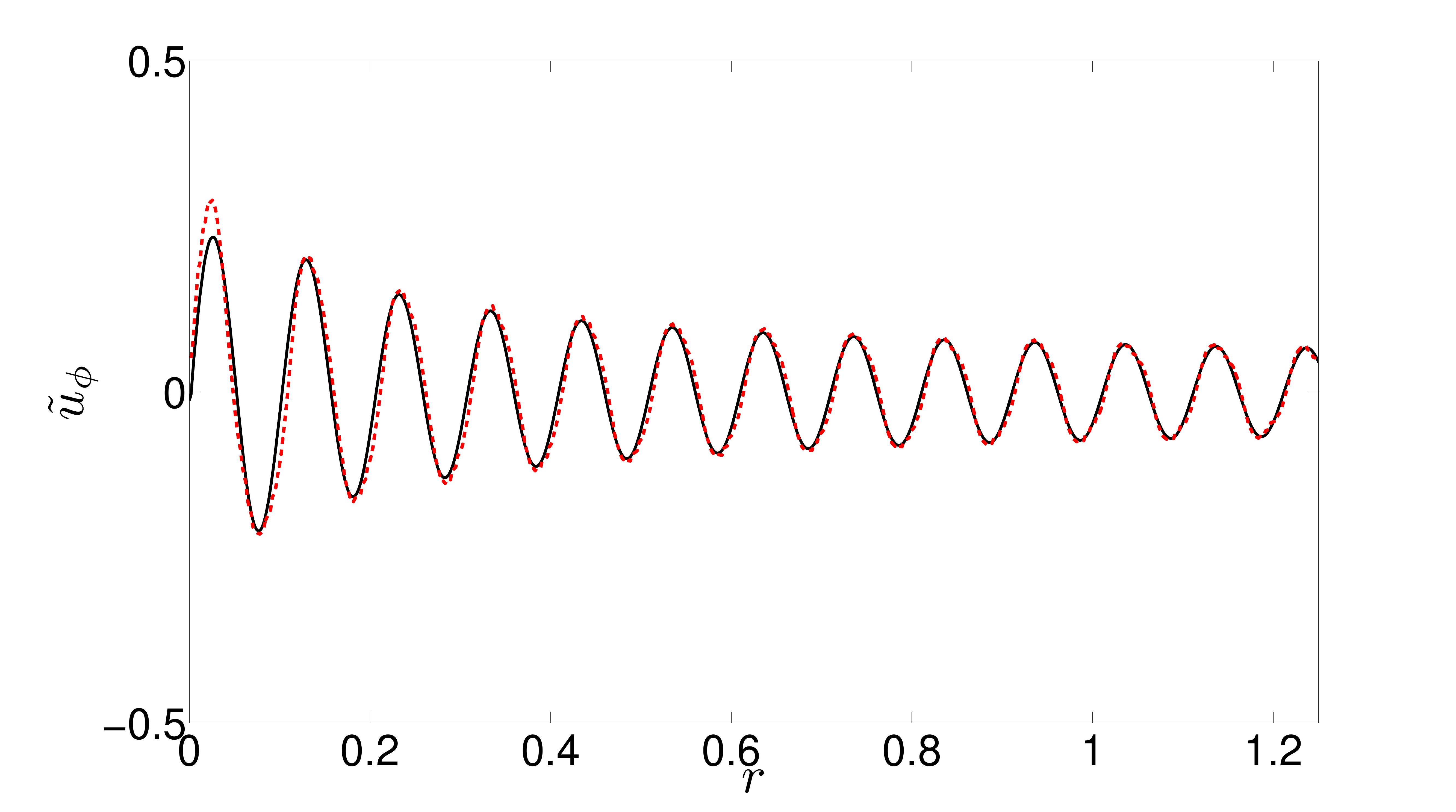}}
    \subfigure{\label{linref2}
      \includegraphics[width=0.41\textwidth]{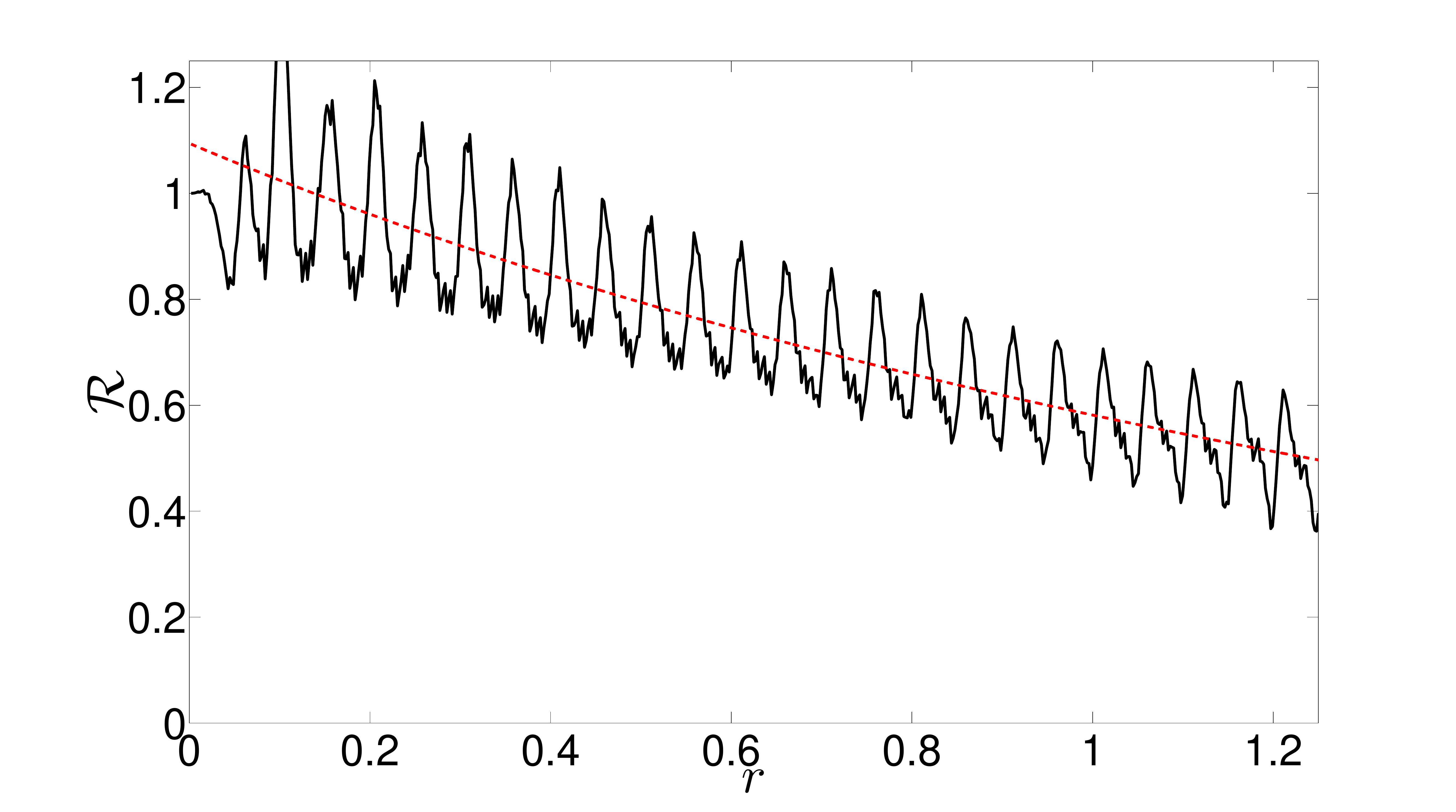}
    }
    \subfigure{
      \includegraphics[width=0.41\textwidth]{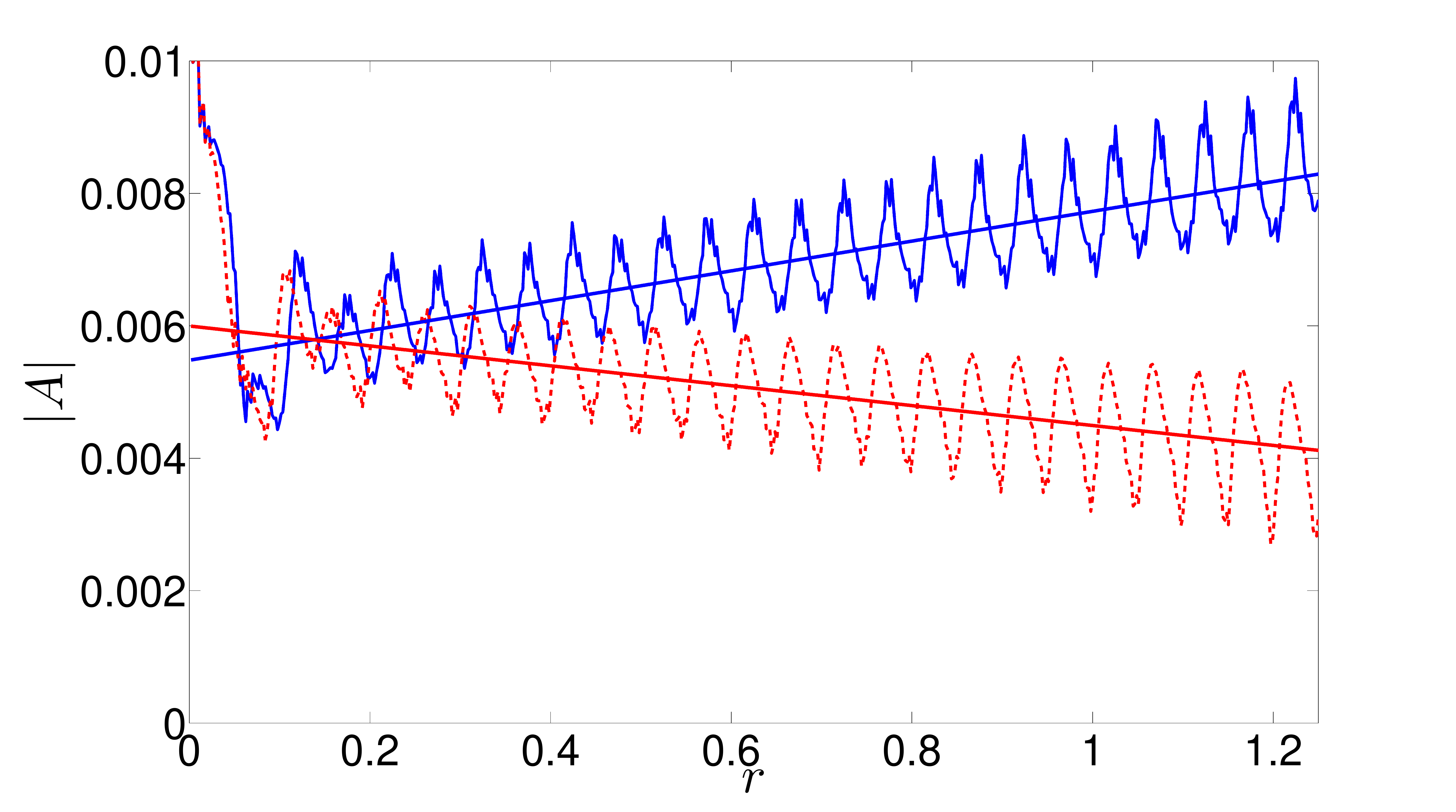}}
  \end{center}
  \caption{Radial velocity along the line $y$-axis
  (top) and azimuthal velocity along the line $y=x$ (middle top)
  in a small-amplitude simulation, after
  standing waves have been set up at $t=36 t_{c}$, with $\mathrm{max}(\tilde{u}_{\phi})
  \sim 0.3$ (solid lines). Also plotted is the reconstructed solution using
  $A_{in}$ and $A_{out}$ obtained using the method described in
  Appendix \ref{refcoeff} (dashed lines). These are well matched everywhere except near the
  centre showing that our decomposition works well for
  these cases. In the bottom right we plot $|A_{in}|$ (solid curve)
  and $|A_{out}|$ (dashed curve) versus radius, together with a linear fit to each curve. 
  The waves damp as they propagate due to viscosity. In the bottom panel
  we plot the reflection coefficient $\mathcal{R}$ versus radius.
  }
  \label{Fig:linearsimtestt45fr10tstep10lowth}
\end{figure}

As the waves approach the centre and reflect, an OW is
produced. As this process continues, the amplitude of the OW matches that of the
IW near the centre. For these low-amplitude cases, the phase change on reflection
is negligible. This means that we have coherent
reflection from the centre, which allows standing waves to be
produced. These waves are stationary in a frame rotating with
$\Omega_{p}$, as inferred from linear theory.
We confirm that our simulations
produce the correct standing wave solution, by plotting an example of a
comparison between the simulation and the wave solution in
Fig.~\ref{Fig:Comparisonwithanalytic}.
We plot the velocity components in two dimensions for an example simulation in which
standing waves have formed for a small-amplitude case with $\mathrm{max}(\tilde{u}_{\phi}) \sim
0.3$ in Fig.~\ref{Fig:2Dlinearsim}.

After a few wave crossing times, the reflection coefficient increases to
values approaching unity throughout the grid, though its value
decreases with radius, shown in
Fig.~\ref{Fig:linearsimtestt45fr10tstep10lowth} for a 
low-amplitude case, with $\mathrm{max}(\tilde{u}_{\phi}) \sim 0.3$.
In this figure we also plot the results
of our IW/OW decomposition in a small-amplitude simulation with a resolution
$1536\times 1536$. Our reconstructed solutions match the data
well except very close to the centre, thus showing 
that our decomposition works well for these cases.

The decay in $\mathcal{R}$ with radius is a result of the nonzero
viscosity, which results in a decay of wave amplitude with
time (and therefore distance from where they are excited). 
The OW has been damped for longer, which results in the amplitude of the OW being smaller than
that of the IW. A simple estimate of the amplitude decay due to
viscosity with propagation from radius $r$ and then reflected back to $r$ again gives 
\begin{eqnarray}
\label{viscousdamping}
\frac{u_{r}}{u_{r,0}} \propto \exp \left(-2\int_{0}^{r} \frac{\nu
  k^{2}}{c_{g,r}} dr\right) \approx \exp \left(-\frac{16 \pi^{3} \nu}{\omega
  \lambda_{r}^{3}}r\right),
\end{eqnarray}
since $k \sim k_{r}$ except near the centre, and $c_{g,r}\simeq \omega
\lambda_{r}/2\pi$ throughout the box. This roughly matches the amplitude decay
between $A_{in}$ and $A_{out}$ at $r=1.2$, implying that the decay in
amplitude is indeed due to viscous damping of the waves. In addition, the regular
oscillations in the amplitudes result from the fact that our exact
solution in the inviscid case is not an exact solution
in the presence of viscosity. This was verified by running a
low-amplitude simulation with $\nu = 0$, in which case the
oscillations disappear.

\begin{figure}
  \begin{center}
    \subfigure{
      \includegraphics[width=0.51\textwidth]{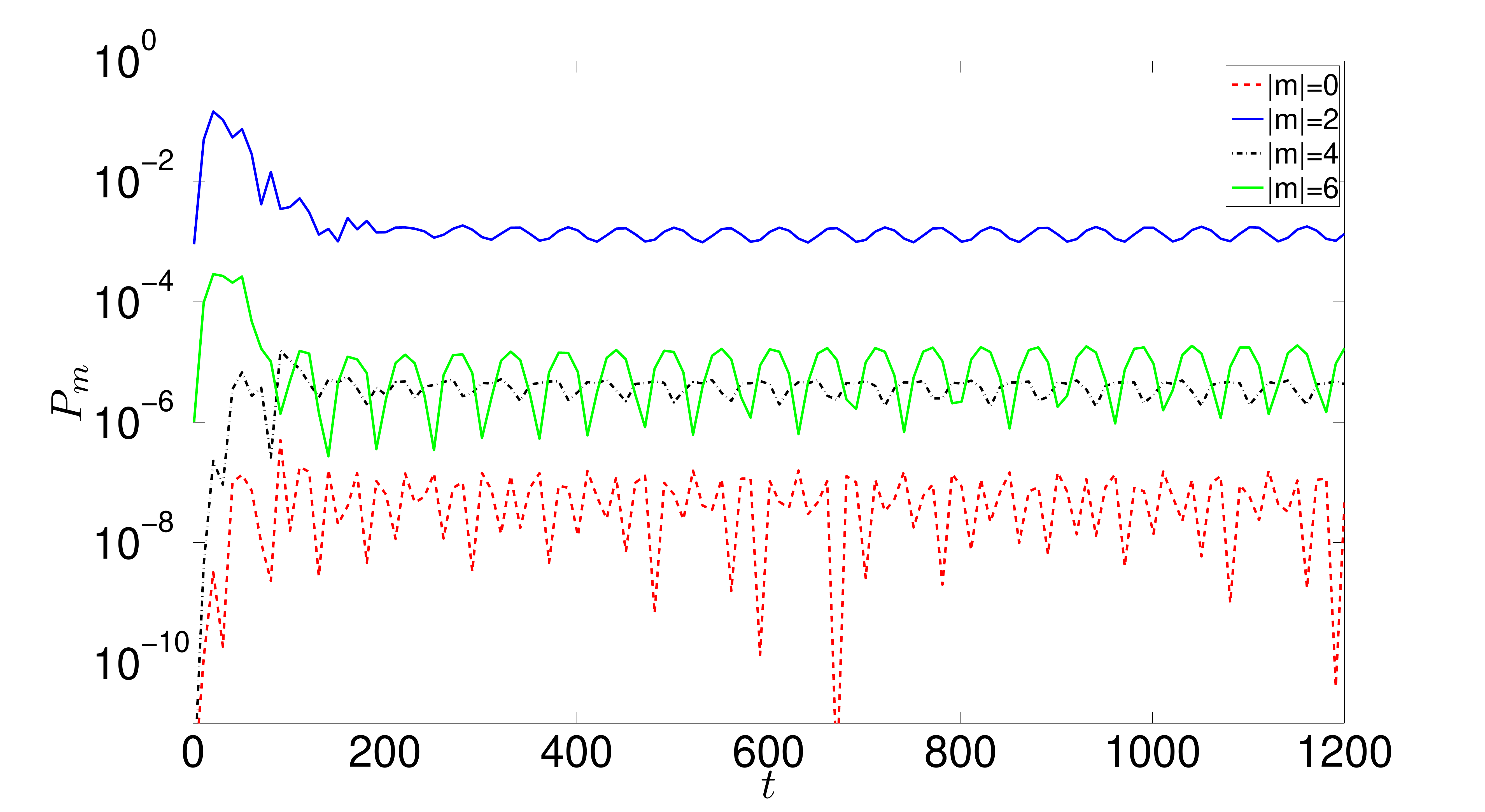}
    }
    \subfigure{
      \includegraphics[width=0.51\textwidth]{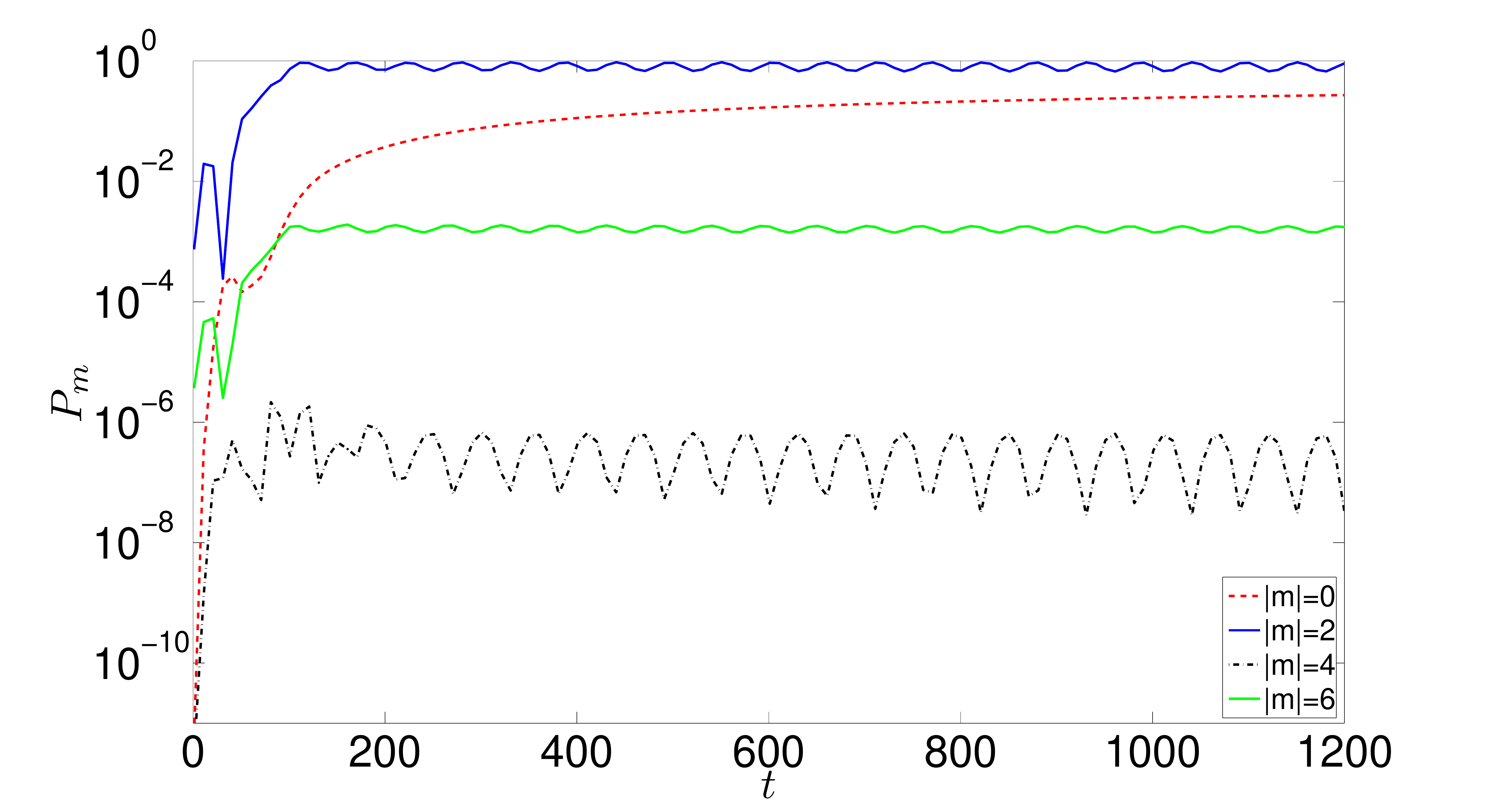}
    }
  \end{center}
  \caption{Temporal evolution of the power spectral density $P_{m}$ at
    $r=0.1$, in the lowest four azimuthal wavenumbers
    $m$ in the solution for $u_{r}$
    (top) and $u_{\phi}$ (bottom), in a low-amplitude simulation with
    $\tilde{u}_{\phi} \sim 0.3$ near the centre, for grid resolution
    $512\times 512$. The solution
    is in the form of $m=2$ waves, and reaches a steady state in the frame
    rotating with $\Omega_{p}$. Growth of $m=0$ is
    nonzero as a result of viscous damping of the waves. No wave breaking
    occurs because $\tilde{u}_{\phi} < 0.5$ in the solution. The $m=6$ components are most likely due to
    errors in the Fourier analysis.
  }
  \label{Fig:FourierAnalysisLIN}
\end{figure}

We find that the wave reflects coherently from the centre when
$\tilde{u}_{\phi} \lesssim 0.5$. Long-term simulations ($t \sim$
several hundred $t_{c}$) do not show the development of any
instabilities that act on waves with $\tilde{u}_{\phi} \lesssim 0.5$,
though there is a slow growth of $m=0$ components of
$\tilde{u}_{\phi}$ in the solution, as can be seen in
Fig.~\ref{Fig:FourierAnalysisLIN}. In this figure, we plot $P_{m}$ for
the first few even wavenumbers in the flow from an example
low-amplitude simulation. Negligible growth in odd $m$-values is
observed, which is consistent with the symmetry of the basic wave 
and the quadratic nonlinearities of the Boussinesq-type system.
The growth in $m=0$ is a result of viscosity, which acts to damp the waves and transfer angular momentum
from $m=2$ to the mean flow. This can distinguished from a process
resulting from nonlinear interactions, because it is found to depend
on $\nu$. However, most importantly, no instability is
observed for waves with $\tilde{u}_{\phi} \lesssim 0.5$.

\section{High-amplitude forcing: wave breaking and critical layer formation}
\label{nonlinearsims}

\begin{figure}
  \begin{center}
    \subfigure{
      \includegraphics[width=0.375\textwidth]{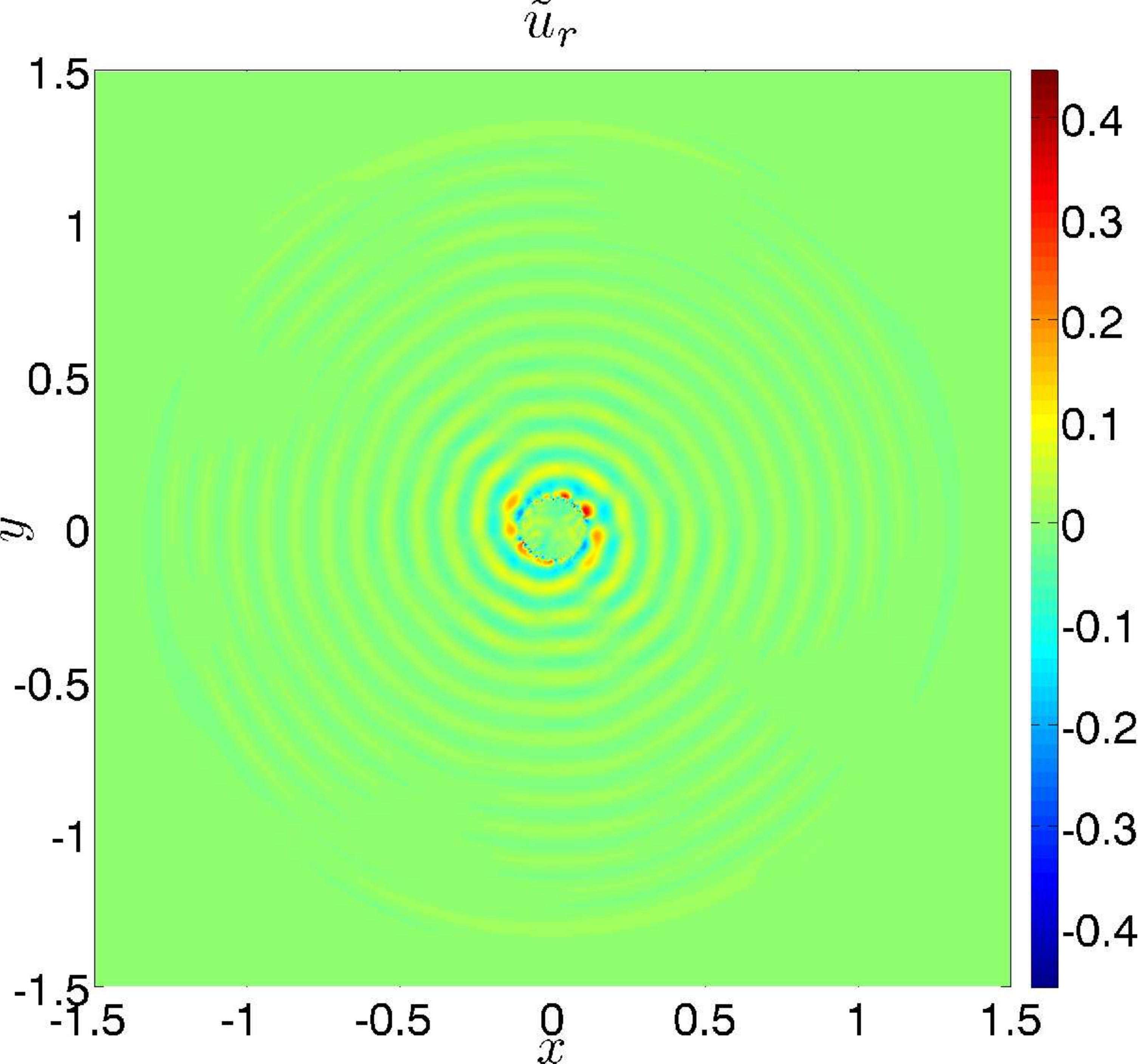}
    }
    \subfigure{
      \includegraphics[width=0.375\textwidth]{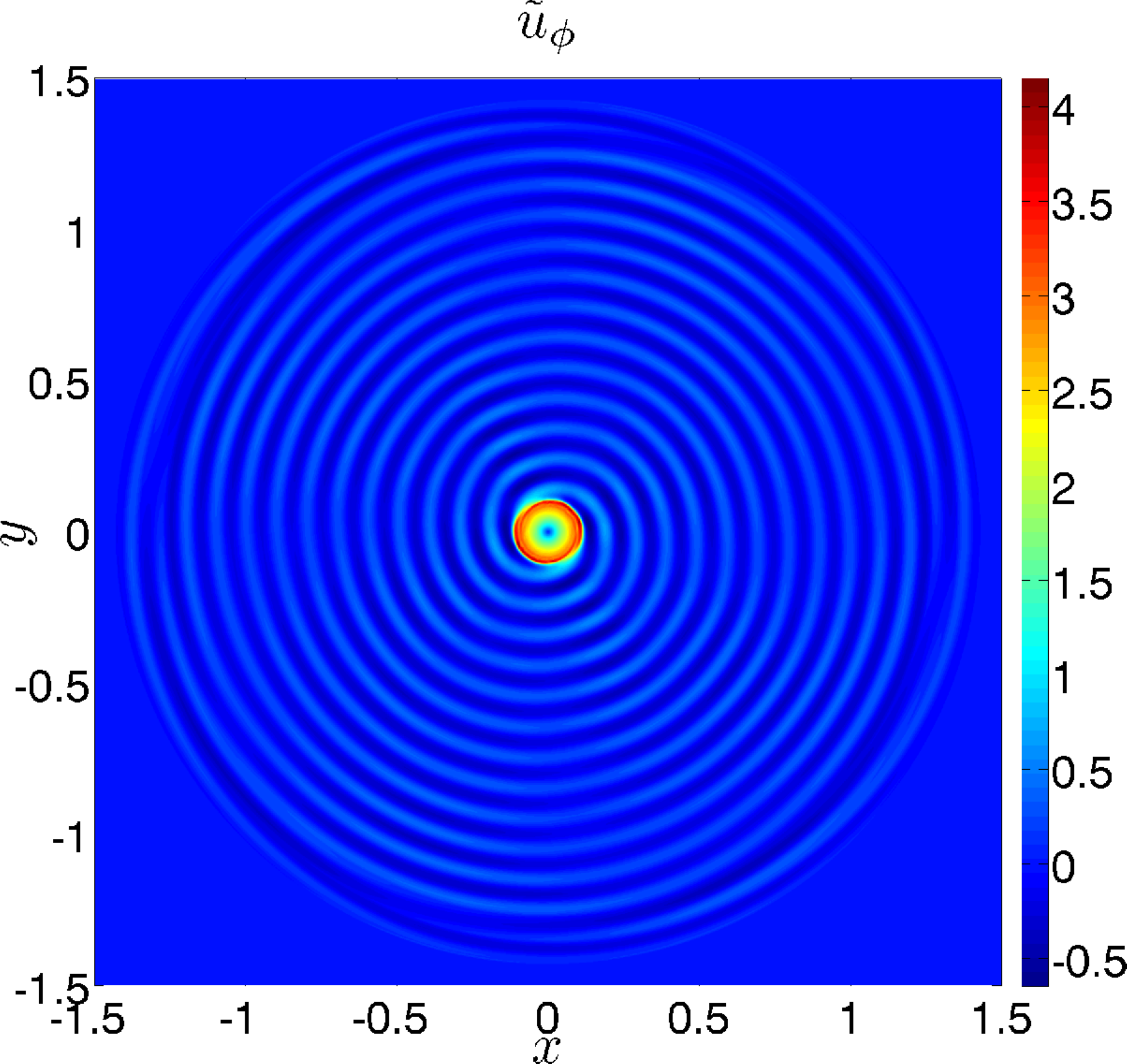}
    }
\end{center}
\caption{The top panel shows the radial velocity in a high-amplitude simulation
  with $\tilde{f}_{r}=15$ at $t=7 t_{c}$, after wave breaking has
  occurred, from a simulation with a resolution $1536\times 1536$. The bottom
  panel shows $\tilde{u}_{\phi}$ at the same time in the simulation.}
\label{2dnonlinear} 
\end{figure}

If we increase the value of $\tilde{f}_{r}$, then the above picture
changes considerably when a critical wave amplitude is exceeded. Once
$\tilde{u}_{\phi} > \tilde{u}_{\phi,crit} \approx 0.5$, wave breaking
occurs near the centre within several wave periods (a few
$2\pi/\omega$), and the outcome of the simulations is very different
from the small-amplitude case. This occurs when the wave overturns the
stratification -- see Eqs.~\ref{convectiveinstability} and
\ref{convinstcrit}.  In Fig.~\ref{2dnonlinear} we plot the 2D velocity
components after wave breaking has occurred in a simulation with
$\tilde{f}_{r} = 15$.

For highly nonlinear forcing, for example with $\tilde{f}_{r} > 20$,
the waves break as they reach the centre with sufficient amplitude
before there has been any significant reflection. For $\tilde{f}_{r}
\sim 10$, the amplitude of the IW alone is insufficient to cause
breaking, and we must wait for reflection at
the centre to produce an OW of comparable amplitude before
$\tilde{u}_{\phi} > \tilde{u}_{\phi,crit}$. Once this critical value
of $\tilde{u}_{\phi}$ is exceeded, the waves break. This is an
irreversible deformation of the otherwise wavy material contours
\citep{Perspectives2000}.

\begin{figure}
  \begin{center}
    \subfigure{
      \includegraphics[width=0.4\textwidth]{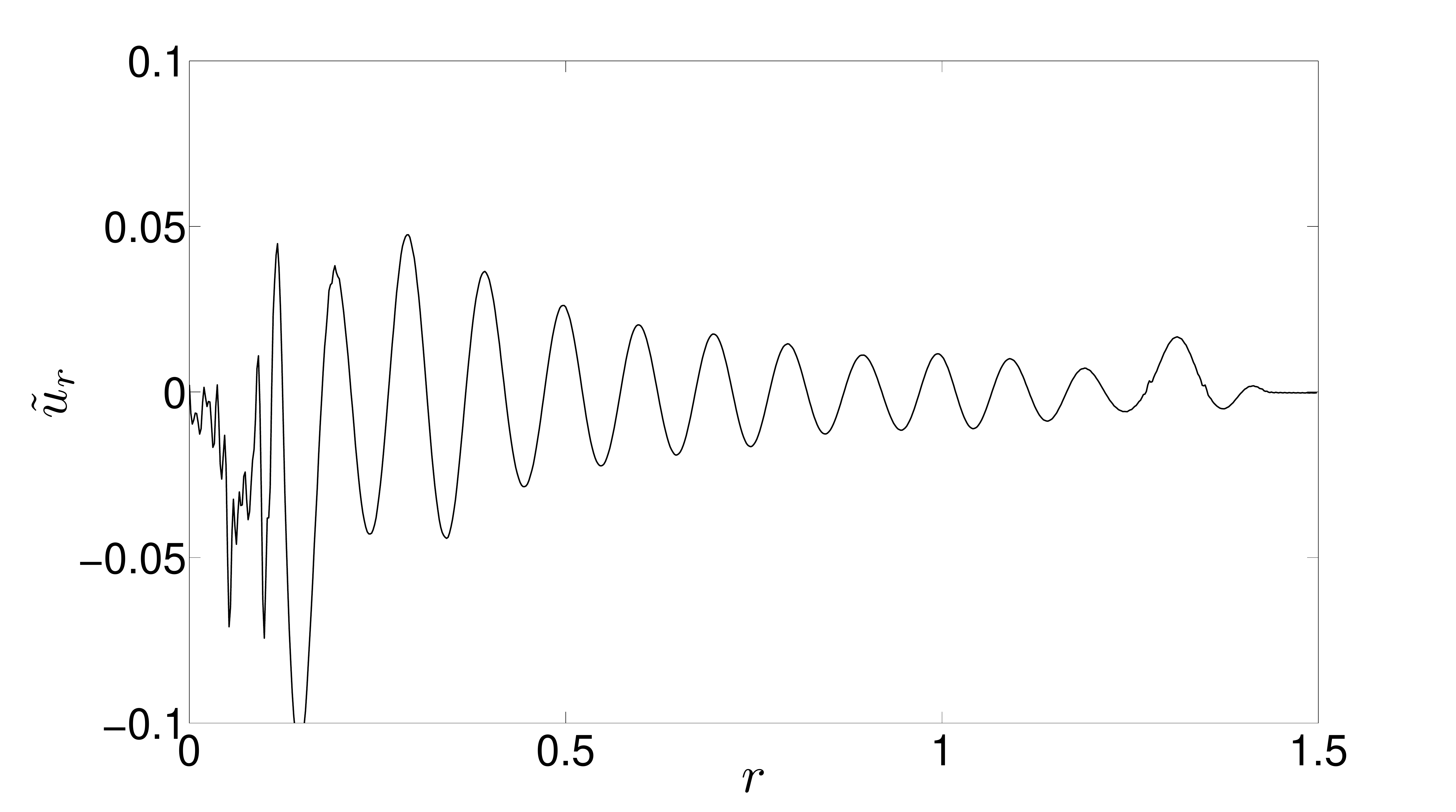}
    }
    \subfigure{
      \includegraphics[width=0.4\textwidth]{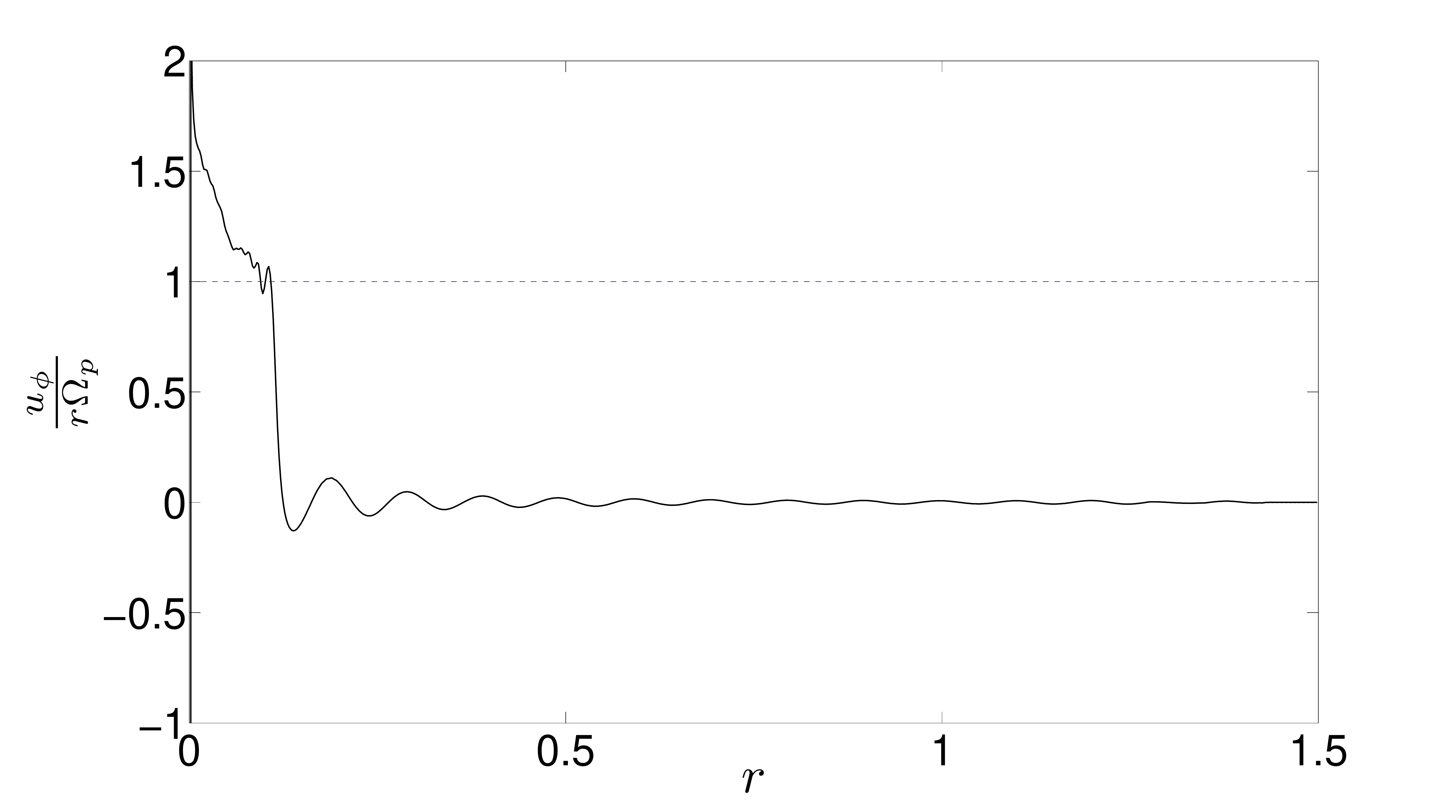}
    }
  \end{center}
  \caption{The top panel shows the radial velocity along $y=x$ for the same
  case as Fig.~\ref{2dnonlinear}. In the region $r \in [0.3,1.2]$ the solution is well
  described by linear waves with $|A_{out}| \ll |A_{in}|$, but not
  near the centre. The bottom
  panel shows the angular velocity of the fluid
  normalised to the angular pattern speed of the forcing
  $\Omega_{p}$ along the $x$-axis, at the same
  time in the simulation. This shows that the central regions after wave
  breaking are spun up to slightly exceed $\Omega_{p}$. The critical
  layer occurs where $u_{\phi}/(r\Omega_{p}) = 1$.}
  \label{Fig:spinupcentre}
\end{figure}

Once breaking occurs, we observe consequent mean flow acceleration
(i.e. growth of $m=0$ components of $u_{\phi}$), as the angular
momentum of the waves is deposited locally where the wave breaks. This
acts to spin up (if $\Omega_{p} > 0$) 
the central regions, which at this stage contain
a sufficiently small fraction of the angular momentum of the star that
their spin can be readily affected by these waves. Once this process
has begun, the central regions spin up to $\sim \Omega_{p}$, and a
critical layer is formed, at which the Doppler-shifted frequency of
the waves goes to zero.
At this location, the azimuthal phase
velocity of the waves would equal that of the local rotation of the
fluid, if they were ever to reach it intact. In reality, as subsequent
IWs approach the critical layer, nonlinearities dominate, and the
waves undergo breaking before they reach it - though see
discussion in \S~\ref{refcoeffresults}. We plot the angular velocity
of the fluid normalised to $\Omega_{p}$ in the bottom panel of Fig.~\ref{Fig:spinupcentre}.

As IWs approach the critical layer, their radial wavelength decreases,
and they slow down, i.e. $\hat{c}_{g,r} \rightarrow 0$ as
$\hat{\omega}\rightarrow 0$. This causes a buildup of wave energy just
above the critical layer, in which nonlinearities become important. In
this thin region, the quadratic nonlinearities produce higher
wavenumber disturbances with even $m$ values -- see
\S~\ref{Fourier}. These are produced by the self-nonlinearity of the
primary IW ($m=2$) as it approaches the critical layer -- these
self-nonlinearities vanish in the absence of a mean flow. These daughter
waves are damped faster than the primary because they have lower
frequencies and therefore shorter radial wavelengths, which is a
result of the theorem proved in \cite{Hasselman1967}. Thus the IW is
irreversibly deformed, and transfers its angular momentum to
either the mean flow or to daughter waves that are then more easily
dissipated by viscosity. A large fraction of the angular momentum of
these waves must be given to the mean flow when these waves
dissipate. This process acts to spin up the fluid just above the
critical layer to $\sim \Omega_{p}$. As subsequent IWs are absorbed by
the critical layer, the spatial extent of the mean flow expands
outwards, i.e. the star is spun up from the inside out. We envisage
that this process will continue until the mean flow encompasses the
bulk of the radiation zone or the planet plunges into the star --
though see \S~\ref{discussion}.  Long-term simulations, lasting for
several hundred $t_{c}$, show that this
appears to be the case.

This picture is analogous to \cite{Goldreich1989}, who propose that
early-type stars in close binaries would spin down (if $\Omega > n$, or spin up if
$\Omega < n$) from the outside in, once a critical layer has formed
near the surface as a result of radiative damping of the waves, where
this effect is strong. In our problem, an instability of the primary
wave, which occurs once the wave overturns the stratification, causes
wave breaking. 
This results in angular momentum
deposition and spin up (for the case in which
$\Omega < n$, spin down if $\Omega > n$) of the central regions, 
which causes the formation of a critical
layer near the centre of a solar-type star. The rate of expansion of the
spatial extent of this region depends on the forcing amplitude; for
larger amplitudes it expands faster.  The critical layer moves
outwards when the dissipation of subsequent IWs deposits sufficient
angular momentum to spin up the fluid to $\sim \Omega_{p}$, and the
spatial extent of the mean flow expands.  In this picture, there is a
front of synchronization which gradually moves outwards (though see
\S~\ref{discussion}).

\subsection{Growth of azimuthal wavenumbers in the flow}
\label{Fourier}

\begin{figure}
  \begin{center}
    \subfigure{
      \includegraphics[width=0.51\textwidth]{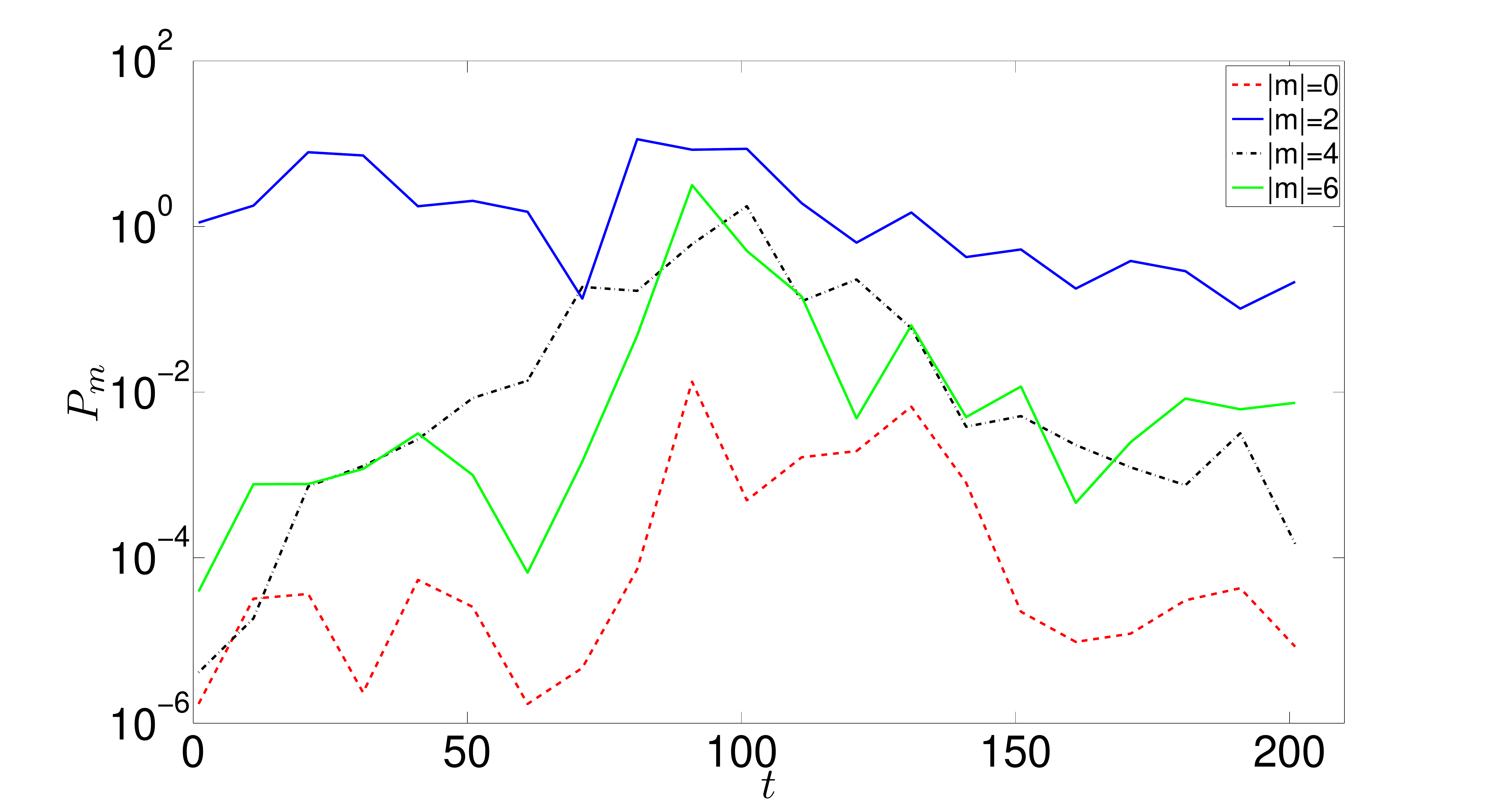}
    }
    \subfigure{
      \includegraphics[width=0.51\textwidth]{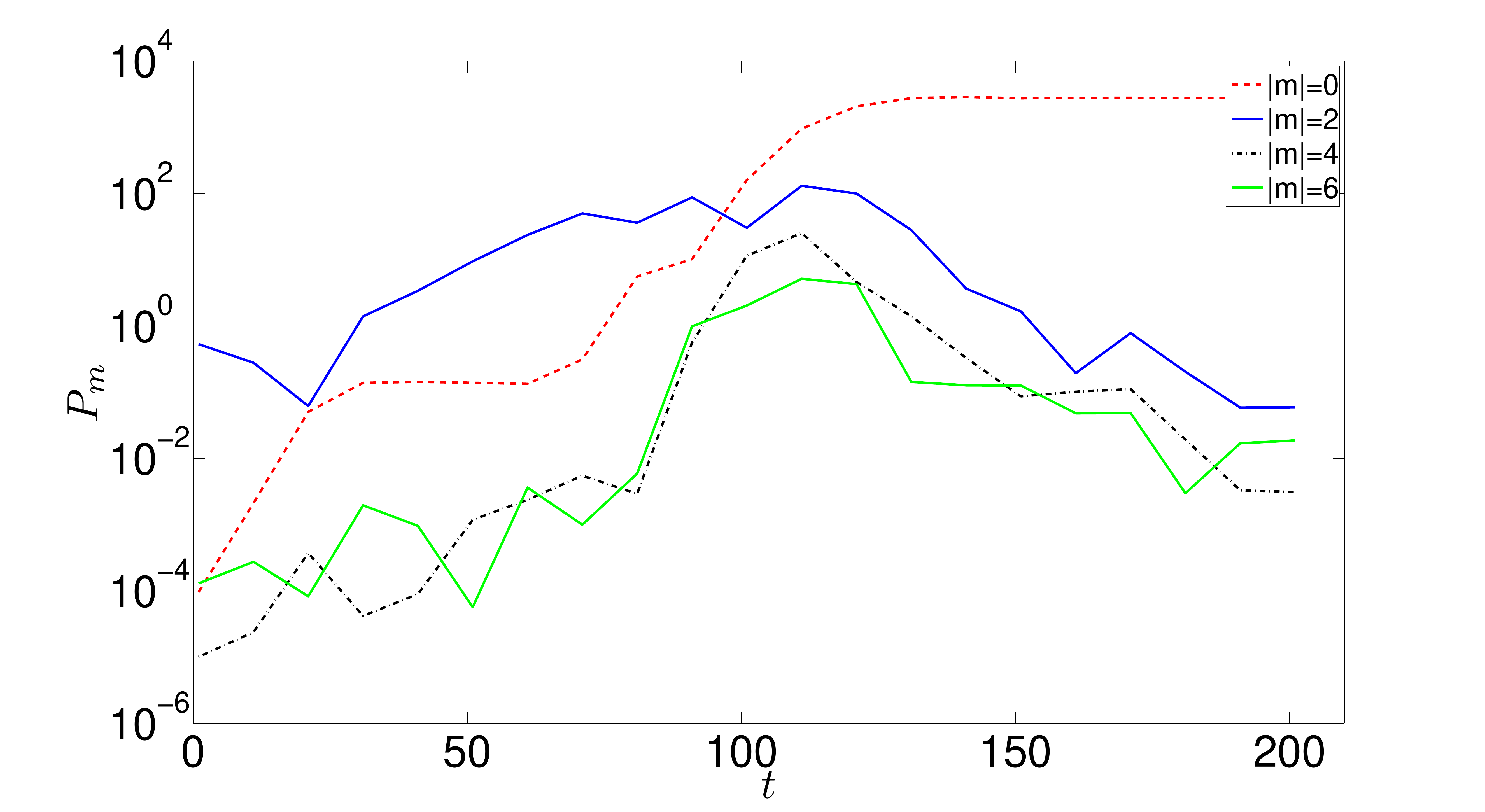}
    }
  \end{center}
  \caption{Temporal evolution of the power
  spectral density $P_{m}$ at $r=0.1$, in the lowest four azimuthal wavenumbers
  $m$ in the solution for $u_{r}$
  (top) and $u_{\phi}$ (bottom), in a simulation with
  $\tilde{f}_{r} = 15$ for resolution $1536\times1536$. The solution is primarily in the form of
  $m=2$ waves, until growth of even-$m$ disturbances
  occurs as the wave breaks. Note that $m=0$ grows strongest for
  $u_{\phi}$, i.e. angular momentum is transferred
  from the primary waves to the mean flow. Once $r=0.1$ is
  located inside the rotating region, $u_{r}$ is primarily composed of
  $|m|=2$.}
  \label{Fig:FourierAnalysis1}
\end{figure}

\begin{figure}
  \begin{center}
    \subfigure{
      \includegraphics[width=0.51\textwidth]{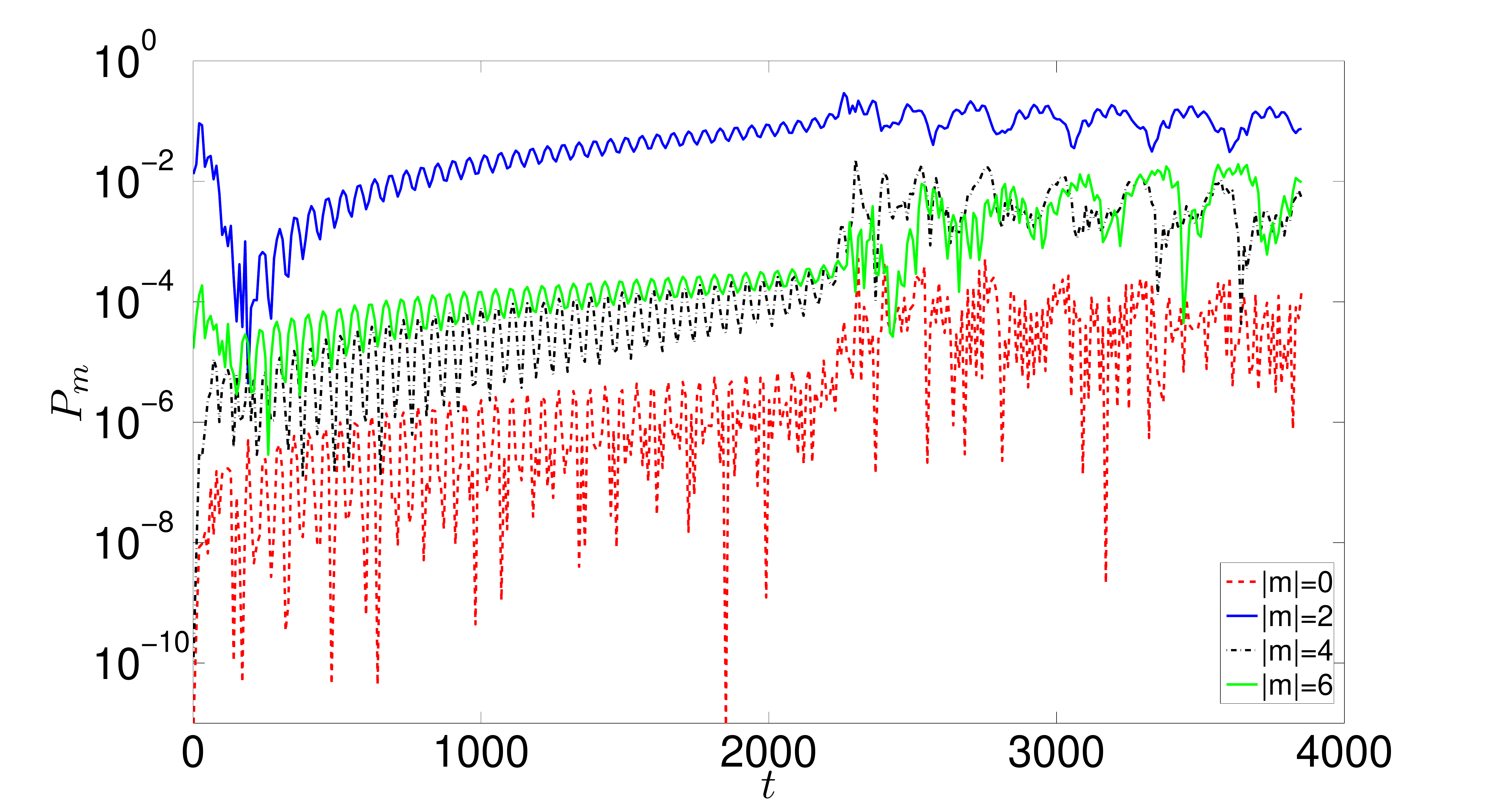}
    }
    \subfigure{
      \includegraphics[width=0.51\textwidth]{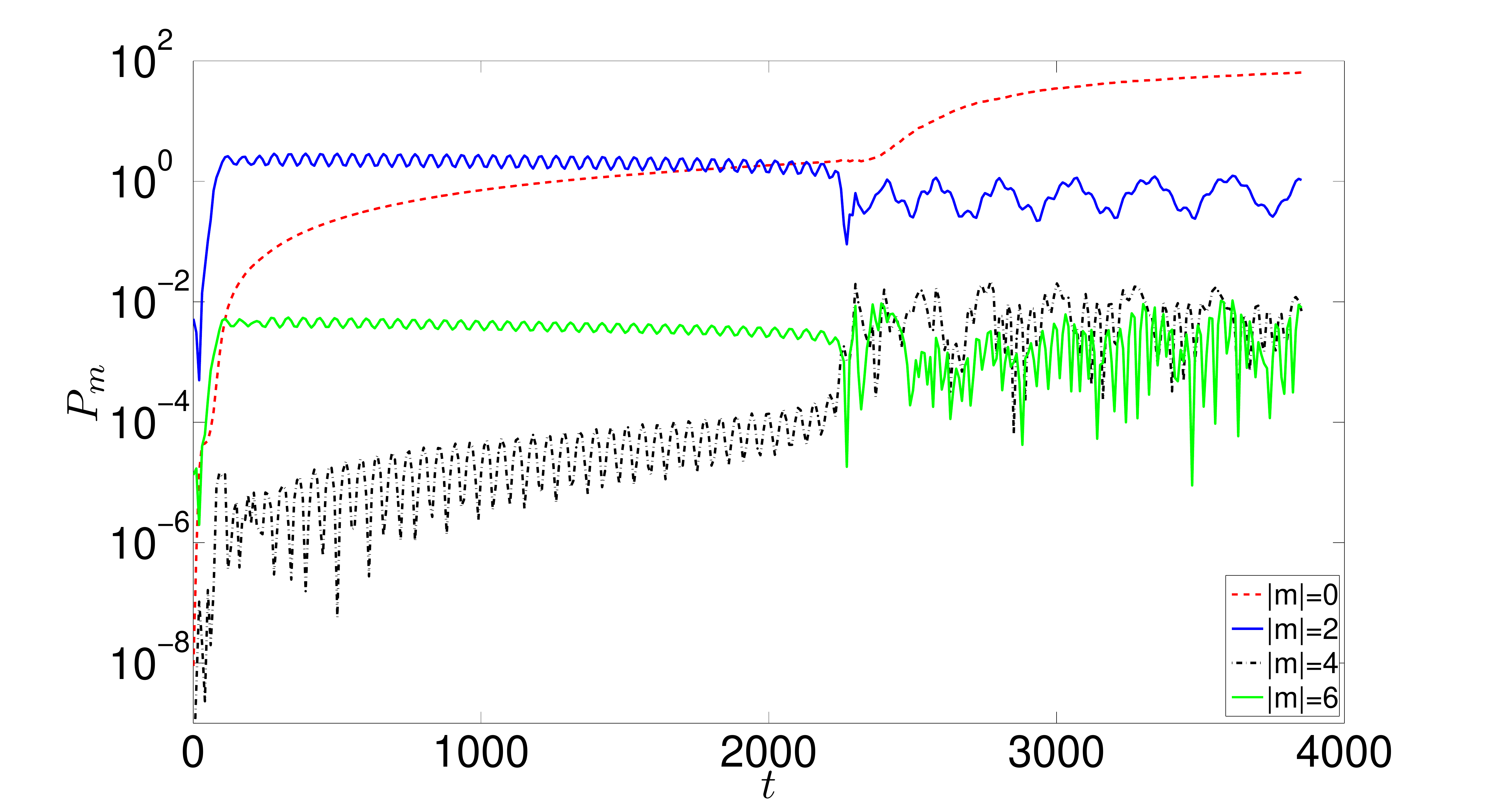}
    }
  \end{center}
  \caption{Temporal evolution of the power spectral density $P_{m}$ at
    $r=0.1$, in the lowest four azimuthal wavenumbers
  $m$ in the solution for $u_{r}$
  (top) and $u_{\phi}$ (bottom), in a simulation with
  $\tilde{f}_{r} = 2.5$ for resolution $512\times 512$. Until $t=2200$, viscous
  damping acts on the waves, transferring angular momentum to the mean
  flow. Once $u_{\phi} > u_{\phi,crit}$ in the
  solution, wave breaking occurs. This occurs at $t\sim 2200$, resulting in a jump
  in the growth of $m=0$, and a drop of energy in $|m|=2$.}
  \label{Fig:FourierAnalysis2}
\end{figure}

We now discuss the results of a spectral analysis of the simulation data so that we
can study the growth of the mean flow ($m=0$) and daughter waves
produced by breaking (other $|m| \ne 2$ wavenumbers). We plot $P_{m}$ for
the first few even wavenumbers for a set of examples in
Figs.~\ref{Fig:FourierAnalysis1} and
\ref{Fig:FourierAnalysis2}. Negligible growth in odd $m$-values is
observed, which is consistent with the symmetry of the basic wave 
and the quadratic nonlinearities (though it is possible in principle for the wave
to be unstable to odd-$m$ perturbations).

At the beginning of the simulations $m=2$ dominates until the primary
wave breaks and transfers angular momentum to the mean flow. When subsequent
waves approach the critical layer the primary IW transfers angular momentum to
higher $m$-value disturbances -- this can be seen from
Fig.~\ref{Fig:FourierAnalysis1} prior to $t\sim 100$, after which the ring
$r=0.1$ is enveloped by the mean flow. After this, $m=0$ dominates
$u_{\phi}$. On the other hand, $u_{r}$ is then primarily in the form of
$|m|=2$ disturbances, which from examination of simulation output,
counter-rotate with the forcing, and have angular pattern speed $-\Omega_{p}$.
The excitation of these waves could
explain the counter-intuitive effect of the most central regions spinning
slightly faster than $\Omega_{p}$, since they carry negative angular
momentum. These waves appear to \textit{reflect} from both the $m=2$
critical layer (though note that these waves do not see this as a
critical layer) and the centre. As these waves approach the $m=2$
critical layer, since they are counter-propagating waves, their frequency is
Doppler-shifted upwards towards $N$, and they undergo total
internal reflection. These waves appear to reflect back and forth from the
critical layer and the centre.

We also note the appearance of oscillations in the energy in $m=6$ in the
solution. This is most likely due to errors in the Fourier analysis,
because we are not sampling the solution with evenly spaced
points. The amplitude of these oscillations is much smaller than that
of the $m=2$ or $m=0$ waves, so should not affect any conclusions
drawn from these results.

We experimented with the forcing amplitude to study cases in which
wave amplitudes were just sufficient to cause breaking after running
the simulation for $\sim 25t_{c}$. The results of our spectral
analysis of the results of such a simulation are plotted in
Fig.~\ref{Fig:FourierAnalysis2}.  We clearly see evidence for viscous
damping in producing growth of energy in $m=0$. This can be
distinguished from the sudden growth which results after the onset of
the instability that leads to wave breaking. The growth of an
instability on the primary wave occurs once the wave overturns the
stratification, when $\tilde{u}_{\phi} > \tilde{u}_{\phi,crit}$ near
the centre, at $t=2200$. After onset, the wave breaks and a critical
layer is formed, leading to the general picture described above. For
this simulation, viscous damping of the wave is responsible for
spinning up the flow, and the formation of a critical layer.

\subsection{Discussion of wave reflection from the critical layer and
  implications}
\label{refcoeffresults}

\begin{figure}
  \begin{center}
    \subfigure{\label{linref1}
      \includegraphics[width=0.4\textwidth]{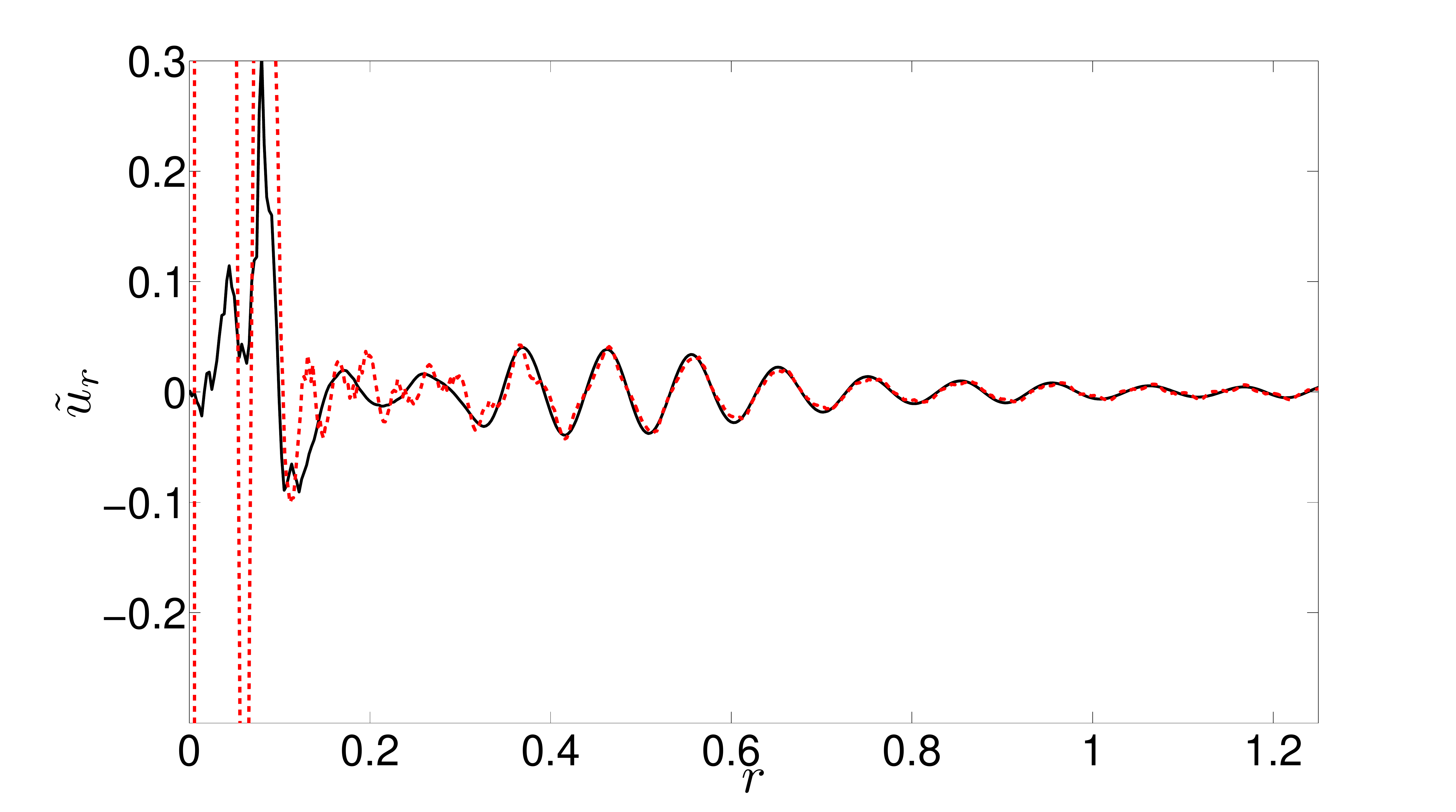}
    }
    \subfigure{
      \includegraphics[width=0.4\textwidth]{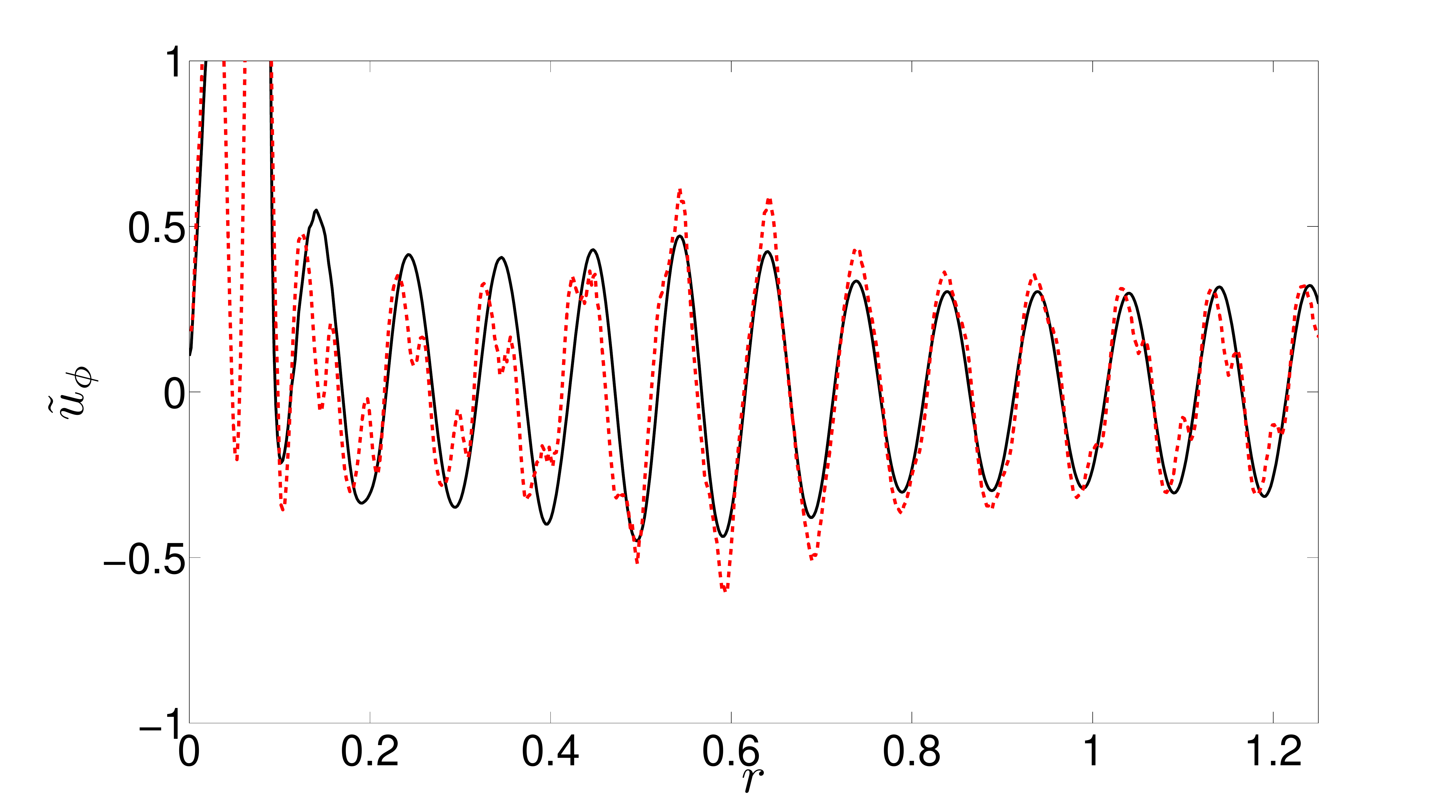}}
    \subfigure{\label{linref2}
      \includegraphics[width=0.4\textwidth]{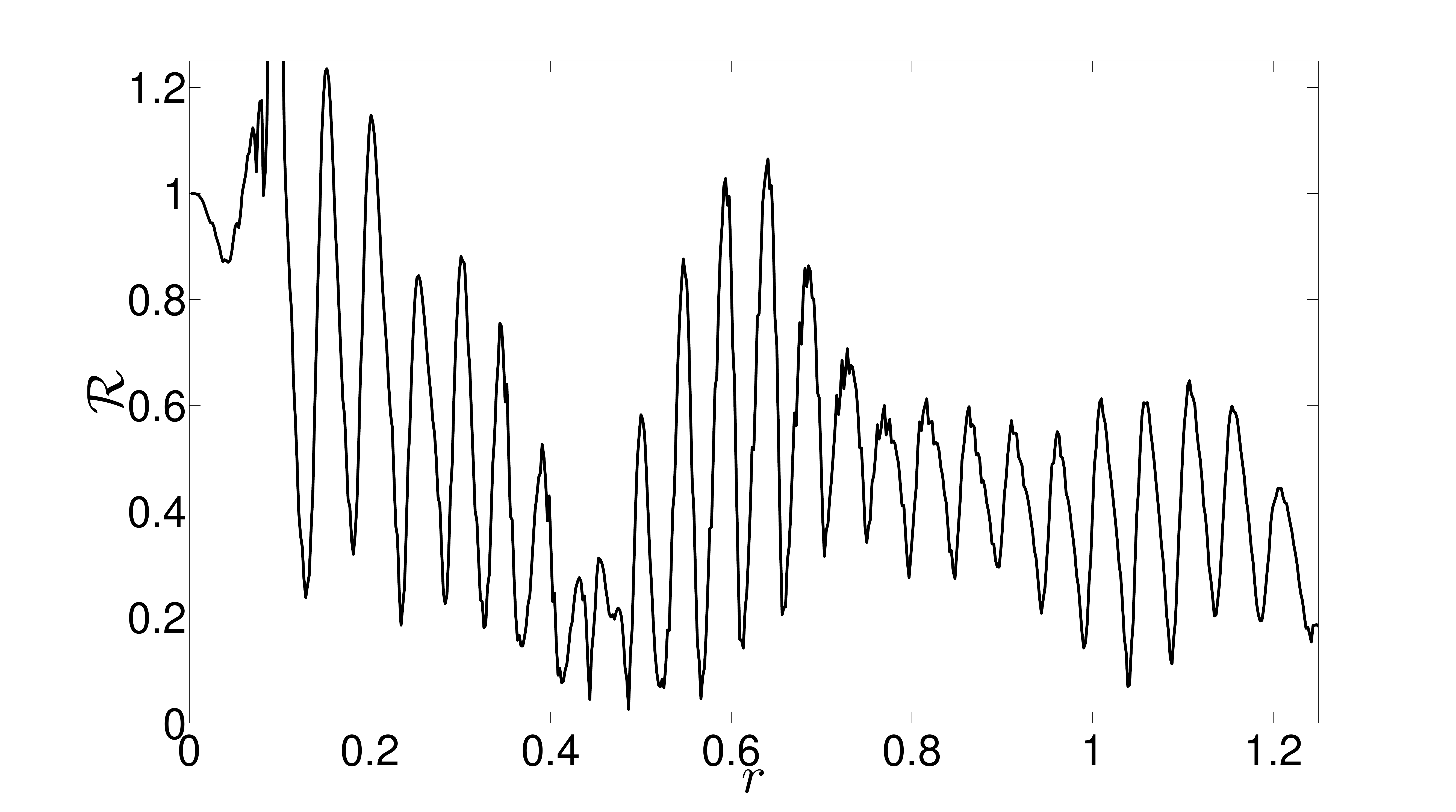}
    }
    \subfigure{
      \includegraphics[width=0.4\textwidth]{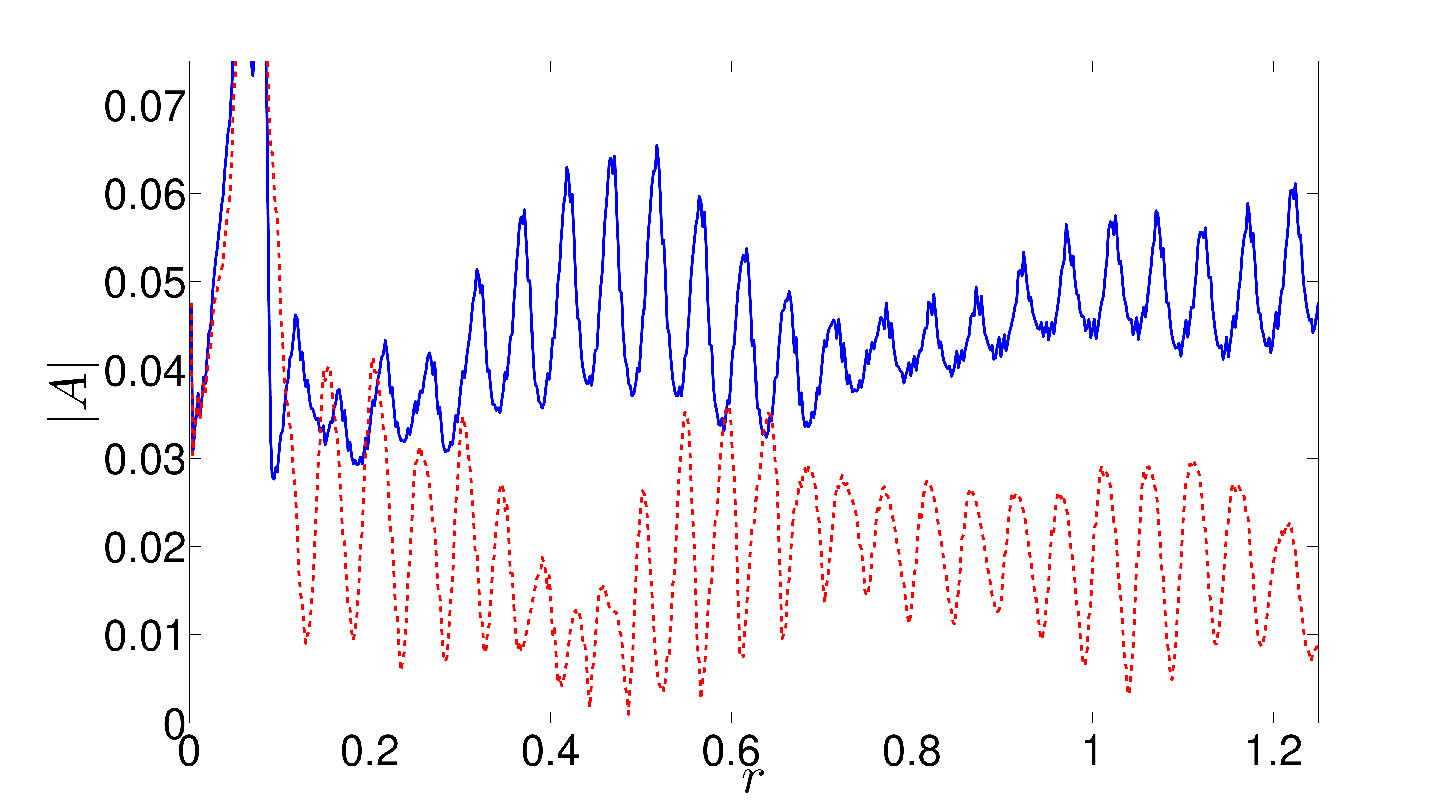}}
  \end{center}
  \caption{Radial velocity along the line $y=x$ (top) and
    azimuthal velocity along the $x$-axis (middle top) in a
    large-amplitude simulation with wave breaking, at $t=7t_{c}$, with
    $\tilde{f}_{r} = 15$. These are plotted (solid lines) together
    with the reconstructed solution using $A_{in}$ and $A_{out}$
    obtained from the method described in Appendix \ref{refcoeff} (dashed
    lines) -- these are well matched for $r \in [0.5,1.2]$. Bottom
    right panel shows $A_{in}$ (top line) and $A_{out}$ (bottom line)
    vs radius. There is significant absorption of IWs near the
    critical layer at $r\sim 0.1$, resulting in $|A_{out}| \ll
    |A_{in}|$. The bottom panel shows the reflection coefficient $\mathcal{R}$ versus radius. $\mathcal{R}
    \ll 1$ in the region where the decomposition works
    well.
  }
  \label{Fig:refcoeffnonlinearsim512fr100t209}
\end{figure}

Once the critical layer has formed, we find that a large fraction of
the IW angular momentum is absorbed near the centre. First, we
confirmed this naively by watching animations of the time dependence
of the velocity components. For both components, the wave pattern
moves outwards, which corresponds to inward propagating IGWs (see
\S~\ref{lineartheory}), so at least a significant fraction of the
solution is in the form of IWs. This is quantified by performing our
IW/OW decomposition. The results of this for a typical simulation are
plotted in Fig.~\ref{Fig:refcoeffnonlinearsim512fr100t209}, where
$\tilde{f_{r}} = 15$. The time has been chosen after the critical
layer has formed, and the mean flow has been accelerated near the
centre. The IW/OW wave decomposition does not work well near the
centre, as we might expect, since here the disturbance is primarily
the mean flow ($m=0$), though there are also other components. Several
wavelengths from the critical layer, in the region $0.5 < r < 1.2$,
the reconstructed wave solution matches the simulation output quite
well. The matching is much noisier than in
Fig.~\ref{Fig:linearsimtestt45fr10tstep10lowth}, since the solution
contains contributions from $m \ne 2$ wavenumbera and $\omega\ne 1$
frequencies in addition to $m=2$, $\omega=1$ waves. 

Fig.~\ref{Fig:refcoeffnonlinearsim512fr100t209} shows that the
amplitude of the IW decays as it propagates towards the centre. There
is significant absorption of wave angular momentum near the critical
layer, as the IW propagates through the mean shear. This results in
$|A_{out}| \ll |A_{in}|$, though the reflection is nonzero so
$|A_{out}| \ne 0$. This leads to $\mathcal{R} \ll 1$ in the region
where the decomposition works well, which implies that most of the
angular momentum in the IWs is absorbed near the centre -- also note
that the energy flux ratio $\propto \mathcal{R}^{2} \ll 1$. In
addition, the phase of the OW is perturbed with respect to the IW,
which inhibits the formation of standing waves. $\mathcal{R}$
oscillates with radius, mainly because the solution is composed of
some $\omega \ne 1$ and $m\ne 2$ components, which are not filtered by
our IW/OW decomposition.

When a single propagating wave approaches a critical layer, the
outcome has previously been found to depend on the ratio of the strength of nonlinear wave-wave
couplings to linear viscous and radiative damping. Nonlinear wave-wave
couplings occur over a timescale \citep{BookerBretherton1967}
\begin{eqnarray}
t_{NL} = O \left(k_{h}^{-\frac{2}{3}} |\partial_{r}U_{h}|^{-\frac{1}{3}} U_{r}^{-\frac{2}{3}}\right),
\end{eqnarray} 
and linear viscous and radiative damping occur over a timescale 
\begin{eqnarray}
t_{L}  = O \left(k_{h}^{-\frac{2}{3}} \nu^{-\frac{1}{3}} |\partial_{r}U_{h}|^{-\frac{2}{3}}\right).
\end{eqnarray} 
The ratio of these terms define the parameter (\citealt{Maslowe1986}; \citealt{Koop1981})
\begin{eqnarray}
\lambda \equiv \frac{t_{NL}}{t_{L}} \sim \left(\frac{\nu |\partial_{r}U_{h}|}{U^{2}_{r}}\right)^{1/3}.
\end{eqnarray}
where $k_{h}$ is the horizontal wavenumber, $\partial_{r}U_{h}$ is the typical shear in the mean flow, $U_{r}$ is a typical
radial velocity in the wave, and $\nu$ is a diffusivity -- which for
the centre of a star is likely to be primarily radiative diffusion
rather than viscosity, so $\nu$ will be primarily the thermal
conductivity $\kappa$.

Nonlinearity acts to promote energy transfer away from the critical
layer through the generation of daughter waves, and linear damping
acts to suppress this resonant wave production.  If $\lambda \gg 1$,
then the time required for nonlinear effects to become important is
long compared with that for linear damping to become important. In
this limit, there is negligible wave reflection, and nearly all
incoming wave energy is absorbed by the critical layer, as predicted
by \cite{Hazel1967}. In the opposite limit, when $\lambda \ll 1$,
nonlinear effects become manifest prior to the time when they become
suppressed by linear damping. In this case, the flow can be extremely
complicated, and nonlinearity in the critical layer region can lead to
wave reflection, with ampltiudes $O\left(\exp
\left(-\pi(Ri-1/4)^{1/2}\right)\right)$ or less
\citep{Breeding1971}. In this limit, nonlinear wave-wave couplings
lead to the generation of many smaller-scale daughter waves, some of
which propagate away from the critical layer, carrying a fraction of
the wave energy \citep{Fritts1979}.  Experiments of internal waves
approaching a critical layer have been performed by \cite{Koop1981}
and \cite{Koop1986}, in which $\lambda = O(1)$. For this value, the
effects of both nonlinearity and linear damping become manifest at
approximately the same time. Their laboratory experiments show that
wave reflection, in the form of daughter waves produced by nonlinear
couplings that propagate away from the critical layer, is suppressed
by viscosity for this value of $\lambda$.

Relating these results to our simulations, we find typical values of
the parameter $\lambda \sim 10^{-1}$ near the critical layer, due to
the explicit viscosity ($\nu = 2 \times 10^{-6}$) added in the code,
since $|\partial_{r}U_{h}| \sim O(10^{1})$ and $U_{r} = \tilde{u}_{r}
\sim 0.1$. In this limit nonlinearities are likely to become important
before viscous diffusion. Since this is the limit in which we would be
expected to find wave reflection, if any occurs at all, and we find
little reflection of waves from the critical layer, it is likely that
most of the IWs are absorbed near to the centre, and not
reflected. In any case, the reflected waves will not have the same
frequency and horizontal wavenumber as the primary wave. Instead,
reflected waves will be in the form of disturbances with smaller frequencies,
and therefore shorter radial wavelengths, as well as higher
$m$-values, as a result of wave-wave coupling
\citep{Hasselman1967}. Such disturbances will be more easily
dissipated by radiative diffusion, since the rate of energy
dissipation $\sim m^{3}/\omega^{4}$.

We conclude that if wave breaking and critical layer
formation occurs, it is probably reasonable to assume that the IWs are
entirely absorbed in the radiation zone, primarily near to the
critical layer. This is inferred from our simulations, as can be seen
in the example in Fig.~\ref{Fig:refcoeffnonlinearsim512fr100t209},
in which $|A_{out}| \ll |A_{in}|$. The result of this is that if wave
breaking and critical layer formation occurs, it is not possible for
global standing modes to develop in the radiation zone. It would then
be appropriate to calculate the tidal dissipation rate using the
method of GD98.

\section{Discussion}
\label{discussion}

\subsection{Orbital evolution of the planetary companion}

The main motivation for our work is to study the tidal $Q^{\prime}$
expected for solar-type stars, and in particular to connect this with
the survival of close-in extrasolar planets. 
If the amplitude of the tide is sufficient to excite IGWs that exceed the critical amplitude
for wave breaking at the centre, then our results find it
highly likely that the wave reflection will be imperfect, and a large
fraction of the incoming wave energy will be absorbed. 
It is then reasonable to estimate the amount of tidal dissipation by assuming
that the incoming waves are completely absorbed. This was first done
by GD98 for a nonrotating solar model, and we here reproduce the
relevant quantities to calculate $Q^{\prime}$ using their approach. 
The energy flux $F$ in ingoing IGWs excited at the
convective/radiative interface (at radius $r=r_{int}$) is (see GD98
Eq. 13)
\begin{eqnarray}
\label{energyflux}
F &=& \beta \mathcal{G} \omega^{\frac{11}{3}},
\end{eqnarray}
where $\omega = 2n$ is the frequency of IGWs excited by a planet in a
circular, nonsynchronous orbit,
\begin{eqnarray}
\beta &=&
\frac{3^{\frac{2}{3}}}{8\pi}\left[\Gamma\left(\frac{1}{3}\right)\right]^{2}
\left[l(l+1)\right]^{-\frac{4}{3}}
\end{eqnarray}
is a constant (for quadrupolar tides $l=2$), and
\begin{eqnarray}
\mathcal{G} &=& \left[\rho r^{5} \left|\frac{dN^{2}}{d \ln
  r}\right|^{-\frac{1}{3}}\left|\frac{\partial \xi_{r}}{\partial
  r}\right|^{2}\right]_{r = r_{int}} = \tilde{\mathcal{G}}\frac{2\pi \Psi^{2}}{\omega_{dyn}{^4}},
\end{eqnarray}
where $\tilde{\mathcal{G}}$ depends on the stellar properties at the interface between radiative and
convective regions, and 
\begin{eqnarray}
\frac{dN^{2}}{d \ln r}(r_{int})&\approx & -100 \omega^{2}_{dyn}, \\
\frac{\partial \xi_{r}}{\partial r}(r_{int}) &\approx & \sigma_{c}
\frac{\Psi}{\omega^{2}_{dyn}},
\end{eqnarray}
and $\sigma_{c}$ is a constant whose value depends primarily on the
thickness of the convection zone, and is equal to -1.2 for the
current Sun. The values of these quantities depend on the
stellar model. The amplitude of the largest tide for a
circular orbit is (e.g. OL04)
\begin{eqnarray}
  \Psi = \sqrt{\frac{3}{5}}\frac{m_{p}}{(m_{\star}+m_{p})}n^{2},
\end{eqnarray}
if the tidal potential has the form $\mathrm{Re}[\Psi r^{2}
  \tilde{P}^{m}_{2}(\cos \theta )e^{i(m\phi - \omega t)}]$, where
$\tilde{P}^{m}_{2}$ corresponds to the Legendre
polynomial of degree 2 normalised over $\theta$ (related to
the spherical harmonic $Y_{2,m}
= (1/\sqrt{2\pi}) \tilde{P}^{m}_{2}e^{im\phi}$).
Converting the energy flux $F$ to an angular momentum flux, and
assuming all wave angular momentum is deposited in the star, we
calculate a torque $T$, using the torque
formula given in \cite{Peale1999}, giving
\begin{eqnarray}
  T = \frac{m}{\omega}F =  \frac{9}{4}\frac{1}{Q^{\prime}_{\star}}\left(\frac{m_{p}}{m_{\star}+m_{p}}\right)^{2}\frac{m_{\star}R_{\star}^{2}}{\omega_{dyn}^{2}}n^{4},
\end{eqnarray}
where $m=2$. Then
\begin{eqnarray}
  Q^{\prime}_{\star} &=&
  \frac{15}{32}\frac{3^{\frac{2}{3}}}{\pi^{\frac{8}{3}}}\frac{1}{\left[\Gamma(\frac{1}{3})\right]^{2}}\frac{1}{\tilde{\mathcal{G}}}m_{\star}R_{\star}^{2}\omega_{dyn}^{2}P^{\frac{8}{3}} \\
  \label{Qfactorcalc}
  &\approx& 1.5 \times 10^{5} \left[\frac{P}{1 \;\mathrm{day}}\right]^{\frac{8}{3}},
\end{eqnarray}
where $P$ is the orbital period, which is twice the tidal period for
the diurnal component of the tide. Note that the exact value depends
on the stellar model adopted -- in particular the value of
$\tilde{\mathcal{G}}$, which is determined from the stellar properties
at the interface between radiative and convective regions, and the
thickness of the convection zone. The given value applies to a stellar
model of the current Sun (see below for a discussion of the variation
in this parameter for other stars). The largest uncertainty is likely
to be in $\left|\frac{dN^{2}}{d \ln r}\right|^{-\frac{1}{3}}$, since
this may depend on the modelling of convective overshoot. However,
uncertainties in $\tilde{\mathcal{G}}$ for the current Sun are much
less than an order of magnitude, so this is a good estimate of
$Q^{\prime}_{\star}$ that results from the mechanism described in this
paper, for our closest star.

We now estimate the orbital and stellar properties required to excite
waves that are sufficiently nonlinear near the centre for breaking to
occur, using the Appendix of OL07\footnote{Note that the ApJ version
of this paper has a misprint in equation (A3). The first term in the
square brackets should have the coefficient $\sqrt{6}$ replaced by
$1/\sqrt{6}$, since the given function does not go to zero as
$x\rightarrow 0$.} and the above energy flux.  The nonlinearity
parameter in OL07 is defined so that for $A > 1$, the wave overturns
the stratification during part of its cycle.  Equating the energy flux
in ingoing waves near the centre to Eq.~\ref{energyflux} gives
\begin{eqnarray}
\label{nonlinearity}
A &=&
\frac{3}{20}\frac{3^{\frac{5}{12}}\Gamma(\frac{1}{3})}{2^{\frac{1}{4}}\pi^{\frac{2}{3}}}\frac{\tilde{G}^{\frac{1}{2}}}{\omega_{dyn}^{2}}
\left(\frac{C^{5}}{\rho_{c}}\right)^{\frac{1}{2}}\left(\frac{m_{p}}{m_{\star}+m_{p}}\right) P^{\frac{1}{6}} \\
&\approx & 0.3 \left(\frac{\tilde{\mathcal{G}}}{\tilde{\mathcal{G}}_{\odot}}\right)^{\frac{1}{2}}\left(\frac{C}{C_{\odot}}\right)^{\frac{5}{2}}\left(\frac{m_{p}}{M_{J}}\right)\left(\frac{M_{\odot}}{m_{\star}}\right)\left(\frac{P}{1 
\;\mathrm{day}}\right)^{\frac{1}{6}},
\end{eqnarray}
where $\tilde{\mathcal{G}}_{\odot} \approx 2 \times 10^{47} \mathrm{kg}\,\mathrm{m}^2\mathrm{s}^{2/3}$ and $C_{\odot} \approx 8 \times 10^{-11}
\mathrm{m}^{-1}\mathrm{s}^{-1}$. If we assume that the primary
instability acting on 3D waves occurs for the minimum amplitude at which the 3D wave solution
overturns the stratification ($A > 1$), which we have found is indeed the case
in 2D, then we have the following criterion for wave breaking:
\begin{eqnarray}
\label{criterion}
\left(\frac{\tilde{\mathcal{G}}}{\tilde{\mathcal{G}}_{\odot}}\right)^{\frac{1}{2}}\left(\frac{C}{C_{\odot}}\right)^{\frac{5}{2}}\left(\frac{m_{p}}{M_{J}}\right)\left(\frac{M_{\odot}}{m_{\star}}\right)\left(\frac{P}{1 
\;\mathrm{day}}\right)^{\frac{1}{6}} \gtrsim 3.3.
\end{eqnarray}

A Jupiter-mass planet in a one-day orbit around the current Sun would
not raise tides of sufficient amplitude near the centre for breaking
to occur, and would likely survive, because it does not satisfy this
criterion. Note, however, that Jupiter, with its period of $P = 4332$d, does
satisfy this criterion. The waves excited by Jupiter in the Sun would
be of very low amplitude, but they are also of very low
frequency. This means that their wavelength is extremely short, so
the energy of these waves would be concentrated into an extremely small
volume near the centre of the star, if they were to reach it. However,
radiative diffusion is certain to damp these waves before they reach
the centre, since they are of such short wavelength. This process
would, in any case, contribute negligibly to the orbital evolution of
Jupiter.

We performed a study of the variation in the parameters
$\tilde{\mathcal{G}}$ and $C$ using
an extensive set of stellar models, with masses $0.5 \leq
m_{\star}/M_{\odot} \leq 1.1$, and ages that represent the range of
main-sequence ages expected for these stars. 
These were provided by J\o rgen Christensen-Dalsgaard,
and were computed using ASTEC \citep{JCD2008}. This involved integrating GD98 Eq.(3)
throughout the convection zone in each of these models, where $N\sim 0$, using a linear
shooting method, to determine $\sigma_{c}$. The results of this study
are that $\tilde{\mathcal{G}} \sim \tilde{\mathcal{G}}_{\odot}$ to within a
factor of 5 for all solar-type stars, throughout the
main-sequence age of each star. This is true even taking into account
the evolution of the position of the interface between convection and
radiation zones, and the resulting change in the density of the star
at the interface. The main
uncertainty in each model is probably $\left|\frac{dN^{2}}{d \ln
  r}\right|$ at the interface, since the slope of $N^{2}$ there is uncertain. However, this is
only raised to the -1/3 power, so this effect probably does not
contribute more than a factor of 2 uncertainty in
$\tilde{\mathcal{G}}$. Taken together with changes in stellar mass and
radius, we find that our estimate of $Q^{\prime}_{\star}$ in
Eq.~\ref{Qfactorcalc} is quite robust, if critical layer formation
occurs at the centre. This is approximately true for all stars within the mass range $0.5 \leq
m_{\star}/M_{\odot} \leq 1.1$, throughout their main-sequence lifetime.

The variation in $A$ is primarily dependent on $C$, since $A \propto
\tilde{\mathcal{G}}^{\frac{1}{2}}C^{\frac{5}{2}}$. The strong dependence on $C$
means that as a star evolves the tide could become nonlinear at a
critical age, since $C$ increases with evolution on the main-sequence (see
Fig.~\ref{Fig:NvsrModelS} for the evolution of $C$ with the age of the
Sun). For a given age, this parameter is found to be larger in more
massive stars, due to their greater central condensation. In addition, stars with
lower metallicity also have a greater central condensation for a
given age, and so have larger $C$ values over stars with higher
metallicity. Over the range of stars considered in this study, $C$ is
found to take values between $0.1-10 C_{\odot}$. This leads to a large
variation in $A$ values, for fixed orbital parameters. As a result,
this parameter is critical in determining
whether wave breaking occurs at the centre. We have stated that a
short-period Jupiter-mass planet does not satisfy Eq.~\ref{criterion}
around the current Sun. However, such a planet around a similar age $1.0 M_{\odot}$ star with
a metallicity $Z=0.01$ will cause wave breaking at the centre, since
$C$ is larger by a factor of 3. Thus,
there is a strong dependence of the breaking criterion on the stellar model, primarily through
the parameter $C$.

Tidal dissipation of the quadrupolar
tide raised in the star leads to evolution of the semi-major axis at
the rate
\begin{eqnarray}
\frac{\dot{a}}{a} = -\frac{9}{2}\left(\frac{R_{\star}}{a}\right)^{5}n \left[\frac{\mathrm{sgn}(2n-2\Omega)}{Q^{\prime}_{\star}}\right].
\end{eqnarray}
The inspiral time for a planet into
the current Sun is then 
\begin{eqnarray}
\label{inspiraltime}
\tau_{a} = -\frac{2}{21}\frac{a}{\dot{a}} \approx 2.3 \; \mathrm{Myr}
\left(\frac{M_{J}}{m_{p}}\right)\left(\frac{M_{\odot}}{m_{\star}}\right)^{\frac{5}{6}}
\left(\frac{R_{\odot}}{R_{\star}}\right)^{5}\left(\frac{P}{1 \;\mathrm{day}}\right)^{7}
\end{eqnarray}
since $\dot{a}/a \propto a^{-21/2}$. Note that the strong frequency
dependence of $Q^{\prime}_{\star}$ means that this mechanism could be very
important for short-period systems. This predicts that a
planet spiralling into its star will undergo rapid acceleration as it
migrates inwards. This is as a consequence not only of the reduction
in semi-major axis, but also the decrease in $Q^{\prime}_{\star}$ as the tidal
frequency increases with the inspiral. It must be noted that simple timescale estimates
do not accurately reflect the evolution if the orbit is eccentric,
inclined, or if the stellar spin is not much slower than the orbit
\citep{Barker2009}, but this estimate shows that this mechanism can be
very efficient in contributing to the tidal evolution of hot Jupiters on
the tightest orbits. Indeed, $\tau_{a} < 5$ Gyr for a
Jupiter-mass planet in an orbit of less than about
three days around the current Sun, if it were to cause wave breaking
near the centre.

We can crudely estimate the maximum orbital period of a planet that
can be pulled into the star by this process by
equating the moments of inertia of the radiation zone (which extends to
radius $R_{RZ}$), to that of the
orbit $\mu a^{2} = r_{g}^{2}m_{\star}$, giving
\begin{eqnarray}
P \simeq 1.8 \;\; \mathrm{days}
\left(\frac{m_{\star}}{M_{\odot}}\right)^{\frac{1}{2}}\left(\frac{R_{RZ}}{0.7R_{\odot}}
\right)\left(\frac{M_{J}}{\mu}\right)^{\frac{1}{2}},
\end{eqnarray}
where $\mu = \frac{m_{p}m_{\star}}{m_{p}+m_{\star}}$ is the reduced
mass and $r^{2}_{g}=0.076$ is the dimensionless radius of gyration for
a polytrope of index 3.
For a planet with such an orbital period, the whole of the radiation
zone must be spun up to cause the planet to completely spiral into the star. Once
the entire radiation zone has spun up to the orbital frequency, this
process becomes ineffective and the
corresponding tidal torque will vanish. However, the tidal torque 
is nonzero due to dissipation of the equilibrium tide by turbulent
convection, and so even if this process stops, it does not guarantee
the survival of the planet. In addition, magnetohydrodynamic coupling
between the convection and radiation zones could act to partially counteract
the spin-up of the interior, if the convection zone is spinning slower than the
orbit. Magnetic braking of the star through the interaction of its
magnetic field with a stellar wind acts to spin down the
convection zone, which would also gradually spin down the radiation zone
through these couplings. If the coupling between the convection and
radiation zones of the star is efficient, then our mechanism would again
become effective. In the case that the coupling timescale balances the
timescale for spin up of the radiation zone 
due to IGW absorption at a critical layer, then the planet would migrate into the star on the
magnetic braking timescale. A detailed study of these effects is not currently
possible, since there are many uncertainties, but this is worthy of
future consideration.

A Jupiter-mass planet on a
$\sim 1$ day orbit will spin up a substantial fraction
of the radiation zone on infall. If the ratio of the orbital moment of
inertia to the spin moment of inertia of the radiation zone $\gtrsim
1$, then the above process alone will be unable to cause the planet to spiral
into the star.

\subsection{Critical layer formation induced by radiative diffusion}

Near the centre of a star, radiative diffusion is the dominant linear
dissipation mechanism. If the waves are sufficiently attenuated by radiative diffusion on their
reflection from the centre, then over a sufficiently long time, they
may be able to spin up the innermost regions of the star to the angular pattern speed of the
tide $\Omega_{p}$, hence producing a critical layer. We have already observed
this process occuring in our simulations in
Fig.~\ref{Fig:FourierAnalysis2}, albeit with viscosity and not
radiative diffusion acting on the waves. Here we provide an
order-of-magnitude estimate of the timescale for this process.

If we assume that the wave is attenuated as it propagates from a
radius $R$, to the centre, and back to $R$ again, by a factor
$e^{-\alpha}$, i.e., that the energy flux is $F e^{-\alpha}$, then the
torque on this region of the star is given by the angular momentum
transferred to the mean flow, and is
\begin{eqnarray}
\label{radtorque}
T = \frac{m}{\omega} F (1-e^{-\alpha}).
\end{eqnarray}
Here we assume that $F$ is that given in Eq.~\ref{energyflux}, which
is probably reasonable when the response is non-resonant. 
The wave attenuation $\alpha$ is given by
\begin{eqnarray}
\alpha \approx 2\int_{0}^{R} \frac{\eta_{rad} k^{2}}{c_{g,r}}dr,
\end{eqnarray}
where $\eta_{rad}$ is the thermal conductivity, $k$ is the
wavenumber. We have $k\sim k_{r}$ except within the last wavelength
from the centre, and the radial group velocity $c_{g,r} \approx \frac{\omega}{k_{r}}$. The
thermal conductivity can be calculated from the properties of the appropriate
stellar model from
\begin{eqnarray}
\eta_{rad} = \frac{16\sigma T^{3}}{3 \kappa \rho^{2} c_{p}},
\end{eqnarray}
where $\sigma$ is the Stefan-Boltzmann constant, $\kappa$ is the
opacity, and $c_{p}$ is the specific heat at constant pressure. We can
take $\eta_{rad}$ to be constant over the inner $\sim 3 \%$ of
a star, to a first approximation, with a value $\approx 16.7
\mathrm{m}^{2} \mathrm{s}^{-1}$ in the case of the Sun. In addition, $k_{r}$ is roughly
constant with radius within this region, though diverges as $r\rightarrow 0$. We can reasonably
estimate
\begin{eqnarray}
\alpha \approx 2 \times 6^{\frac{3}{2}}
\eta_{rad}\frac{C^{3}}{\omega^{4}}R \approx 4\times 10^{-4}
\left(\frac{R}{R_{\odot}}\right)\left(\frac{P}{1 \mathrm{d}}\right)^{4}
\end{eqnarray}
where $R$ is the size of the region spun up by this process. 
This is because $k_{r} \sim
\frac{\sqrt{l(l+1)} N}{r \omega}$, if $\omega \ll N$, and $N
= C r$. We also note that the tidal frequency $\omega = 2\left(\frac{2\pi}{P}\right)$.
This means that the attenuation by radiative diffusion within the
innermost few percent of the star, is small for waves excited by
planets on one-day orbits. However, the wavelength of the waves
becomes shorter for longer period orbits, so their attenuation by
radiative diffusion is more efficient. Note that $\alpha
\gtrsim 1$ only when $P \gtrsim 8$ days, even when radiative diffusion over the entire radiation zone is
considered (in fact GD98, who made a more accurate
calculation by including the radial dependence of the wavenumber and
diffusivity, found that $\alpha \gtrsim 1$ for $P \gtrsim 11.6$ days). This means that the
waves excited by planets on short-period orbits with $P \lesssim 3$ days, whose
survival could be threatened by the process of wave breaking, will not be
significantly attenuated in traversing the radiation zone, according to linear
theory. It is therefore appropriate to ask whether they will break on reaching
the stellar centre.

To a first approximation, the central $\sim 3 \%$ of a star can be modelled as a uniform
density sphere, with the central density $\rho_{c}$. Its moment of
inertia is $I = \frac{4}{5}\pi R^{5}\rho_{c}$. Using
Eq.~\ref{radtorque}, we have the following differential equation
describing the spin evolution of the central regions: 
\begin{eqnarray}
I\frac{d\Omega}{dt} = \frac{m}{\omega} F (1-e^{-\alpha}).
\end{eqnarray}
Assuming that the system evolves slowly, which is probably true until
breaking occurs, we can take $F$, $\omega$, and $\alpha$, to be constant in time as the central
regions are spun up. This allows a straightforward solution, 
giving the resulting timescale to spin up the region of the
central wavelength to $\Omega_{p}$, in
the case of Jupiter orbiting the Sun with a one-day period (in which
$R\sim 0.01R_{\odot}$), of
\begin{eqnarray}
 t_{SU} = \frac{4}{5}\frac{\pi R^{5} \rho_{c}
  \Omega_{p}^{2}}{F(1-e^{-\alpha})} \approx 3 \; \mathrm{Myr}.
\end{eqnarray}
Note that this timescale strongly depends on the size of the region
that is spun up. When $\alpha \ll 1$, which is
valid for $P \lesssim 10$ days, $t_{SU} \propto P^{5/3}$.

In this estimate we have used Model S of the current Sun, so this value only applies to our star. 
However, we have repeated this calculation using the same set of
stellar models discussed in the previous section, with masses in the
range $0.5 \leq m_{\star}/M_{\odot} \leq 1.1$, and find that $t_{SU}
\lesssim 1$ Gyr for each of these models, for Jupiter orbiting the
star with a one-day period. 

This striking estimate indicates that $\textit{all}$ gas giants on
short-period orbits around G or K stars could eventually cause the formation of a critical
layer near the centre of the star, given sufficient time $\lesssim
O(1)$ Gyr. Once this has formed, we have found
in our simulations in \S \ref{Results}, that it is reasonable to assume
that the ingoing wave angular momentum flux is entirely absorbed near
the centre. Hence, our estimate
of $Q_{\star}^{\prime}$ in the previous section could apply to
$\textit{all}$ slowly rotating G and K stars. The effects of rotation
will complicate matters if the tidal frequency is less than twice the
spin frequency, as studied in OL07.

Unlike the mechanism of nonlinear wave breaking, which is the main
subject of this paper, the formation of a critical layer by radiative
diffusion requires the progressive spin-up of the region of the
central wavelength by a very gradual deposition of angular momentum.
This process could be interrupted by other mechanisms of angular
momentum transport that resist the development of differential
rotation, such as hydrodynamic instabilities or magnetic stresses. A
consideration of such effects will be required in order to determine
whether this process operates in reality.

\subsection{Long-term evolution of the radiation zone}

As the planet migrates inwards due to the IGWs it excites being
absorbed at a
critical layer, $\omega$ increases, since
$\dot{n} > 0$. If we assume that the tidal frequency increases from
$\omega_{1}$ at a time $t_{1}$ to $\omega_{2} >
\omega_{1}$ at a slightly later time $t_{2}$, then since $\omega_{2} >
\omega_{1}$, the waves
excited at $t_{2}$ no longer see the critical layer for
$\omega_{1}$ waves. If the change in frequency is sufficiently
small, we would still expect significant attenuation by
the shear as the waves approach the critical layer for $\omega_{1}$ waves. This would
transfer angular momentum from the waves to the mean flow, and may
spin up the region near the original critical layer to the pattern
speed for $\omega_{2}$ waves, hence producing a critical layer
for these waves. If this process occurs, then a weak radial differential
rotation profile could be set up in the radiation zone, with
$\frac{d\Omega(r)}{dr}>0$. Similarly, if the planet migrates outwards, then a
profile with $\frac{d\Omega(r)}{dr}<0$ could be set up. If the flow
does not have time to adjust to the change in frequency of
the forcing, and cannot spin up sufficiently to produce a critical layer for
$\omega_{2}$ waves, then the dissipation rate may be
reduced. However, it seems plausible that the change in the orbit will
be gradual enough so that IGWs reaching the centre will be significantly
attenuated. This is because their radial wavelengths get Doppler-shifted
downwards by the shear, making them more susceptible to radiative damping.

We have so far restricted our investigation to $m=2$ waves. If the
orbit of the planet is eccentric or inclined with respect to the
stellar equator, then IGWs with other $m$-values could be
excited. If these break, then they could each have their own critical
layer. It would be interesting to study the effects of tidal forcing
with several different frequencies and $m$-values to see how this
would affect the reflection of the different waves. In addition,
\cite{Rogers2006} found that IGWs excited by turbulent
convection at the top of the radiation zone, which had $1 < m < 15$
and typical frequencies $\omega \sim 20 \mu$Hz, could reach the centre with sufficient
amplitude to undergo breaking and spin up the central regions. If this
commonly occurs in solar-type stars, then we might expect the core
to be differentially rotating, even in the absence of tidal forcing. For most
$m$-values the resulting spin frequency of the central regions is probably
slower than that of the relevant pattern speed for tides raised by a close-in planet,
so we expect that this would minimally attenuate any tidally excited
low-$m$ waves. Nevertheless, the interactions of multiple waves near the centre
warrants further study.

\section{Conclusions}

We have presented a study of the fate of internal gravity waves approaching the centre of
a solar-type star, primarily using two-dimensional numerical simulations. A train of
internal gravity waves excited at the interface between the convection and radiation zones
propagates towards the centre. These waves break if they
reach the centre with steepness sufficient to overturn the
stratification, which in 2D corresponds to $\tilde{u}_{\phi} > \tilde{u}_{\phi,crit} \sim
0.5$. Once this occurs, nonlinear effects cause the subsequent
formation of a critical layer, as the waves transfer their angular 
momentum to the mean flow, bringing the central regions of the star
into corotation with the
tidal forcing. This acts as an absorbing barrier for subsequent ingoing
waves, which continue to be absorbed near
the critical layer, resulting in an expansion of the spatial extent of
the mean flow. The general picture of this process is that the star
spins up (or down) from the inside out, until either the planet has
plunged into the star, or the radiation zone of the star has spun up (or down) to match that
of the evolving orbit. 

In 2D, wave breaking occurs near the centre if $\tilde{u}_{\phi} \gtrsim
0.5$. The amplitude of gravity waves excited by
tidal forcing have this value near the centre of the current Sun if
\begin{eqnarray}
\left(\frac{m_{p}}{M_{J}}\right)\left(\frac{P}{1 
\;\mathrm{day}}\right)^{\frac{1}{6}} \gtrsim 3.3.
\end{eqnarray}
However, this value depends strongly on the stellar model, in
particular the value of $\frac{d N}{dr}$ at the centre, which varies
between different stars, and with main-sequence age. It is likely that
a short-period Jupiter-mass planet will not excite waves with sufficient amplitude
to cause breaking near the centre of the current Sun, since it does
not satisfy this criterion. There is also a dependence on the stellar
properties at the interface between convection and radiation zones
(see Eq.~\ref{criterion}). However, the uncertainties in these
parameters are expected to be markedly less than an order of
magnitude, given a stellar mass and approximate age (and metallicity).

By decomposing the numerical solutions into an ingoing and outgoing
wave, we find that \textit{if wave breaking occurs} most of the
angular momentum of the ingoing wave is absorbed near the centre, and
is not reflected. This has very important implications for tidal
dissipation in solar-type stars. We find that the assumption of GD98
and OL07, that tidally excited internal gravity waves approaching the
centre a star are not coherently reflected, and do not produce
standing modes, is appropriate. Neglecting the effects of rotation, we
use the calculations of GD98 (with the minor corrections of OL07) to estimate the modified tidal quality
factor that results from dissipation of these waves near the
centre. Its value is
\begin{eqnarray}
Q^{\prime}_{\star} \approx 1.5 \times 10^{5} \left[\frac{P}{1 \;\mathrm{day}}\right]^{\frac{8}{3}},
\end{eqnarray}
for the current Sun. This mechanism produces efficient tidal dissipation over a continuous
range of tidal frequencies, once wave breaking occurs. If wave breaking does not occur, the
wave is perfectly reflected from the centre, and global standing modes
can be set up in the radiation zone. In this case, efficient dissipation
only occurs when the tidal frequency becomes resonant with a global standing
mode, but this contributes negligibly to the
dissipation, since the system moves rapidly through these resonances
\citep{Terquem1998}.

From studying an extensive set of stellar models, this
estimate of $Q^{\prime}_{\star}$ is found to vary by no more than a
factor of 5 between all main-sequence stars, with masses in
the range $0.5-1.1 M_{\odot}$, at any stage in their main-sequence
lifetime. This means that our estimate is quite robust, and is
likely to apply to all G and K stars within the mass range $0.5-1.1
M_{\odot}$, at any stage in their main-sequence lifetime (as long as
they do not possess a convective core).

The strong frequency dependence of $Q^{\prime}_{\star}$ implies a rapid ($\sim
1$ Myr) and accelerating
inward migration of planets on the tightest
orbits around solar-type stars, if these planets excite waves
with sufficient amplitudes to cause breaking. 
These planets can spiral completely into their stars if their orbital
moment of inertia is smaller than the spin moment of inertia of the
radiation zone. If the planet has a larger moment of inertia than that of
the radiation zone, then the planet will spin up the entire radiation
zone, and this process will become ineffective in the absence of any
competing effects. Consequently, we predict that fewer 
hot Jupiters which satisfy the breaking criterion, 
with orbital periods of less than 2-3 days, 
will be found around solar-type stars. Planets with masses much larger
than Jupiter could transfer sufficient angular momentum to the
radiation zone before they are engulfed by the star. Further evolution
would then depend on the strength of the coupling between the radiation and
convection zones, and on magnetic braking of the star through the
interaction of its magnetic field with a stellar wind. The outcome
that results from the interaction of these effects is uncertain.

Our work is an extension of nonlinear mechanism for tidal dissipation
proposed by GD98, and contributes several new results. While their
model was applied to the circularisation of close binary stars, we
(following OL07) focus on the inward tidal migration of short-period
extrasolar planets. We have performed numerical simulations, which
clearly determine the criterion for wave breaking in two dimensions.
Furthermore, we have found that the resulting spin-up of the central
regions of the star leads to an ongoing tidal torque as incoming waves
are absorbed in a critical layer that migrates outwards. Our work
allows a calculation of the tidal quality factors in an extensive
range of stellar models when this process occurs. It also raises
issues for future investigation, particularly concerning processes of
angular momentum transport in the radiation zone and their interaction
with the wave breaking process.

Of the very close-in hot Jupiters currently observed, those around
WASP-12 \citep{Hebb2008}, WASP-18 \citep{Hellier2009} and OGLE-TR-56-b
\citep{Sasselov2003}, the host stars all have convective cores. This
means that the mechanism outlined in this paper will not work in such
stars, regardless of the mass of the planet. This is because the
convective cores would reflect internal gravity waves and prevent them from
reaching the centre, where they can break. For such stars, the tidal
dissipation is found to be weak \citep{Barker2009}, potentially
explaining the survival of those planets. 

A recently discussed hot Jupiter is CoRoT-2b \citep{Gillon2009}, for
which Spitzer IR observations have been reported. This system has a
$m_{p} \approx 3.5 M_{J}$ planet in a $P = 1.74$ d orbit around a star
with a radiative core, with mass $m_{\star} \approx 0.96M_{\odot}$.
The age of this star has been inferred to be young, due to the high Li
I abundance, its rapid rotation, and the strong Ca II H \& K lines,
though there are significant uncertainties. However, best estimates
give the main-sequence age $\lesssim 300$ Myr. At this young age, we
expect the nonlinearity of the gravity waves approaching the centre of
the star to be weak (see Eq.~\ref{nonlinearity}), since $\frac{d
  N}{dr}$ at the centre is likely to be small compared with that of
the current Sun (see Fig.~\ref{Fig:NvsrModelS}). This means that the
process that we have discussed is unable to operate, and based on
these considerations, we would expect $Q_{\star}^{\prime}$ to be
large, providing a potential explanation for the survival of this
planet.

In addition, a recently accepted paper reported the discovery of the
latest candidate for the shortest period transiting planet, around the
G-type star WASP-19 \citep{Hebb2010}. This planet has mass $m_{p} = 1.15
M_{J}$, in an orbit of $P = 0.78$ d, around a star of mass $m_{\star}
\approx 0.95 M_{\odot}$, and hence will contain a radiative core.
The stellar age is poorly constrained, but we find that the gravity
waves excited by this planet will not have sufficient amplitude to
cause wave breaking at the centre of the star, for all reasonably aged
stellar models of a similar mass star. This means that the planet will
not be subject to accelerating tidal decay through this process,
perhaps explaining its survival. Constraints on the tidal
$Q_{\star}^{\prime}$ for this system would then give us information on
alternative mechanisms of tidal dissipation, such as the dissipation
of the equilibrium tide by turbulent convection.

We have identified a further process by which efficient tidal
dissipation of internal gravity waves could occur. If the waves are
of too low amplitude to initiate breaking, the weak deposition of
angular momentum through radiative damping of the waves, can spin up
the region of the central wavelength over a timescale of millions of
years, until a critical layer is formed and the mechanism discussed
previously can continue. This process could provide efficient tidal
dissipation in solar-type stars perturbed by less massive companions.
However, it may be prevented by hydrodynamic or magnetorotational
instabilities, or by other effects that resist the development of
differential rotation near the centre of the star.

This work has pointed out the importance of nonlinear effects near the
centre of a solar-type star in contributing to tidal
dissipation. Linear theories of tides (e.g. OL04; OL07;
\citealt{SW2002}; \citealt{WS2002}) must include a correct
parametrisation of the effects of internal wave breaking near the
centre (like that in OL07 and GD98), if they are to correctly determine the efficiency of tidal
dissipation in solar-type stars. So far, this work has been only
two-dimensional, has included only a
single component of the tidal potential, neglected any
rotation profile in the radiation zone, as well as any possible
influence of magnetic fields. In future work, we plan to study the
stability of the nonlinear gravity wave solution in
\S~\ref{lineartheory}, to understand the initial breaking process in
more detail. In addition, three-dimensional effects could be studied in
an extension of our current simulations. This appears to be a
promising mechanism of tidal dissipation, which warrants further study.

\section*{Acknowledgments}
We would like to thank Geoffroy Lesur for kindly providing a version
of his Snoopy code, and with help modifying and running the code on
our problem, as well as in visualisation of the results. In addition, 
we are grateful to Michael McIntyre for pointing out useful
references, J\o rgen Christensen-Dalsgaard for kindly taking the
time to provide a set of
stellar interior models and helping with their interpretation, and
finally the referee, Jean-Paul Zahn, for a careful reading of the manuscript, and for his
helpful suggestions that have improved the clarity of several points in
the paper. A.J.B would like to thank STFC for a research
studentship.

\appendix

\section[]{Reflection coefficient calculation}
\label{refcoeff}

Since we have an exact analytic solution to the Boussinesq-type
problem (see \S\ref{lineartheory}), we can can deconstruct the numerical solution
into a sum of ingoing and outgoing waves. Doing this enables us to
quantify the amount of angular momentum absorbed as these waves approach and/or
reflect from the centre. The approach we use is now described.

A solution can be decomposed into the sum of an
ingoing and outgoing wave as (in the dimensionless units of \S \ref{lineartheory})
\begin{eqnarray}
\psi_{in}(r,\xi) &=& [J_{m}(r) + iY_{m}(r)]e^{i\xi}, \\
\psi_{out}(r,\xi) &=& [J_{m}(r) - iY_{m}(r)]e^{i \xi}, \\
\psi(r,\xi) &=& \mathrm{Re}\left[A_{in}\psi_{in}(r,\xi) + A_{out}\psi_{out}(r,\xi)\right],
\end{eqnarray}
where $\xi = m\phi - \omega t$. The velocity corresponding components
are
\begin{eqnarray}
u_{r}(r,\xi) &=& \mathrm{Re}\left[\frac{im}{r}A_{in}\psi_{in}(r,\xi) +
  \frac{im}{r}A_{out}\psi_{out}(r,\xi)\right] \\
 &=& \mathrm{Re}\left[v_{r}(r,\xi) + w_{r}(r,\xi)\right], \\
u_{\phi}(r,\xi) &=&
  \mathrm{Re}\left[-A_{in}\partial_{r}\psi_{in}(r,\xi) 
    -A_{out}\partial_{r}\psi_{out}(r,\xi)\right] \\
&=& \mathrm{Re}\left[v_{\phi}(r,\xi) +
  w_{\phi}(r,\xi)\right],
\end{eqnarray}
for general $m \in \mathbb{Z}$. From now on we choose $m=2$, which
corresponds with that of our forcing. These are steady, but
non-axisymmetric, in the frame rotating with $\Omega_{p}=\omega/m$. We assume that the
amplitudes are locally independent of $r$, so that we can ignore their radial
derivatives. Since the amplitudes $A_{in}$ and $A_{out}$ are in
general complex, we require 4 pieces of information from the
simulations to be able to calculate them. We choose to calculate these
from $u_{r}$ and $u_{\phi}$ given from the simulations at two
different azimuthal positions, around the same radial ring. We repeat
this process for 8 pairs of points around the ring, and take the mean
of these values to calculate $A_{in}$ and $A_{out}$. This involves solving 
\begin{eqnarray}
\mathrm{A}\mathbf{x} = \mathbf{b}
\end{eqnarray}
where $\mathbf{x} = (A_{in},A_{out})$ (which has 4 components for complex
amplitudes) and $\mathbf{b} =
(u_{r}(r,\xi_{1}),u_{\phi}(r,\xi_{1}),
u_{r}(r,\xi_{2}),u_{\phi}(r,\xi_{2}))$,
and the matrix
\begin{eqnarray}
\begin{tiny}
A = \left( 
\begin{array}{cccc}
\mathrm{Re}[v_{r}(r,\xi_{1})] &
\mathrm{Im}[v_{r}(r,\xi_{1})] &
\mathrm{Re}[w_{r}(r,\xi_{1})] &
\mathrm{Im}[w_{r}(r,\xi_{1})] \\
%%%%%%%%%%%%%%%%%%%%%%
\mathrm{Re}[v_{\phi}(r,\xi_{1})] &
\mathrm{Im}[v_{\phi}(r,\xi_{1})] &
\mathrm{Re}[w_{\phi}(r,\xi_{1})] &
\mathrm{Im}[w_{\phi}(r,\xi_{1})] \\
%%%%%%%%%%%%%%%%%%%%%%%%%
\mathrm{Re}[v_{r}(r,\xi_{2})] &
\mathrm{Im}[v_{r}(r,\xi_{2})] &
\mathrm{Re}[w_{r}(r,\xi_{2})] &
\mathrm{Im}[w_{r}(r,\xi_{2})] \\
%%%%%%%%%%%%%%%%%%%%%%%%%
\mathrm{Re}[v_{\phi}(r,\xi_{2})] &
\mathrm{Im}[v_{\phi}(r,\xi_{2})] &
\mathrm{Re}[w_{\phi}(r,\xi_{2})] &
\mathrm{Im}[w_{\phi}(r,\xi_{2})] \\
\end{array} \right)
\end{tiny}
\end{eqnarray}
for any two points with $\phi = \xi_{1}$ and $\phi = \xi_{2}$. If we
choose $\xi_{2}-\xi_{1} = \frac{(2n+1)\pi}{4}$ for $n \in \mathbb{Z}$,
this matrix is non-singular for all radii, and is related to the
Wronskian. This is calculated for all radial rings in the grid. We
verified this method on an analytic standing wave solution
(Eq.\ref{analytic}) using Mathematica, and wrote a Matlab code to
read in Snoopy/ZEUS data and compute $A_{in},A_{out}$ and the
reflection coefficient $\mathcal{R}$, which is defined in
Eq.~\ref{refcoeffdef}.  For perfect standing waves, $A_{in} =
A_{out}$, and $\mathcal{R}=1$. If the ingoing wave is entirely
absorbed at the centre, then $\mathcal{R}=0$.

The main reason for this approach is that it enables us to compute all
four unknowns (the real and imaginary parts of the complex amplitudes)
for each $r$, which requires $u_{r}$ and $u_{\phi}$ for two
azimuths. Integrating azimuthally over a ring to eliminate $m\ne2$
components, would leave only two unknowns, $u_{r}$ and $u_{\phi}$ --
which is not sufficient to determine the amplitudes. The disadvantage
of our approach is that $m\ne2$ components also contribute to the
amplitudes. This problem can be ignored if we trust the computed values of
$\mathcal{R}$, only where the solution is well described by the
linear solution i.e. far from the wave breaking and forcing regions.

When the solution has different frequency and wavenumber components
than $\omega=1$, $m=2$, the solution cannot be simply decomposed into
an ingoing and an outgoing $m=2$ wave solution with an amplitude that
is roughly independent of radius. In which case, the amplitudes of the
waves that are calculated from this method if the solution is
\textit{not} predominantly $m=2$ can oscillate wildly with radius,
because the solution is not a simple $m=2$ wave solution. If the
frequency (and hence radial wavelength) of the wave is different from
that of the chosen ingoing/outgoing wave solution, then we will also
not be able to match the solution exactly. One way of gauging how well
our decomposition into a single ingoing/outgoing wave is to plot the
solution reconstructed from the calculated $A_{in},A_{out}$ against
the numerical solution from Snoopy/ZEUS output. If these are very
different then we cannot make definitive conclusions about the
solution from these values.

\section[]{ZEUS comparison}
\label{ZEUS}
We confirm that the results are not dependent on the numerical method
by reproducing the basic results using a stripped down version of
ZEUS--2D\footnote{which has kindly been made freely available by
J.Stone at http://www.astro.princeton.edu/$\sim$jstone/zeus.html}
\citep{Stone1992}. ZEUS solves the equations of ideal compressible
hydrodynamics, using a simple Eulerian method based on
finite-differences, implemented using a covariant formalism, enabling
the use of non-cartesian orthogonal coordinate systems. For our
problem we solve the problem using cylindrical ($r,\phi$) coordinates,
which are the most natural to use for our problem. However, the
coordinate singularity at the origin requires that we cut out a small
region at the centre, on which we impose reflecting boundary
conditions. What may seem an advantage of this coordinate system, that
is the higher resolution at the centre, which is automatically present
when we use a uniform grid in $r$ and $\phi$, requires very short
timesteps when the resolution is increased, as a result of the CFL
stability constraint. This becomes prohibitive as we increase the
resolution of the grid to above $100\times 150$ in $r,\phi$
respectively, so only preliminary low resolution runs were performed
using this code. This is also because this code solves the
compressible equations, and therefore resolves sound waves. For our
problem we require a ratio of sound speed to radial group velocity of
gravity waves $\chi = c_{s}/c_{g,r}\sim 6\times 10^{3}$ ($\chi^{-1}$
is a measure of the importance of effects of compressibility), in
order to reproduce an equivalent set-up to that used in the Snoopy
code above, so most of the computational time is spent resolving sound
waves. In the paper we only analyse the Snoopy results, since they are
at a much higher resolution, but here describe our problem set-up in
ZEUS, for completeness. We use a circularly symmetric parabolic
density stratification, $\rho(r) = \rho_{0} - \rho_{2} r^{2}$ and
calculate the pressure ($p$) profile from hydrostatic equilibrium. We
solve the equations
\begin{eqnarray*}
  && D\rho = -\rho \nabla \cdot \mathbf{u} \\
 && D\mathbf{u} = -\frac{1}{\rho}\nabla p + \mathbf{g} + \Bigg\{
  \begin{array}{ll} 0, & r_{inner} \leq r < r_{force}, \\
    \mathbf{f}, & r_{force} \leq r < r_{damp}, \\
    \mathbf{d}, & r_{damp} \leq r < r_{box}, \\
  \end{array} \\
 && D \left(\frac{e}{\rho}\right) = -\frac{p}{\rho} \nabla \cdot
  \mathbf{u},
\end{eqnarray*} 
where $e=(\gamma-1)p$ is the specific internal energy of the gas,
$\mathbf{f} = - f_{r}\cos(2\phi-\omega t) \, \mathbf{e}_{r}$, and
$\mathbf{d} = - d(r)\mathbf{u}$. We choose $\gamma = 5/3$, as
appropriate for a monatomic ideal gas. Radial gravity has been
implemented as a source term in the radial momentum equation as
$\mathbf{g} = -g_{1}r\mathbf{e}_{r}$.  Both the inner and outer
boundaries have reflecting boundary conditions, and we also implement
a linear frictional damping in a region adjacent to the boundary to
prevent the reflection of (most of) the outgoing wave energy from the
outer boundary. A parabolic smoothing function $d(r) =
\left(\frac{r-r_{damp}}{r_{box}-r_{damp}}\right)^{2}$ is used in the
damping terms. We choose $r_{box} = 1.0$, $r_{damp} = 0.9$, $r_{force}
= 0.85$ and $r_{inner} = 0.01$. In the code we specify
$\rho_{0},\rho_{2},\omega,\lambda_{r} \, \& \, \tilde{f}_{r}$; the
other relevant parameters are calculated from these. Choosing
$\rho_{0} = 1.0,\rho_{2} = 0.1,\omega = 1.0,\lambda_{r} = 0.1$ and a
suitable value for $\tilde{f}_{r}$ is sufficient to fully specify the
problem.

A minimum value of $\chi = 6285$ is found from these initial
conditions. Such a high value is required for the wavelength of the
gravity waves to be $\lambda_{r} \simeq 0.1$, which allows $\sim 8$
wavelengths to be resolved within the grid. This value is not much smaller
than that appropriate at the centre of a solar-type ($\chi \sim
10^{4}-10^{5}$). We set up the initial conditions in such a way to
minimise this value given the above input parameters.

Calculations were performed in an inertial frame, though the results
were interpreted in a frame rotating with the angular pattern speed of
the tide, $\Omega_{p} = \omega/2$. In this rotating frame, the linear
wave solution is steady, which allows the instability to be easily
recognised as departures from a steady state.

With this resolution, there are some numerical errors near the inner
boundary. This results from the fact that we only remove a small
region near the centre, which is comparable with the size of a grid
cell. In addition, the code has no explicit viscosity or
thermal conduction, so we have less control over the scales of
dissipation, than in Snoopy. Nevertheless, ZEUS reproduces the same
basic results as the Snoopy code, which
indicates both that the effects of nonzero compressibility are not important in this problem,
and that our basic results are not dependent on the numerical method.

\setlength{\bibsep}{0pt}
\bibliography{tidbib}
\bibliographystyle{mn2e}
\end{document}